\documentclass[journal]{IEEEtran}
\usepackage{epsbox}
\usepackage[dvips]{graphicx,color}
\usepackage{latexsym}
\usepackage{amsmath}
\usepackage{amssymb}
\usepackage{array}
\begin{document}
\newcommand{\lcov}{
                     convex }

\newcommand{\cars}{s}
\newcommand{\sigNi}{\sigma_{N_i}^2}
\newcommand{\sigN}{\sigma_{N}^2}

\newcommand{\cef}{c_i}
\newcommand{\cefsq}{c_i^2}

\newcommand{\Iset}{L}
\newcommand{\Ntn}{\mbox{\boldmath $N$}}
\newcommand{\lNtn}{\mbox{\scriptsize\boldmath $N$}}
\newcommand{\Nitn}{\mbox{\boldmath $N$}_i}
\newcommand{\lNitn}{\mbox{\scriptsize\boldmath $N$}_i}
\newcommand{\Nost}{N_{0,t}}
\newcommand{\Nast}{N_{1,t}}
\newcommand{\Nlst}{N_{L,t}}

\newcommand{\tinbNs}{\tilde{N}
                    }
\newcommand{\tiNs}{\tilde{\mbox{\boldmath $N$}}
                  }
\newcommand{\tilNs}{\tilde{\mbox{\scriptsize\boldmath$N$}}
                   }
\newcommand{\Xotn}{\mbox{\boldmath $X$}_0}
\newcommand{\lXotn}{\mbox{\scriptsize\boldmath $X$}_0}
\newcommand{\Xost}{X_{0,t}}

\newcommand{\Xatn}{\mbox{\boldmath $X$}_1}
\newcommand{\Xast}{X_{1,t}}

\newcommand{\Xbtn}{\mbox{\boldmath $X$}_2}
\newcommand{\Xbst}{X_{2,t}}

\newcommand{\Xltn}{\mbox{\boldmath $X$}_L}
\newcommand{\Xlst}{X_{L,t}}
\newcommand{\Xlsn}{X_{L,n}}

\newcommand{\Xitn}{\mbox{\boldmath $X$}_i}
\newcommand{\lXitn}{\mbox{\scriptsize\boldmath $X$}_i}
\newcommand{\Xisa}{X_{i,1}}
\newcommand{\Xisb}{X_{i,2}}
\newcommand{\Xist}{X_{i,t}}
\newcommand{\Xisn}{X_{i,n}}

\newcommand{\Xtn}{\mbox{\boldmath $X$}}

\newcommand{\hatXotn}{\hat{\mbox{\boldmath $X$}}_0}
\newcommand{\hatXost}{\hat{X}_{0,t}}

\newcommand{\tiXs}{
                  {\mbox{\boldmath $Y$}}
                  }
\newcommand{\tilXs}{
                   {\mbox{\scriptsize\boldmath $Y$}}
                   }

\newcommand{\Zatn}{\mbox{\boldmath $Z$}_1}
\newcommand{\lZatn}{\mbox{\scriptsize\boldmath $Z$}_1}
\newcommand{\Zbtn}{\mbox{\boldmath $Z$}_2}
\newcommand{\lZbtn}{\mbox{\scriptsize\boldmath $Z$}_2}

\newcommand{\rdf}{J}  

\newcommand{\Dff}{-}  
\newcommand{\coS}{S^{\rm c}}

\newcommand{\piso}{\pi(1)} 
\newcommand{\pisi}{\pi(i)} 
\newcommand{\pisiadd}{\pi({i+1})} 
\newcommand{\pisimo}{\pi({i-1})} 
\newcommand{\pisj}{\pi(j)} 
\newcommand{\pisjmo}{\pi({j-1})} 
\newcommand{\pisjadd}{\pi({j+1})} 
\newcommand{\piss}{\pi(s)} 
\newcommand{\pissmo}{\pi({s-1})} 
\newcommand{\pios}{\pi} 

\newcommand{\pibi}{\pi(S_i)} 
\newcommand{\picobi}{\pi(S_{i}^{\rm c})}
\newcommand{\pibimo}{\pi(S_{i-1})} 
\newcommand{\picobimo}{\pi(B_{i-1}^{\rm c})}
\newcommand{\pibj}{\pi(S_j)} 
\newcommand{\picobj}{\pi(S_{j}^{\rm c} )}

\newcommand{\pibs}{\pi(S_L)} 
\newcommand{\pibsmo}{\pi(S_{L-1})} 

\newcommand{\piLam}{{\pi, L}} 

\newcommand{\maho}{many-help-one } 
\newcommand{\Mho}{Many-help-one } 
\newcommand{\oho}{one-helps-one } 

\newcommand{\D}{\mbox{\rm d}} %
\newcommand{\E}{\mbox{\rm E}} %

\newcommand{\conv}{\mbox{\rm conv}} %

\newcommand{\rsub}{\empty}

\newcommand{\EP}[1]
{
\ts \frac{1}{2\pi{\rm e}}{\rm e}^{\scriptstyle \frac{2}{n}h(#1)}
}
\newcommand{\CdEP}[2]
{
\ts \frac{1}{2\pi{\rm e}}{\rm e}^{\frac{2}{n}
h(\scriptstyle #1|\scriptstyle #2)}
}

\newcommand{\MEq}[1]{\stackrel{
{\rm (#1)}}{=}}
\newcommand{\MLeq}[1]{\stackrel{
{\rm (#1)}}{\leq}}
\newcommand{\MGeq}[1]{\stackrel{
{\rm (#1)}}{\geq}}
\newcommand{\MsL}[1]{\stackrel{
{\rm (#1)}}{<}}
\newcommand{\MgL}[1]{\stackrel{
{\rm (#1)}}{>}}
\newcommand{\MSub}[1]{\stackrel{
{\rm (#1)}}{\subseteq}}
\newcommand{\MSup}[1]{\stackrel{
{\rm (#1)}}{\supseteq}}
\newcommand{\ML}[1]{\stackrel{
{\rm (#1)}}{<}}
\newcommand{\MG}[1]{\stackrel{
{\rm (#1)}}{>}}
\newcommand{\MPreq}[1]{\stackrel{
{\rm (#1)}}{\preceq}}
\newcommand{\MSueq}[1]{\stackrel{
{\rm (#1)}}{\succeq}}

\arraycolsep 0.5mm

\newcommand{\bfig}{\begin{figure}[t]}
\newcommand{\efig}{\end{figure}}
\setcounter{page}{1}
\newenvironment{indention}[1]{\par
\addtolength{\leftskip}{#1}\begingroup}{\endgroup\par}
%
\newcommand{\namelistlabel}[1]{\mbox{#1}\hfill} 
\newenvironment{namelist}[1]{%
\begin{list}{}
{\let\makelabel\namelistlabel
\settowidth{\labelwidth}{#1}
\setlength{\leftmargin}{1.1\labelwidth}}
}{%
\end{list}}
%

%
\newcommand{\bc}{\begin{center}}  %
\newcommand{\ec}{\end{center}}
\newcommand{\befi}{\begin{figure}[h]}  %
\newcommand{\enfi}{\end{figure}}
\newcommand{\bsb}{\begin{shadebox}\begin{center}}   %
\newcommand{\esb}{\end{center}\end{shadebox}}
\newcommand{\bs}{\begin{screen}}     %
\newcommand{\es}{\end{screen}}
\newcommand{\bib}{\begin{itembox}}   %
\newcommand{\eib}{\end{itembox}}
\newcommand{\bit}{\begin{itemize}}   %
\newcommand{\eit}{\end{itemize}}
\newcommand{\defeq}{\stackrel{\triangle}{=}}
\newcommand{\qed}{\hbox{\rule[-2pt]{3pt}{6pt}}}
\newcommand{\beq}{\begin{equation}}
\newcommand{\eeq}{\end{equation}}
\newcommand{\beqa}{\begin{eqnarray}}
\newcommand{\eeqa}{\end{eqnarray}}
\newcommand{\beqno}{\begin{eqnarray*}}
\newcommand{\eeqno}{\end{eqnarray*}}
\newcommand{\ba}{\begin{array}}
\newcommand{\ea}{\end{array}}
\newcommand{\vc}[1]{\mbox{\boldmath $#1$}}
\newcommand{\lvc}[1]{\mbox{\scriptsize\boldmath $#1$}}

\newcommand{\wh}{\widehat}
\newcommand{\wt}{\widetilde}
\newcommand{\ts}{\textstyle}
\newcommand{\ds}{\displaystyle}
\newcommand{\scs}{\scriptstyle}
\newcommand{\vep}{\varepsilon}
\newcommand{\rhp}{\rightharpoonup}
\newcommand{\cl}{\circ\!\!\!\!\!-}
\newcommand{\bcs}{\dot{\,}.\dot{\,}}
\newcommand{\eqv}{\Leftrightarrow}
\newcommand{\leqv}{\Longleftrightarrow}
\newtheorem{co}{Corollary} 
\newtheorem{lm}{Lemma} 
\newtheorem{Ex}{Example} 
\newtheorem{Th}{Theorem}
\newtheorem{df}{Definition} 
\newtheorem{pr}{Property} 
\newtheorem{pro}{Proposition} 
\newtheorem{rem}{Remark} 

\newcommand{\lcv}{convex } 

\newcommand{\hugel}{{\arraycolsep 0mm
                    \left\{\ba{l}{\,}\\{\,}\ea\right.\!\!}}
\newcommand{\Hugel}{{\arraycolsep 0mm
                    \left\{\ba{l}{\,}\\{\,}\\{\,}\ea\right.\!\!}}
\newcommand{\HUgel}{{\arraycolsep 0mm
                    \left\{\ba{l}{\,}\\{\,}\\{\,}\vspace{-1mm}
                    \\{\,}\ea\right.\!\!}}
\newcommand{\huger}{{\arraycolsep 0mm
                    \left.\ba{l}{\,}\\{\,}\ea\!\!\right\}}}
\newcommand{\Huger}{{\arraycolsep 0mm
                    \left.\ba{l}{\,}\\{\,}\\{\,}\ea\!\!\right\}}}
\newcommand{\HUger}{{\arraycolsep 0mm
                    \left.\ba{l}{\,}\\{\,}\\{\,}\vspace{-1mm}
                    \\{\,}\ea\!\!\right\}}}

\newcommand{\hugebl}{{\arraycolsep 0mm
                    \left[\ba{l}{\,}\\{\,}\ea\right.\!\!}}
\newcommand{\Hugebl}{{\arraycolsep 0mm
                    \left[\ba{l}{\,}\\{\,}\\{\,}\ea\right.\!\!}}
\newcommand{\HUgebl}{{\arraycolsep 0mm
                    \left[\ba{l}{\,}\\{\,}\\{\,}\vspace{-1mm}
                    \\{\,}\ea\right.\!\!}}
\newcommand{\hugebr}{{\arraycolsep 0mm
                    \left.\ba{l}{\,}\\{\,}\ea\!\!\right]}}
\newcommand{\Hugebr}{{\arraycolsep 0mm
                    \left.\ba{l}{\,}\\{\,}\\{\,}\ea\!\!\right]}}
\newcommand{\HUgebr}{{\arraycolsep 0mm
                    \left.\ba{l}{\,}\\{\,}\\{\,}\vspace{-1mm}
                    \\{\,}\ea\!\!\right]}}

\newcommand{\hugecl}{{\arraycolsep 0mm
                    \left(\ba{l}{\,}\\{\,}\ea\right.\!\!}}
\newcommand{\Hugecl}{{\arraycolsep 0mm
                    \left(\ba{l}{\,}\\{\,}\\{\,}\ea\right.\!\!}}
\newcommand{\hugecr}{{\arraycolsep 0mm
                    \left.\ba{l}{\,}\\{\,}\ea\!\!\right)}}
\newcommand{\Hugecr}{{\arraycolsep 0mm
                    \left.\ba{l}{\,}\\{\,}\\{\,}\ea\!\!\right)}}

\newcommand{\hugepl}{{\arraycolsep 0mm
                    \left|\ba{l}{\,}\\{\,}\ea\right.\!\!}}
\newcommand{\Hugepl}{{\arraycolsep 0mm
                    \left|\ba{l}{\,}\\{\,}\\{\,}\ea\right.\!\!}}
\newcommand{\hugepr}{{\arraycolsep 0mm
                    \left.\ba{l}{\,}\\{\,}\ea\!\!\right|}}
\newcommand{\Hugepr}{{\arraycolsep 0mm
                    \left.\ba{l}{\,}\\{\,}\\{\,}\ea\!\!\right|}}

\newenvironment{jenumerate}
	{\begin{enumerate}\itemsep=-0.25em \parindent=1zw}{\end{enumerate}}
\newenvironment{jdescription}
	{\begin{description}\itemsep=-0.25em \parindent=1zw}{\end{description}}
\newenvironment{jitemize}
	{\begin{itemize}\itemsep=-0.25em \parindent=1zw}{\end{itemize}}
\renewcommand{\labelitemii}{$\cdot$}

\newcommand{\iro}[2]{{\color[named]{#1}#2\normalcolor}}
\newcommand{\irr}[1]{{\color[named]{Red}#1\normalcolor}}
\newcommand{\irg}[1]{{\color[named]{Green}#1\normalcolor}}
\newcommand{\irb}[1]{{\color[named]{Blue}#1\normalcolor}}
\newcommand{\irBl}[1]{{\color[named]{Black}#1\normalcolor}}

\newcommand{\irp}[1]{{\color[named]{Yellow}#1\normalcolor}}
\newcommand{\irO}[1]{{\color[named]{Orange}#1\normalcolor}}
\newcommand{\irBr}[1]{{\color[named]{Purple}#1\normalcolor}}
\newcommand{\IrBr}[1]{{\color[named]{Purple}#1\normalcolor}}
%
\title{
Many-Help-One Problem for Gaussian Sources with a Tree Structure
on Their Correlation
}
%
\author{Yasutada~Oohama
\thanks{Manuscript received xxx, 20xx; revised xxx, 20xx.}%
\thanks{Y. Oohama is with the Department of Information Science and 
Intelligent Systems, University of Tokushima, 
2-1 Minami Josanjima-Cho, Tokushima 770-8506, Japan.}
}
\markboth{
}
{Oohama: Gaussian Distributed Source Coding Systems}
%


\maketitle

\begin{abstract}
In this paper we consider the separate coding problem for $L+1$ 
correlated Gaussian memoryless sources. We deal with the case where $L$ 
separately encoded data of sources work as side information at the 
decoder for the reconstruction of the remaining source. The 
determination problem of the rate distortion region for this system is 
the so called many-help-one problem and has been known as a highly 
challenging problem. The author determined the rate distortion region in 
the case where the $L$ sources working as partial side information are 
conditionally independent if the remaining source we wish to reconstruct 
is given. This condition on the correlation is called the CI condition. 
In this paper we extend the author's previous result to the case where 
$L+1$ sources satisfy a kind of tree structure on their correlation. We 
call this tree structure of information sources the TS condition, which 
contains the CI condition as a special case. In this paper we derive an 
explicit outer bound of the rate distortion region when information 
sources satisfy the TS condition. We further derive an explicit 
sufficient condition for this outer bound to be tight.
In particular, we determine the sum rate 
part of the rate distortion region for the case where information 
sources satisfy the TS condition. 
For some class of Gaussian sources with the TS condition we
derive an explicit recursive formula of this sum rate part. 
\end{abstract}

\IEEEpeerreviewmaketitle

\section{Introduction}
In multi-user source networks separate coding systems of correlated
information sources are significant from both theoretical and practical
point of view. The first fundamental result on those coding systems
was obtained by Slepian and Wolf \cite{sw}. They considered 
a separate source coding system of two correlated information 
sources. Those two sources are separately encoded and sent 
to a single destination, where the decoder reconstruct 
the original sources. 

In the above source coding system, we can consider the situation, where
the decoder wishes to reproduce one of two sources. We call this source
the {\it primary source}. In this case the remaining source that we call
the {\it auxiliary source} works as a partial side information at the
decoder for the reconstruction of the primary source. Wyner \cite{w0},
Ahlswede and K\"orner \cite{ak} determined the admissible rate region
for this system, the set that consists of a pair of transmission rates
for which the primary source can be decoded with an arbitrary small error
probability.

We can naturally extend the system studied by Wyner, Ahlswede and 
K\"orner to the one where there are several separately encoded data of 
auxiliary sources serving as side informations at the decoder. The 
determination of the admissible rate region for this system is called 
the \maho problem. In this sense Wyner, Ahlswede and K\"orner solved the 
so called \oho problem. The \maho problem has been known as a highly 
challenging problem. 
To date, partial solutions given by K\"orner 
and Marton \cite{km}, Gelfand and Pinsker \cite{gp}, 
Oohama \cite{oh4},\cite{oh7}, and Tavildar {\it et al.} \cite{tvw} 
are known. 

Gelfand and Pinsker \cite{gp} studied an interesting case of 
the \maho problem. They determined the admissible rate region 
in the case, where the auxiliary sources are {\it conditionally 
independent} if the primary source is given. We hereafter say 
the above correlation condition on the information sources 
the CI condition. 

In Oohama \cite{oh4}, the author extended the \maho problem studied by 
Gelfand and Pinsker \cite{gp} to a continuous case. He considered the 
\maho problem for $L+1$ correlated memoryless Gaussian sources, where 
$L$ auxiliary sources work as partial side information at the decoder for 
the reconstruction of the primary source. The mean square error was 
adopted as a distortion criterion between the decoded output and the 
original primary source output.  The rate distortion region was defined 
by the set of all transmission rates for which the average distortion 
can be upper bounded by a prescribed level. In \cite{oh4}, the author 
determined the rate distortion region when information sources satisfy 
the CI condition. This result contains the author's previous works for 
Gaussian \oho problem \cite{oh1} and Gaussian CEO problem \cite{oh2}.    

The problem still remains open for Gaussian sources with general 
correlation. Pandya {\it et al.} \cite{pdya0} studied the general case 
and derived an outer bound of the rate distortion region using some 
variant of bounding technique the author \cite{oh1} used to prove the 
converse coding theorem for Gaussian \oho problem. However, their 
bounding method was not sufficient to provide a tight result.

In Oohama \cite{oh7}, the author extended the result of \cite{oh4}. 
He considered a case of correlation on Gaussian sources, where $L+1$ 
sources satisfy a kind of tree structure on their correlation. The author 
called this tree structure of information sources the TS condition. 
The TS condition contains the CI condition as a special case. 
In \cite{oh7}, the author derived an explicit outer bound of the 
rate distortion region for Gaussian sources satisfying TS condition. 
Furthermore, he had shown that for $L=2$, this outer bound coincides 
with the rate distortion region. The author also presented a sufficient 
condition for the outer bound to coincide with the rate distortion region. 

Subsequently, Tavildar {\it et al.} \cite{tvw} extended the TS
condition to a binary Gauss Markov tree structure condition. They
studied a characterization of the rate distortion region for Gaussian
source with the complete binary tree structure and succeeded in it. To
derive their result, they made the full use of the complete binary
tree structure of the source.  They further determined the rate
distortion region for Gaussian sources with general tree structure.

%

In Oohama \cite{oh7}, the analysis for matching condition of the rate
distortion region and the derived outer bound was not sufficient, so
that the author could not realize that there exists a part of the
rate distortion region where the outer bound derived by him coincides
with the rate distortion region.  In this paper we give a further
analysis on matching condition for the outer bound derived by Oohama
\cite{oh7} to coincide with the rate distortion region and derive a
condition much stronger than the matching condition in \cite{oh7}.
Through this analysis we obtain an insight on a way of examining the
sum rate part of the rate distortion region to show that for Gaussian
sources with the TS condition the minimum sum rate part of the outer
bound given by Oohama \cite{oh7} is tight. This result implies that in
Oohama \cite{oh7}, the author had already obtained an explicit
characterization of the sum rate part of the rate distortion region
before the work by Tavildar {\it et al.} \cite{tvw}.  On this optimal
sum rate we derive its explicit recursive formula for some class of
Gaussian sources with the TS condition.  Our formula contains the
result of Oohama \cite{oh2} for Gaussian CEO problem as a special
case.

The rest of this paper is organized as follows. 

In Section II, we present a problem formulation 
and state the previous works.

In Section III, we give our main result. We first derive an explicit 
outer bound of the rate distortion region when information sources 
satisfy the TS condition. This outer bound is essentially the same as 
the author's previous outer bound in \cite{oh7}, but it has a form more 
suitable than the previous one for analysis of a matching condition. 
Using the derived outer bound, we presented an 
explicit sufficient condition for the outer bound to coincide with the 
inner bound. 

In Section IV, we investigate the sum rate part of the rate distortion
region.  We show that for the outer bound in this paper and that in
\cite{oh7}, their sum rate parts coincide with the sum rate part of
the inner bound. Hence, in the case where information sources satisfy
the TS condition, we establish an explicit characterization of the
sum rate part of the rate distortion region. This optimal sum rate has
a form of optimization problem. For some class of the Gaussian source
with the TS condition, we solve this optimization problem to establish
an explicit recursive formula of the optimal sum rate. 

In Section V, we give the proofs of the results. Finally, in Section VI, 
we conclude the paper. 

\section{Problem Statement 
and Previous results} 

In this section we state the problem formulation and previous results. 
We first state some notations used throughout this paper. Let 
$\Phi=\{1,2,\cdots, |\Phi|\}$ and ${\cal A}_i, i \in \Phi $ 
be arbitrary sets. Consider a random variable $A_i, i\in \Phi$ 
taking values in ${\cal A}_i$. We write $n$ direct product of 
${\cal A}_i$ as 
$
{\cal A}_i^n \defeq 
\underbrace{{\cal A}_i\times\cdots \times {\cal A}_i}_{n}\,.
$ 
Let a random vector consisting of $n$ independent copies 
of the random variable $A_i$ be denoted by 
$
{\vc A}_i=A_{i,1}A_{i,2}$ $\cdots A_{i,n}.
$
We write an element of ${\cal A}_i^n$ as   
$
{\vc a}_i=a_{i,1}a_{i,2}$ $\cdots a_{i,n}.
$
Let $S$ be an arbitrary subset of $\Phi$. 
Let $A_S$ and ${\vc A}_S$ denote random 
vectors $(A_i)_{i\in S}$ and $({\vc A}_i)_{i\in S}$, 
respectively.   
Similarly, let $a_S$ 
denote a vector $(a_i)_{i\in S}$. 
When $S=\{k,k+1,\cdots,l \}$, we also 
use the notation ${A}_k^{l}$ for ${A}_S$ and use 
similar notations for other vectors or random variables. 
When $k=1$, we sometimes omit subscript 1. 
Throughout this paper all logarithms 
are taken to the natural.

\subsection{Formal Statement of the Problem}

Let $X_i, i=0,1,2,\cdots,L$ be correlated zero mean Gaussian random 
variables taking values in real lines ${\cal X}_i$. 
Let $\Lambda=\{1,2,$ $\cdots,L\}$. The CI condition Oohama 
\cite{oh4} treated corresponds to the case where ${X}_1,{X}_2,\cdots, 
{X}_L$ are independent if $X_0$ is given. In this paper we deal with the 
case where $X_1,$ $\cdots,{X}_L$ have some correlation 
when $X_0$ is given. Let $\left\{(\Xost,\Xast,\cdots, 
\Xlst)\right\}_{t=1}^{\infty}$be a stationary memoryless multiple 
Gaussian source. For each $t=1,$ $2,\cdots$, $(X_{0,t},X_{1,t},\cdots,$ 
$\!X_{L,t})\,$ obeys the same distribution as $(X_0,$ $\!X_1,\cdots$, 
$\!X_L)\,$. 

\bfig
\setlength{\unitlength}{1.05mm}
\begin{picture}(80,56)(8,-4)
\put(10,40){\framebox(6,6){$X_0$}}
\put(10,25){\framebox(6,6){$X_1$}}
\put(13,15){$\vdots$}
\put(10, 5){\framebox(6,6){$X_L$}}

\put(21,46){${\Xotn}$}
\put(16,43){\vector(1,0){13}}

\put(21,31){${\Xatn}$}
\put(16,28){\vector(1,0){13}}

\put(23,19){$\vdots$}

\put(21,11){${\Xltn}$}
\put(16,8){\vector(1,0){13}}

\put(29,40){\framebox(6,6){$\varphi_0$}}
\put(37,46){$\varphi_0({\Xotn})$}

\put(29,25){\framebox(6,6){$\varphi_1$}}
\put(37,31){$\varphi_1({\Xatn})$}

\put(32,15){\vdots}
\put(40,19){\vdots}

\put(29,5){\framebox(6,6){$\varphi_L$}}
\put(37,11){$\varphi_L({\Xltn})$}

\put(35,43){\line(1,0){15}}
\put(50,43){\vector(1,-1){15}}

\put(35,28){\vector(1,0){30}}

\put(35,8){\line(1,0){15}}
\put(50,8){\vector(3,4){15}}

\put(65,25){\framebox(6,6){$\psi$}}
\put(71,28){\vector(1,0){10}}
\put(82,27){$\hatXotn$}
\end{picture}
\vspace*{-5mm}
\caption{
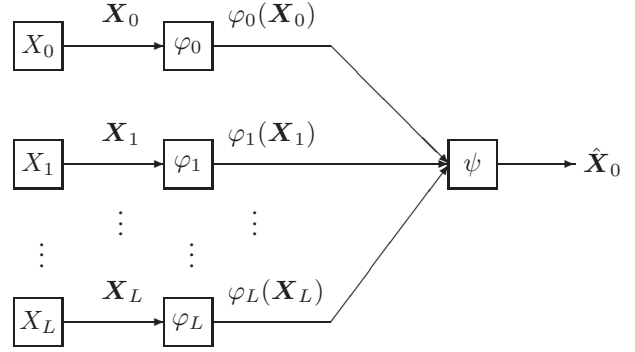
Communication system 
with $L$ side informations at the decoder.
}\label{fig:Fig1}
\vspace*{-7mm}
\efig

The multiterminal source coding system treated in this paper 
is depicted in Fig. \ref{fig:Fig1}. For each $i=0,1,\cdots,L $, 
the data sequence $\Xitn$ is separately 
encoded to $\varphi_i(\Xitn)$ 
by encoder function $\varphi_i$.
The encoded data $\varphi_i(\Xitn), i=0,1,\cdots,L $ are sent 
to the information processing center, where the decoder
observes them and outputs the estimation $\hatXotn$ of $\Xotn$ 
by using the decoder function $\psi$. The encoder 
functions $\varphi_i\,,i=0,1, 
\cdots, L$ are defined by 
\beq
\varphi_i: {\cal X}_i^n  \to {\cal M}_i=
\left\{1,2,\cdots, M_i\right\}  
\label{eqn:enc}
\eeq
and satisfy rate constraints 
\beq
\frac{1}{n}\log M_i \leq R_i+ \delta
\label{eqn:rate1}
\eeq
where $\delta$ is an arbitrary prescribed positive number. 
The decoder function $\psi$ is defined by  
\beq
\psi: {\cal M}_0 \times {\cal M}_1 \times \cdots\times {\cal M}_L 
\to {\cal X}_0^n\,.
\label{eqn:dec}
\eeq

Denote by ${\cal F}_{\delta}^{(n)}(R_0,$ $\!R_1,\cdots, R_L)$ the set
that consists of all the $(L+2)$ tuple of encoder and decoder
functions $(\varphi_0,\varphi_1,\cdots,$ $\!\varphi_L, \psi)$
satisfying (\ref{eqn:enc})-(\ref{eqn:dec}). 
Let $d(x,\hat{x})$ $=(x-\hat{x})^2$, $(x,\hat{x})\in {\cal X}_0^2 $ 
be a square distortion measure. For $\Xotn$ and
its estimation $\hatXotn=\psi(\varphi_0(\Xotn), \varphi_1(\Xatn),$
$\cdots,\varphi_L(\Xltn))$, define the average distortion by
\beqno
\Delta(\Xotn,\hatXotn)
\defeq \frac{1}{n} \sum_{t=1}^n \E d(X_{0,t},\hat{X}_{0,t})\,.
\eeqno
For a given $D>0$, the rate vector $(R_0,R_1,\cdots, R_L)$ is 
{\it admissible} if for any positive $\delta>0$ and any $n$ with $n\geq
n_0(\delta)$, there exists
$(\varphi_0,\varphi_1,\cdots,\varphi_L,\psi)\in$ $\!{\cal F}_{\delta}^{(n)} 
(R_0,$ $\!R_1,\cdots, R_L)$ such that
$\Delta(\Xotn,\hatXotn) \leq D + \delta $. Let ${\cal R}_{\Iset}(D)$ denote
the set of all the admissible rate vector. 
Our aim is to characterize ${\cal R}_{\Iset}(D)$ in an explicit form.
On a form of ${\cal R}_{\Iset}(D)$, we have a particular interest 
in its sum rate part. To examine this quantity, define 
$$ 
R_{{\rm sum}, L}(D,R_0)\defeq \min_{(R_0,R_1,\cdots,R_L)\in {\cal R}_{L}(D)}
\left\{\sum_{i=1}^{L}R_i \right\}\,.
$$
To determine $R_{{\rm sum}, L}(D, R_0)$ in an explicit form 
is also of our interest. 

By the rate-distortion theory for single Gaussian 
sources, when 
$R_0\geq \frac{1}{2}\log^+[\frac{\sigma_{X_0}^2}{D}]$, 
$R_1=R_2=$$\cdots=R_L=0$ is admissible. 
Here $\log^{+}a= \max\{ \log a, 0\}$. 
Hence, we have 
\beqno
& &{\cal R}_L(D)\cap
\left\{
R_0\geq \ts \frac{1}{2}\log^+[\frac{\sigma_{X_0}^2}{D}]
\right\}
\\
&=&\left\{(R_0,R_1,\cdots,R_L):\right.
\ba[t]{l}
R_0
\geq \frac{1}{2}\log^+[\frac{\sigma_{X_0}^2}{D}]\\ 
\left. R_i\geq 0, i\in \Lambda
\right\}.
\ea
\eeqno
Throughout this paper we assume that $D\leq\sigma_{X_0}^2$ 
and $R_0<$ $\frac{1}{2}\log[\frac{\sigma_{X_0}^2}{D}]$.

\subsection{Tree Structure of Gaussian Sources}

In this subsection we explain the tree structure of Gaussian source 
which is an important class of correlation. 
Consider the case where the $L+1$ random variables 
$X_0,X_1,\cdots, X_L$ satisfy the following
correlations: 
\beqa
&& \left.
\ba{l} 
Y_0=X_0,
\vspace{1mm}\\
Y_{l}=Y_{l-1} + Z_{l}, 1\leq l \leq L, 
\vspace{1mm}\\
X_l=Y_{l}+N_l, 1\leq l \leq L-1, 
\vspace{1mm}\\
X_L=Y_{L}, N_L=Z_L,
\ea
\right\}
\label{eqn:tsaa}
\eeqa 
where $Z_i, i \in \Lambda$  are $L$ independent Gaussian 
random variables with mean 0 and variance $\sigma_{Z_i}^2$
and $N_i, i=1,2, $ $\cdots, L-1$ are $L-1$ independent Gaussian random 
variables with mean 0 and variance $\sigNi$. 
We assume that $Z^{L}$ is independent of $X_0$ and that 
$N^{L-1}$ is independent of $X_0$ and $Z^{L}$. 
We can see that the above $(X_0,X_1,\cdots, X_L)$ has a kind of 
tree structure(TS). We say that the source $(X_0,X_1,\cdots, X_L)$ 
satisfies the TS condition when it satisfies (\ref{eqn:tsaa}). 
The TS condition contains the CI condition as a special 
case by letting $\sigma_{Z_i}, i=1,2,\cdots, L-1$ be zero. 
Let $S$ be an arbitrary subset of $\Lambda$. The TS condition 
is equivalent to the condition that for $S\subseteq \Lambda$, 
the random variables $X_S,(X_0,Z^{L-1}),X_{\coS}$ 
form Markov chains $X_S \to (X_0,Z^{L-1}) \to X_{\coS}$ 
in this order. The TS and CI conditions 
in the case of $L=4$ are shown in Fig. \ref{fig:Fig2} 
and \ref{fig:Fig3}, respectively.
\bfig
\setlength{\unitlength}{1.05mm}
\begin{picture}(65,52)(2,0)

\put(41,19){\vector(-3,-2){12}} 
\put(41,19){\vector(3,-2){12}} 
\put(41,   19.25){\line(0,1){1.5}}
\put(40.25,20   ){\line(1,0){1.5}}
\put(41,   20   ){\circle{2.0}}
\put(46,20){\vector(-1,0){4}}
\put(46,19){$Z_3$}
%
\put(53,    9.25){\line(0,1){1.5}}
\put(52.25,10   ){\line(1,0){1.5}}
\put(53,   10   ){\circle{2.0}}
\put(58,10){\vector(-1,0){4}}
\put(58, 9){$N_4$}
\put(53,9){\vector(0,-1){4}}
\put(51.5,2){$X_4$}
%
\put(29,    9.25){\line(0,1){1.5}}
\put(28.25,10   ){\line(1,0){1.5}}
\put(29,   10   ){\circle{2.0}}
\put(34,10){\vector(-1,0){4}}
\put(30,15){$Y_3$} \put(48,15){$Y_3$}
\put(34,9){$N_3$}
\put(29,9){\vector(0,-1){4}}
\put(27.5,2){$X_3$}
%
\put(29,29){\vector(-3,-2){12}} 
\put(29,29){\vector(3,-2){12}} 
\put(29,   29.25){\line(0,1){1.5}}
\put(28.25,30   ){\line(1,0){1.5}}
\put(29,   30   ){\circle{2.0}}
\put(34,30){\vector(-1,0){4}}
\put(34,29){$Z_2$}
%
%
%

\put(17,   19.25){\line(0,1){1.5}}
\put(16.25,20   ){\line(1,0){1.5}}
\put(17,   20   ){\circle{2.0}}
\put(22,20){\vector(-1,0){4}}
\put(18,25){$Y_2$}\put(36,25){$Y_2$}
\put(22,19){$N_2$}
\put(17,19){\vector(0,-1){4}}
\put(15.5,12){$X_2$}
%
\put(17,45){\vector(0,-1){4}}
\put(15.5,46.5){$X_0$}
%
\put(17,39){\vector(-3,-2){12}} 
\put(17,39){\vector(3,-2){12}} 
\put(17,   39.25){\line(0,1){1.5}}
\put(16.25,40   ){\line(1,0){1.5}}
\put(17,   40   ){\circle{2.0}}
\put(22,40){\vector(-1,0){4}}
\put(22,39){$Z_1$}
%
\put(5,   29.25){\line(0,1){1.5}}
\put(4.25,30   ){\line(1,0){1.5}}
\put(5,   30   ){\circle{2.0}}
\put(10,30){\vector(-1,0){4}}
 \put(6,35){$Y_1$} \put(24,35){$Y_1$}
\put(10,29){$N_1$}
\put(5,29){\vector(0,-1){4}}
\put(3.5,22){$X_1$}
\end{picture}
\vspace*{-2mm}
\caption
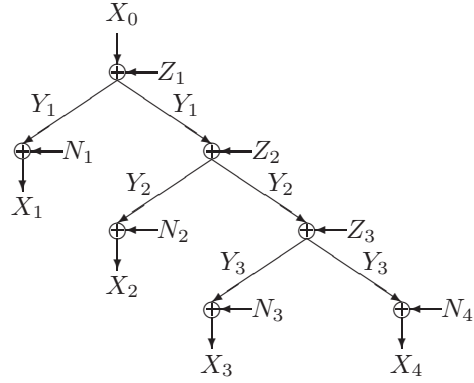
{TS condition in the case of $L=4$.}
\label{fig:Fig2}
\vspace*{-5mm}
\efig

\bfig
\setlength{\unitlength}{1.05mm}
\begin{picture}(65,52)(2,0)

\put(41,19){\vector(-3,-2){12}} 
\put(41,19){\vector(3,-2){12}} 
\put(41,   19.25){\line(0,1){1.5}}
\put(40.25,20   ){\line(1,0){1.5}}
\put(41,   20   ){\circle{2.0}}
\put(46,20){\vector(-1,0){4}}
\put(46,19){$Z_3=0$}
%
\put(53,    9.25){\line(0,1){1.5}}
\put(52.25,10   ){\line(1,0){1.5}}
\put(53,   10   ){\circle{2.0}}
\put(58,10){\vector(-1,0){4}}
\put(58, 9){$N_4$}
\put(53,9){\vector(0,-1){4}}
\put(51.5,2){$X_4$}
%
\put(29,    9.25){\line(0,1){1.5}}
\put(28.25,10   ){\line(1,0){1.5}}
\put(29,   10   ){\circle{2.0}}
\put(34,10){\vector(-1,0){4}}
\put(30,15){$X_0$} \put(48,15){$X_0$}
\put(34,9){$N_3$}
\put(29,9){\vector(0,-1){4}}
\put(27.5,2){$X_3$}
%
\put(29,29){\vector(-3,-2){12}} 
\put(29,29){\vector(3,-2){12}} 
\put(29,   29.25){\line(0,1){1.5}}
\put(28.25,30   ){\line(1,0){1.5}}
\put(29,   30   ){\circle{2.0}}
\put(34,30){\vector(-1,0){4}}
\put(34,29){$Z_2=0$}
%
%
%

\put(17,   19.25){\line(0,1){1.5}}
\put(16.25,20   ){\line(1,0){1.5}}
\put(17,   20   ){\circle{2.0}}
\put(22,20){\vector(-1,0){4}}
\put(18,25){$X_0$}\put(36,25){$X_0$}
\put(22,19){$N_2$}
\put(17,19){\vector(0,-1){4}}
\put(15.5,12){$X_2$}
%
\put(17,45){\vector(0,-1){4}}
\put(15.5,46.5){$X_0$}
%
\put(17,39){\vector(-3,-2){12}} 
\put(17,39){\vector(3,-2){12}} 
\put(17,   39.25){\line(0,1){1.5}}
\put(16.25,40   ){\line(1,0){1.5}}
\put(17,   40   ){\circle{2.0}}
\put(22,40){\vector(-1,0){4}}
\put(22,39){$Z_1=0$}
%
\put(5,   29.25){\line(0,1){1.5}}
\put(4.25,30   ){\line(1,0){1.5}}
\put(5,   30   ){\circle{2.0}}
\put(10,30){\vector(-1,0){4}}
 \put(6,35){$X_0$} \put(24,35){$X_0$}
\put(10,29){$N_1$}
\put(5,29){\vector(0,-1){4}}
\put(3.5,22){$X_1$}
\end{picture}
\vspace*{-2mm}
\caption
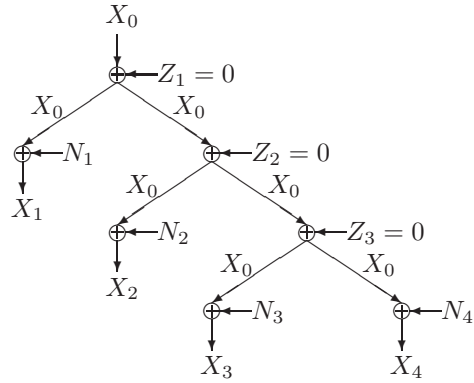
{CI condition in the case of $L=4$.}
\label{fig:Fig3}
\vspace*{-5mm}
\efig

\subsection{
Previous Results 
}

In this subsection we state the previous results on 
the determination problem of ${\cal R}_L(D)$. 
Let ${U}_i, i=0,1,\cdots, L$ be random variables 
taking values in real lines ${\cal U}_i$. 
For $S\subseteq \Lambda$, define 
$$
{\cal G}(D)\defeq
\ba[t]{l}
\left\{(U_0,U^L) \right. : 
  (U_0,U^L)\mbox{ is a Gaussian random }
  \vspace{1mm}\\
  \ba[t]{l}
  \mbox{vector that satisfies}
  \vspace{1mm}\\
  U^L\to X^L \to X_0 \to U_0 
  \vspace{1mm}\\
  U_S\to X_{S\cup \{0\}} \to X_{\coS}\to U_{\coS} 
  \vspace{1mm}\\
  \mbox{for any }S \subseteq \Lambda \,\mbox{ and }
  {\rm E}[X_0-\tilde{\psi}(U_0, U^L)]^2\leq D
  \vspace{1mm}\\
  \mbox{for some linear mapping }
  \tilde{\psi}:{\cal U}_0\times {\cal U}^L
  \to {\cal X}_0\,. 
  \left. \right\}\,,
  \ea
\ea
$$
where $\coS \defeq \Lambda-S$. Let 
$$
\pi=\left(
\ba{ccccc}
    1 &\cdots&    i &\cdots&    L\\ 
\pi(1)&\cdots&\pi(i)&\cdots&\pi(L)
\ea
\right)
$$
be an arbitrary permutation on $\Lambda$ and $\Pi$ be a set of
all permutations on $\Lambda$. For $S \subseteq \Lambda$, 
we set
$
\pi(S)\defeq\{\pi(i)\}_{i\in S}\,.
$
Define $L$ subsets 
$S_i, i=1,2,\cdots, L$ of $\Lambda$ by
$S_{i}\defeq $
$\{i,i+1,$ 
$\cdots, L\}\,.$ Set
\beqno
\tilde
{\cal R}_{\piLam}(D)
&\defeq&\ba[t]{l}
\left\{(R_0,R_1,\cdots, R_L) \right. : \mbox{There exists a random}
\vspace{1mm}\\
  \ba[t]{l}
  \mbox{vector }(U_0,U^L)\in {\cal G}(D)
  \mbox{ such that }
  \vspace{1mm}\\
  R_0 \geq I(X_0;U_0|U^L)
  \vspace{1mm}\\
  R_{\pisi}\geq I(X_{\pisi};U_{\pisi}|
            U_{\picobi})
  \vspace{1mm}\\
  \mbox{$\quad$ for }i=1,2, \cdots, L 
  \left. \right\}\,,
  \ea
\ea\\
\tilde
{\cal R}_{\Iset}^{{(\rm in)}}(D)&\defeq &
\conv\left\{
     \bigcup_{\pi\in \Pi}\tilde{\cal R}_{\piLam}^{({\rm in})}(D)
     \right\}\,,
\eeqno
where $\conv\{A\}$ denotes a convex hull of the set $A$. 
Then, we have the following. 
\begin{Th}[Oohama \cite{oh4}]\label{th:direct}
For Gaussian sources with general correlation 
$$\tilde{\cal R}_{\Iset}^{{(\rm in)}}(D)
\subseteq {\cal R}_{\Iset}(D)\,.$$ For Gaussian sources 
with the CI condition the inner bound 
$\tilde{\cal R}_{\Iset}^{{(\rm in)}}(D)$ 
is tight, that is 
$$
\tilde{\cal R}_{\Iset}^{{(\rm in)}}(D)
={\cal R}_{\Iset}(D)\,.
$$
\end{Th}

The above inner bound $\tilde{\cal R}_{L}^{({\rm in})}(D)$ can be
regarded as a variant of the inner bound which is well known as the
inner bound of Berger \cite{bt} and Tung \cite{syt}. 
Theorem \ref{th:direct} contains the solution that 
Oohama \cite{oh1} obtained to the \oho
problem for Gaussian sources as a special case. 
When $R_0=0$, the second result of Theorem \ref{th:direct} 
has some implications for the Gaussian CEO problem studied 
by Viswanathan and Berger \cite{vb} 
and Oohama \cite{oh2} and source coding problem for multiterminal 
communication systems with a remote source investigated
by Yamamoto and Itoh \cite{yam0} and Flynn and Gray \cite{fg}.

The notion of TS condition for Gaussian sources was first introduced
by Oohama \cite{oh7}. Tavildar {\it et al.} \cite{tvw} extended the TS
condition to a binary Gauss Markov tree structure condition.  They
studied a full characterization of the rate distortion region 
for Gaussian sources with a binary tree structure.
In the next section we shall state the results of 
Tavildar {\it et al.} \cite{tvw} and compare them 
with our results.

%
%

\section{Results on the Rate Distortion Region}

In this section, we state our main results on inner and outer bounds 
of ${\cal R}_L(D)$ in the case where $(X_0,X_1, $ $\cdots, X_L)$ 
satisfies the TS condition. 

\subsection{Definition of Functions and their Properties}

In this subsection we define several functions which are 
necessary to describe our results and present their properties.
Let $r_i,i\in \Lambda$ be nonnegative numbers. 
Define the sequence of nonnegative functions 
$\{f_l(r_{l}^{L})\}_{l=1}^{L-1}$
$\cup\{f_0(r^L)\}$ 
by the following recursion:
\beq
\left.
\ba{l}
f_{L-1}(r_{L-1}^L)=
\frac{1-{\rm e}^{-2r_{L-1}}}{\sigma_{N_{L-1}}^2}
+\frac{1-{\rm e}^{-2r_{L}}}{\sigma_{N_{L}}^2}\,,
\vspace{1mm}\\
f_{l}(r_{l}^{L})=
   \hspace*{-1mm}\ba[t]{l}
   \frac{f_{l+1}(r_{l+1}^{L})}
        {1+ \sigma_{Z_{l+1}}^2 f_{l+1}(r_{l+1}^{L}) }
   +\frac{1-{\rm e}^{-2r_{l}}}{ \sigma_{N_{l}}^2}\,,
   \vspace{1mm}\\
    L-2 \geq l \geq 1\,,
   \ea
\\
f_0(r^L)=\ts \frac{f_{1}(r^{L})}{1+\sigma_{Z_{1}}^2f_{1}(r^{L})}\,.
\ea
\right\} 
\label{eqn:recur0}
\eeq
Next, we define the sequence of nonnegative functions
$$
\{g_l(D,r_0)\}_{l=0,1} \cup \{g_l(D,r_0,r^{l-1})\}_{l=2}^{L-1} 
$$
by the following recursion:
\beq
\hspace*{-1mm}\left.
\ba{l}
g_0(D,r_0)=
           \frac{{\rm e}^{-2r_0}}{D}-\frac{1}{\sigma_{X_0}^2}
           \,,
\vspace{1mm}\\
g_{1}(D, r_0)=\frac{g_{0}(D,r_0)}{1-\sigma_{Z_1}^2g_{0}(D,r_0)}\,,
\vspace{1mm}\\
g_{l+1}(D, r_0, r^{l})\\
= \hspace*{-1mm}\ba[t]{l}
   \frac{
   \left[
   g_{l}(D,r_0,r^{l-1})- 
   \frac{1}{\sigma_{N_{l}}^2}\left(1-{\rm e}^{-2r_{l}}\right)
   \right]^{+}
   }
   {1-\sigma_{Z_{l+1}}^2
    \left[
    g_{l}(D,r_0,r^{l-1})-
    \frac{1}{\sigma_{N_{l}}^2}\left(1-{\rm e}^{-2r_{l}}\right)
    \right]^{+}
    }\,,
   \vspace{1mm}\\
    1 \leq l \leq L-2\,,
   \ea
\ea
\right\}
\eeq
where $[a]^{+}=\max\{a,0\}\,.$ Let ${\cal B}_L(D)$ be the set of 
all nonnegative vectors $r_0^L$ that satisfy
$$
f_0(r^L) \geq g_0(D,r_0)= 
\ts \frac{{\rm e}^{-2r_0}}{D}-\frac{1}{\sigma_{X_0}^2}\,.
$$
Let $\partial{\cal B}_L(D)$ be the boundary of ${\cal B}_L(D)$, 
that is, the set of all nonnegative vectors $r_0^L$ that satisfy
$$
f_0(r^L)=g_0(D,r_0)=
\ts \frac{{\rm e}^{-2r_0}}{D}-\frac{1}{\sigma_{X_0}^2}\,.
$$
We can easily show that the functions we have defined 
satisfy the following property.
\begin{pr}{\label{pr:pr01}
$\quad$
\begin{itemize}
\item[{\rm a)}] For each $i \in \Lambda$, $f_0(r^L)$ is 
a monotone increasing function of $r_i$. 
For each $1\leq l\leq L$ and for each $i=l,l+1, \cdots, L$, 
$f_l(r_l^L)$ is a monotone increasing function of $r_i$.
\item[{\rm b)}] 
For each $2\leq l\leq L-1$ and for each $i=0,1,\cdots,l-1$, 
$g_l(D,r_0,r^{l-1})$ is a monotone decreasing function of $r_i$.
\item[{\rm c)}] If $r_0^L\in {\cal B}_L(D)$, then, 
for $0\leq l\leq L-1\,,$
$$g_l(D,r_0,r^{l-1}) \leq f_l(r_l^L)\,.$$
In the above $L$ inequalities the equalities simultaneously hold 
if and only if $r_0^L\in \partial{\cal B}_L(D)\,.$
\end{itemize}
}\end{pr}

Define 
\beqno
F({r^L}) &\defeq &
\prod_{l=1}^{L-1}
\left[1+\sigma_{Z_l}^2 f_l( r_l^{L} )\right]\,,
\\
G(D,r_0,r^{L-2})
&\defeq &
\prod_{l=1}^{L-1}
\left[1+\sigma_{Z_l}^2 g_l(D,r_0,r^{l-1})\right]\,.
\eeqno
For $S\subseteq \Lambda$, define  
$$
f_0(r_S) \defeq \left. f_0(r^L)\right|_{r_{\coS}={\lvc 0}}\,,
\quad
F(r_S) \defeq \left. F(r^L)\right|_{r_{\coS}={\lvc 0}}\,.
$$
We can easily show that the functions $F({r^L})$
and $G(D,r_0,$ $r^{L-2})$ satisfy the following 
property.
\begin{pr}{
\label{pr:prz001z}
$\quad$
\begin{itemize}
\item[{\rm a)}] For each  $i \in S$, 
$F(r_S)$ is a monotone increasing function of $r_i$.
\item[{\rm b)}] For each $i=0,1,\cdots, L-2$, 
$G(D,r_0,r^{L-2})$ is a monotone decreasing 
function of $r_i$. 
\item[{\rm c)}] If $r_0^L\in {\cal B}_L(D)$, then 
$$ 
G(D,r_0,r^{L-2}) \leq F(r^L)\,.
$$
The equality holds if and only if 
$r_0^L\in \partial{\cal B}_L(D)\,.$
\end{itemize}
}\end{pr}

For $D>0, r_i \geq 0, i\in \Lambda$ 
and $S\subseteq \Lambda$, define 
\beqno
& &{\rdf}_S(D,r_0,r^{L-2},r_S|r_{\coS})
\\
&\defeq & \frac{1}{2} \ts \log^{+}
        \left[
\frac{
G(D,r_0,r^{L-2})
              }
              {F(r_{\coS})}
\cdot
\frac{\sigma_{X_0}^2{\ds {\rm e}^{-2r_{0}}}}
     {\left\{1+ \sigma_{X_0}^2 f_0(r_{\coS})\right\}D}
\cdot{\ds \prod_{i\in S}{\rm e}^{2r_i}}
        \right]\,,
\\
& &K_S(r_S|r_{\coS})
\\
&\defeq & \frac{1}{2} \ts \log
        \left[
\frac{F(r^L)}{F(r_{\coS})}
\cdot
\frac{1+ \sigma_{X_0}^2 f_0(r^L)}
{1+ \sigma_{X_0}^2 f_0(r_{\coS})}
\cdot
       {  
       {\ds \prod_{i\in S}{\rm e}^{2r_i}}}
        \right]\,.
\eeqno
We can show that for $S\subseteq \Lambda$, 
$K_S(r_S|$ $r_{\coS})$ and $J_S(D,r_0,$ $r^{L-2},r_S|r_{\coS})$ 
satisfy the following two properties.
\begin{pr}{
\label{pr:prz01z}
$\quad$
\begin{itemize}
\item[{\rm a)}] If $r_0^L\in {\cal B}_L(D)$, then, for any 
$S\subseteq \Lambda$, 
$$
J_S(D,r_0,r^{L-2},r_S|r_{\coS})
\leq K_S(r_S|r_{\coS})\,.
$$
The equality holds when $r_0^L\in \partial{\cal B}_L(D)$.
\item[{\rm b)}] Suppose that $r^L\in {\cal B}_L(D)$. 
If $\left. r^L\right|_{r_S={\lvc 0}}$ still belongs to 
${\cal B}_L(D)$, then, 
\beqno
& &\left.J_S(D,r_0,r^{L-2},r_S|r_{\coS})
\right|_{r_S={\lvc 0}}
=\left. K_S(r_S|r_{\coS})\right|_{r_S={\lvc 0}}
\\
& &=0\,.
\eeqno
\end{itemize}
}\end{pr}

\begin{pr}\label{pr:matroid}{\rm 
Fix $r^L\in {\cal B}_L(D)$. For $S \subseteq \Lambda$, set 
\beqno
\rho_S&=&\rho_S(r_S|r_{\coS})\defeq J_S(D,r_0,r^{L-2},r_S|r_{\coS})\,.
\eeqno
By definition it is obvious that $\rho_S,S \subseteq \Lambda$ are 
nonnegative. We can show that
$\rho \defeq \{\rho_S\}_{S \subseteq \Lambda}$ 
satisfies the followings:
\begin{itemize}
\item[{\rm a)}] $\rho_{\emptyset}=0$. 
\item[{\rm b)}] 
$\rho_A\leq \rho_B$ for $A\subseteq B\subseteq \Lambda$.  
\item[{\rm c)}] $\rho_A+\rho_B \leq \rho_{A \cap B}+\rho_{A\cup B}\,.$
\end{itemize}
In general $(\Lambda,\rho)$ is called a {\it co-polymatroid} 
if the nonnegative function $\rho$ on $2^{\Lambda}$ satisfies 
the above three properties. Similarly, we set   
\beqno
\tilde{\rho}_S&=&\tilde{\rho}_S(r_S|r_{\coS})\defeq K_S(r_S|r_{\coS})\,,
\quad\tilde{\rho}=\left\{\tilde{\rho}_S\right\}_{S \subseteq \Lambda}\,.
\eeqno
Then, $(\Lambda,\tilde{\rho})$ also has the same three properties 
as those of $(\Lambda,\rho)$ and becomes a co-polymatroid. 
}\end{pr}
 
\subsection{Results}

In this subsection we present our results on inner and outer 
bounds of ${\cal R}_L(D)$. In the previous work \cite{oh7}, 
we derived an outer bound of ${\cal R}_L(D)$. We denote 
this outer bound by $\hat{\cal R}_L^{{(\rm out)}}(D)$.
According to \cite{oh7}, $\hat{\cal R}_L^{{(\rm out)}}(D)$ 
is given by
\beqno
& &\hat{\cal R}_L^{{(\rm out)}}(D)
\vspace{1mm}\\
&=&\Bigl\{(R_0,R^L):\Bigr.
\ba[t]{l}
\mbox{There exists a nonnegative vector }
\vspace{1mm}\\
(r_0,r^L)\mbox{ such that }
\vspace{1mm}\\
R_0\geq r_0 \geq 
       \frac{1}{2} \ts \log^{+}
        \left[\frac{\sigma_{X_0}^2}
              {\left\{
              1+ \sigma_{X_0}^2 f_0(r^L) 
              \right\}
              D}
        \right]\,,
\vspace{1mm}\\
R_i\geq r_i \mbox{ for any }i\in \Lambda,
\vspace{1mm}\\
R_0+\ds \sum_{i\in S}R_i
\vspace{1mm}\\
\geq \ds \frac{1}{2} \log^{+}
\left[{\ts
\frac{G(D,r_0,r^{L-2})\sigma_{X_0}^2}
{
F(r_{\coS})\left\{1+ \sigma_{X_0}^2 f_0(r_{\coS})\right\}D}
}
\right]
\vspace{1mm}\\
\quad +\ds \sum_{i=1}^Lr_i
\vspace{1mm}\\
\left. \mbox{ for any }S\subseteq \Lambda\,.\right\}\,.
\ea
\eeqno
Set 
\beqno
{\cal R}_L^{({\rm out})}(D,r_0^L)
&\defeq&
\ba[t]{l}
  \left\{(R_0,R_1,\cdots,R_L)\right.:
  \\ 
  \ba[t]{rcl}
  R_0 &\geq &r_0\,, 
  \vspace{1mm}\\ 
  \ds \sum_{i \in S} R_i 
  &\geq & {J}_{S}\left(D,r_0,r^{L-2},r_S|r_{\coS}\right)\,,
  \ea
  \vspace{1mm}\\
  \mbox{ for any }S \subseteq \Lambda\,. \left. \right\}\,,
\ea
\nonumber\\
{\cal R}_L^{({\rm in})
}(r_0^L)
&\defeq&
\ba[t]{l}
  \left\{(R_0,R_1,\cdots,R_L)\right.:\\ 
  \ba[t]{rcl}
  \ds R_0 &\geq&r_0\,, 
  \vspace{1mm}\\
  \ds \sum_{i \in S} R_i 
  &\geq&{K}_{S}\left(r_S|r_{\coS}\right)\,,
  \ea
  \vspace{1mm}\\
  \mbox{ for any }S \subseteq \Lambda\,. 
  \left. \right\}\,,
\ea
\nonumber\\
{\cal R}_{L}^{({\rm out})}(D)
&\defeq& 
\bigcup_{r_0^L \in {\cal B}_L(D)} 
{\cal R}_L^{({\rm out})
}(D,r_0^L)\,, 
\nonumber\\
{\cal R}_L^{({\rm in})}(D)
&\defeq & \bigcup_{r_0^L \in 
          {\cal B}_L(D)}
{\cal R}_L^{({\rm in})}(r_0^L)
\,. 
\eeqno
Our main result is as follows. 
\begin{Th}\label{th:converse} 
For Gaussian sources with the TS condition  
\beqno
{\cal R}_{\Iset}^{{(\rm in)}}(D)
   \subseteq \tilde{\cal R}_{\Iset}^{{(\rm in)}}(D)
   &\subseteq&{\cal R}_{\Iset}(D)
\nonumber\\ 
&\subseteq &\hat{\cal R}_{\Iset}^{{(\rm out)}}(D)
\subseteq {\cal R}_{\Iset}^{{(\rm out)}}(D)\,.
\eeqno
\end{Th}

Proof of this theorem will be given in Section V. 
The inclusion
$
{\cal R}_{\Iset}(D)\subseteq 
\hat{\cal R}_{\Iset}^{{(\rm out)}}(D)
$
and an outline of proof of this inclusion was given 
in Oohama \cite{oh7}. Furthermore, by Theorem \ref{th:direct}, 
we have $\tilde{\cal R}_L^{({\rm in})}(D) \subseteq {\cal R}_L(D)$. 
Hence, it suffices to show 
$\hat{\cal R}_L^{{(\rm out)}}(D) 
\subseteq {\cal R}_L^{\rm (out)}(D)$
and ${\cal R}_L^{({\rm in})}(D) 
\subseteq \tilde{\cal R}_L^{({\rm in})}(D)$ 
to prove Theorem \ref{th:converse}.
Proofs of those two inclusions will be given
in Section V.
We can directly prove ${\cal R}_L(D)
\subseteq {\cal R}_L^{\rm (out)}(D)$
in a manner similar to that of Oohama \cite{oh7}. For 
the detail of the direct proof of 
${\cal R}_L(D)\subseteq {\cal R}_L^{\rm (out)}(D)$, 
see Appendix B.

An essential difference between 
${\cal R}_L^{({\rm out})}(D)$ and ${\cal R}_L^{({\rm in})}(D)$  
is the difference between 
$J_S(D,r_0,$ $r^{L-2},r_S|r_{\coS})$ 
in the definition of ${\cal R}_L^{({\rm out})}(D)$
and $K_S(r_S|r_{\coS})$ in the definition 
of ${\cal R}_L^{({\rm in})}(D)$. 
By Property \ref{pr:prz01z} part a) 
and the definitions of ${\cal R}_L^{({\rm out})}(D,r_0^L)$ and 
${\cal R}_L^{({\rm in})}($ $r_0^L)$, if 
$r_0^L\in\partial{\cal B}_L(D)$, then, 
$$
{\cal R}_L^{({\rm out})}(D,r_0^L)
={\cal R}_L^{({\rm in})}(r_0^L)\,. 
$$
This gap suggests a possibility that in some cases those 
two bounds match. In the following we present 
a sufficient condition for ${\cal R}^{{(\rm out)}}_L(D)$ 
$\subseteq$ ${\cal R}_L^{({\rm in})}($$D)\,.$ 
We consider the following condition on $G(D,r_0,$ $r^{L-2})$.

{\it Condition: }For each $l=1,2,$ $\cdots, L-2$, 
${\rm e}^{2r_l}G(D,r_0,r^{L-2})$ is a monotone 
increasing function of $r_l$.

We call the above condition the MI condition. 
The following is our main result on a matching condition 
on inner and outer bounds. 
\begin{lm}\label{lm:region1a} For Gaussian sources with the 
TS condition if $G(D,r_0,r^{L-2})$ satisfies the MI condition, 
then, 
$$
{\cal R}_{\Iset}^{(\rm out)}(D) 
\subseteq {\cal R}_{\Iset}^{({\rm in})}(D)\,.
$$
\end{lm}

Proof of this lemma is given in Section V. Note that 
when $L=2$ or $\sigma_{Z_l}=0,$ for 
$l=2,3,\cdots, L-1$ under the TS condition,
we have 
$$
G(D,r_0,r^{L-2})=1+\sigma_{Z_1}^2g_1(D,r_0)\,,
$$
which satisfies the MI condition. 
TS conditions in the case of $L=2$ 
and the case of $L=3, Z_2=0$ is shown in Fig. \ref{fig:Fig3z}. 
Note that those two conditions are different 
from the CI condition.
Combining Lemma \ref{lm:region1a} 
and Theorem \ref{th:converse}, we establish the following.
\begin{Th}\label{th:exregion} For Gaussian sources with 
the TS condition 
$${\cal R}_2^{(\rm in)}(D)={\cal R}_2(D)
=\hat{\cal R}_2^{(\rm out)}(D) 
={\cal R}_2^{(\rm out)}(D)\,.
$$
Furthermore, if $G(D,r_0,r^{L-2})$ satisfies the MI condition, 
then, 
$$      {\cal R}_L^{({\rm in})}(D)
       ={\cal R}_L(D)
       =\hat{\cal R}_L^{{(\rm out)}}(D)
       ={\cal R}_L^{({\rm out})}(D) \,.$$
\end{Th} 

In Oohama \cite{oh7}, the equality 
${\cal R}_2(D)=\hat{\cal R}_2^{(\rm out)}(D)$ 
was stated without complete proof. We can see that this 
equality can be obtained by Theorem \ref{th:converse}, 
Lemma \ref{lm:region1a}, and the fact that the MI condition 
holds for $L=2$.    
\bfig
\setlength{\unitlength}{1.05mm}
\begin{picture}(83.5,52)(2,0)
%
\put(29,   29.25){\line(0,1){1.5}}
\put(28.25,30   ){\line(1,0){1.5}}
\put(29,   30   ){\circle{2.0}}
\put(34,30){\vector(-1,0){4}}
\put(34,29){$N_2$}
\put(29,29){\vector(0,-1){4}}
\put(27.5,22){${X_2}$} 
%
\put(17,45){\vector(0,-1){4}}
\put(15.5,46.5){${X_0}$}
%
\put(17,39){\vector(-3,-2){12}} 
\put(17,39){\vector(3,-2){12}} 
\put(17,   39.25){\line(0,1){1.5}}
\put(16.25,40   ){\line(1,0){1.5}}
\put(17,   40   ){\circle{2.0}}
\put(22,40){\vector(-1,0){4}}
\put(22,39){$Z_1$}
%
\put(5,   29.25){\line(0,1){1.5}}
\put(4.25,30   ){\line(1,0){1.5}}
\put(5,   30   ){\circle{2.0}}
\put(10,30){\vector(-1,0){4}}
\put(6,35){$Y_1$} \put(24,35){$Y_1$}
\put(10,29){${N_1}$}
\put(5,29){\vector(0,-1){4}}
\put(3.5,22){${X_1}$}
%
%
\put(54,45){\vector(0,-1){4}}
\put(52.5,46.5){${X_0}$}
%
\put(54,39){\vector(-3,-2){12}} 
\put(54,39){\vector(3,-2){12}} 
\put(54,   39.25){\line(0,1){1.5}}
\put(53.25,40   ){\line(1,0){1.5}}
\put(54,   40   ){\circle{2.0}}
\put(59,40){\vector(-1,0){4}}
\put(59,39){$Z_1$}
%
\put(42,   29.25){\line(0,1){1.5}}
\put(41.25,30   ){\line(1,0){1.5}}
\put(42,   30   ){\circle{2.0}}
\put(47,30){\vector(-1,0){4}}
 \put(43,35){$Y_1$} \put(61,35){$Y_1$}
\put(47,29){${N_1}$}
\put(42,29){\vector(0,-1){4}}
\put(40.5,22){${X_1}$}
%
\put(66,29){\vector(-3,-2){12}} 
\put(66,29){\vector(3,-2){12}} 
\put(66,   29.25){\line(0,1){1.5}}
\put(65.25,30   ){\line(1,0){1.5}}
\put(66,   30   ){\circle{2.0}}
\put(71,30){\vector(-1,0){4}}
\put(71,29){$Z_2=0$}
%
\put(54,   19.25){\line(0,1){1.5}}
\put(53.25,20   ){\line(1,0){1.5}}
\put(54,   20   ){\circle{2.0}}
\put(59,20){\vector(-1,0){4}}
\put(55,25){$Y_1$}
\put(73,25){$Y_1$}
\put(59,19){${N_2}$}
\put(54,19){\vector(0,-1){4}}
\put(52.5,12){${X_2}$}
\put(78,   19.25){\line(0,1){1.5}}
\put(77.25,20   ){\line(1,0){1.5}}
\put(78,   20   ){\circle{2.0}}
\put(83,20){\vector(-1,0){4}}
\put(83,19){${N_3}$}
\put(78,19){\vector(0,-1){4}}
\put(76.5,12){${X_3}$}
\end{picture}
\vspace*{-8mm}
\caption
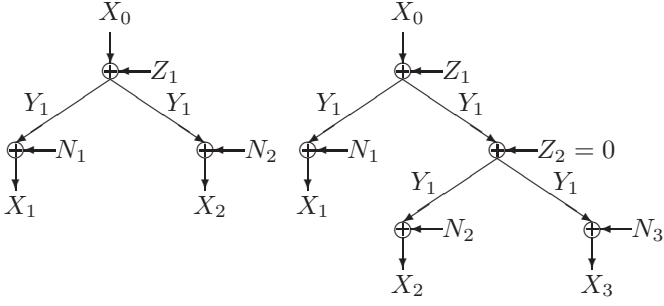
{TS conditions in the case of $L=2$ 
 and the case of $L=3$ and $Z_2=0$.}
\label{fig:Fig3z}
\vspace*{-2mm}
\efig

Next, we present a sufficient condition for $G(D,r_0,r^{L-2})$ 
to satisfy the MI condition. Let $\{f_j^{*}\}_{j=1}^{L-1}$ be a 
sequence of positive numbers defined by the following 
recursion: 
\beq
\left.
\ba{l}
f_{L-1}^{*}=
\frac{1}{\sigma_{N_{L-1}}^2}+\frac{1}{\sigma_{N_{L}}^2}\,,
\vspace{1mm}\\
f_{l}^{*}=
   \hspace*{-1mm}
   \ba[t]{l}
   \frac{f_{l+1}^{*}}
        {1+ \sigma_{Z_{l+1}}^2 f_{l+1}^{*}}
   +\frac{1}{ \sigma_{N_{l}}^2}\,,
    L-2 \geq l \geq 1\,.
   \ea
\ea
\right\} 
\eeq
By definition it is obvious that 
$
f_l(r_{l}^L)\leq f_l^{*}\,. 
$
Then, we have the following proposition.
\begin{pro}\label{pro:pro1}
If 
\beqa
\hspace*{-4mm}& &\sum_{k=l}^{L-2}
{\ts \frac{\sigma_{Z_{k+1}}^2}{\sigma_{N_{l}}^2}}
\left(
1+\sigma_{Z_{k+1}}^2f_{k+1}^{*}
\right)\hspace*{-1mm}
\prod_{j=l+1}^{k}
\left(1+\sigma_{Z_{j}}^2f_{j}^{*}\right)^2
\leq 1
\label{eqn:xzx}
\eeqa
hold for $l=1,2,\cdots, L-2$, then, 
$G(D,r_0,r^{L-2})$ satisfies the MI condition. 
\end{pro}

Proof of this proposition will be given in Appendix A. 
It can be seen from this proposition that for $L\geq 3$, 
the MI condition holds for relatively small values of 
$\sigma_{Z_l}, l=2, \cdots, L-1$. In particular, 
when $L=3$, the sufficient condition given by 
(\ref{eqn:xzx}) is  
$$
\ts \frac{\sigma_{Z_2}^2}{\sigma_{N_1}^2}
\left\{1+{\sigma_{Z_2}^2}
  \left(
       {\ts \frac{1}{\sigma_{N_2}^2}
      +\frac{1}{\sigma_{N_3}^2}
       }
 \right)
\right\}\leq 1\,.
$$
Solving the above inequality with respect 
to $\sigma_{Z_2}^2$, we have
$$
\sigma_{Z_2}^2 \leq \frac{2}
{1+\sqrt{1+ 4{\sigma_{N_1}^2}
        \left(
        {\ts  \frac{1}{\sigma_{N_2}^2}
        +\frac{1}{\sigma_{N_3}^2}
        }
       \right)}}\cdot \sigma_{N_1}^2\,.
$$
The TS condition in the case of $L=3$ is shown 
in Fig. \ref{fig:Fig4}.
\bfig
\setlength{\unitlength}{1.05mm}
\begin{picture}(65,52)(2,0)
\put(41,   19.25){\line(0,1){1.5}}
\put(40.25,20   ){\line(1,0){1.5}}
\put(41,   20   ){\circle{2.0}}
\put(46,20){\vector(-1,0){4}}
\put(46,19){${N_3}$}
\put(41,19){\vector(0,-1){4}}
\put(39.5,12){${X_3}$}
%
%
%
\put(29,29){\vector(-3,-2){12}} 
\put(29,29){\vector(3,-2){12}} 
\put(29,   29.25){\line(0,1){1.5}}
\put(28.25,30   ){\line(1,0){1.5}}
\put(29,   30   ){\circle{2.0}}
\put(34,30){\vector(-1,0){4}}
\put(34,29){$Z_2$}
%
\put(17,   19.25){\line(0,1){1.5}}
\put(16.25,20   ){\line(1,0){1.5}}
\put(17,   20   ){\circle{2.0}}
\put(22,20){\vector(-1,0){4}}
\put(18,25){$Y_2$}\put(36,25){$Y_2$}
\put(22,19){${N_2}$}
\put(17,19){\vector(0,-1){4}}
\put(15.5,12){${X_2}$}
%
\put(17,45){\vector(0,-1){4}}
\put(15.5,46.5){${X_0}$}
%
\put(17,39){\vector(-3,-2){12}} 
\put(17,39){\vector(3,-2){12}} 
\put(17,   39.25){\line(0,1){1.5}}
\put(16.25,40   ){\line(1,0){1.5}}
\put(17,   40   ){\circle{2.0}}
\put(22,40){\vector(-1,0){4}}
\put(22,39){$Z_1$}
%
\put(5,   29.25){\line(0,1){1.5}}
\put(4.25,30   ){\line(1,0){1.5}}
\put(5,   30   ){\circle{2.0}}
\put(10,30){\vector(-1,0){4}}
 \put(6,35){$Y_1$} \put(24,35){$Y_1$}
\put(10,29){${N_1}$}
\put(5,29){\vector(0,-1){4}}
\put(3.5,22){${X_1}$}
\end{picture}
\vspace*{-8mm}
\caption
{TS condition in the case of $L=3$.}
\label{fig:Fig4}
\vspace*{-5mm}
\efig

\subsection{Binary Tree Structure Condition}


As a correlation property of Gaussian source Tavildar {\it et al.}
\cite{tvw} introduced a binary Gauss Markov tree structure condition.
They studied a full characterization of the rate distortion region for
Gaussian sources with this binary tree structure. In this subsection
we describe their result and compare it with our results.

We first explain the binary tree structure introduced by them.  Let
$k$ be a positive integer. We consider the case where $L=2^k$.  Let
$N_{i}^{(j)}\,, 1\leq i\leq 2^j, 1\leq j \leq k\,,$ be zero mean
independent Gaussian random variables with variance
$\sigma_{N_{i}^{(j)}}^2$.  Those $2^{k+1}-2$ random variables are
independent of $X_0$.  Define the sequence of Gaussian random
variables $\{Y_i^{(j)}$ $\}_{1\leq i\leq 2^j, 0\leq j \leq k}$ by the
following recursion:
\beq
\left.
\ba{rcl}
Y_{1}^{(0)}&=&X_0\,,\\
Y_{i}^{(j)}&=&Y_{\lceil \frac{i}{2}\rceil}^{(j-1)}+N_{i}^{(j)}\,,\\
& &\mbox{ for }1\leq i\leq 2^j, 0\leq j \leq k\,,\\
X_i&=&Y_{i}^{(k)}\,,
\mbox{ for }1\leq i\leq 2^k\,,
\ea
\right\}
\label{eqn:zaazz0}
\eeq 
where $\lceil a \rceil$ stands for the smallest integer not below
$a$. We say that for $L=2^k$ the Gaussian source $(X_0,$$X_1,$
$\cdots, X_L)$ satisfies the binary tree structure (BTS) condition
when it satisfies (\ref{eqn:zaazz0}). The binary tree structure in the
case of $k=2$ and $L=2^k=4$ is shown in Fig. \ref{fig:Fig5}.  In this
example, let $\sigma_{N_1^{(2)}}\to\infty$ and $N_2^{(2)}=0$.  Then,
$X_1$ becomes independent of $(X_2,X_3,X_4)$ and $(X_2,X_3,X_4)$ has
the same correlation property as the TS condition in the case of $L=3$
and $Z_1=0$.  The BTS condition in this case is shown in
Fig. \ref{fig:Fig6z}.  In general the set of Gaussian sources
satisfying the TS condition and $Z_1=0$ can be embedded into the set
of Gaussian sources satisfying BTS condition.
\bfig
\setlength{\unitlength}{1.05mm}
\begin{picture}(65,52)(-8,0)
%
\put(17,43){\vector(0,-1){4}}
\put(15.5,44.5){${X_0}$}
\put(17,39){\vector(-3,-2){12}} 
\put(17,39){\vector(3,-2){12}} 
%
\put(5,35){$X_0$} 
\put(5,29){\vector(-1,-1){8}} 
\put(5,29){\vector(3,-4){6}} 
\put(5,   29.25){\line(0,1){1.5}}
\put(4.25,30   ){\line(1,0){1.5}}
\put(5,   30   ){\circle{2.0}}
\put(10,30){\vector(-1,0){4}}
\put(10,29){$N_1^{(1)}$}
%
\put(-5,24){$Y_1^{(1)}$}
\put(-3,   19.25){\line(0,1){1.5}}
\put(-3.75,20   ){\line(1,0){1.5}}
\put(-3,   20   ){\circle{2.0}}
\put(1,20){\vector(-1,0){3}}
\put(1,19){$N_1^{(2)}$}
\put(-3,19){\vector(0,-1){4}}
\put(-5.5,12){${X_1}$}
%
\put(10,24){$Y_1^{(1)}$}
\put(11,   19.25){\line(0,1){1.5}}
\put(10.25,20   ){\line(1,0){1.5}}
\put(11,   20   ){\circle{2.0}}
\put(15,20){\vector(-1,0){3}}
\put(15,19){$N_2^{(2)}$}
\put(11,19){\vector(0,-1){4}}
\put(9.5,12){$X_2$}
%
\put(24,35){$X_0$}
\put(29,29){\vector(-3,-4){6}} 
\put(29,29){\vector(1,-1){8}} 
\put(29,   29.25){\line(0,1){1.5}}
\put(28.25,30   ){\line(1,0){1.5}}
\put(29,   30   ){\circle{2.0}}
\put(34,30){\vector(-1,0){4}}
\put(34,29){$N_2^{(1)}$}
%
\put(20,24){$Y_2^{(1)}$}
\put(23,   19.25){\line(0,1){1.5}}
\put(22.25,20   ){\line(1,0){1.5}}
\put(23,   20   ){\circle{2.0}}
\put(27,20){\vector(-1,0){3}}
\put(27,19){$N_3^{(2)}$}
\put(23,19){\vector(0,-1){4}}
\put(21.5,12){$X_3$}
%
\put(35,24){$Y_2^{(1)}$}
\put(37,   19.25){\line(0,1){1.5}}
\put(36.25,20   ){\line(1,0){1.5}}
\put(37,   20   ){\circle{2.0}}
\put(41,20){\vector(-1,0){3}}
\put(41,19){$N_4^{(2)}$}
\put(37,19){\vector(0,-1){4}}
\put(35.5,12){$X_4$}
\end{picture}
\vspace*{-8mm}
\caption
{BTS condition in the case of $L=4$.}
\label{fig:Fig5}
\vspace*{-5mm}
\efig
%
\bfig
\setlength{\unitlength}{1.05mm}
\begin{picture}(65,52)(-8,0)
%
\put(17,43){\vector(0,-1){4}}
\put(15.5,44.5){${X_0}$}
\put(17,39){\vector(-3,-2){12}} 
\put(17,39){\vector(3,-2){12}} 
%
\put(5,35){$X_0$} 
\put(5,29){\vector(0,-1){4}}
\put(3.5,22){${X_2}$} 
\put(5,   29.25){\line(0,1){1.5}}
\put(4.25,30   ){\line(1,0){1.5}}
\put(5,   30   ){\circle{2.0}}
\put(10,30){\vector(-1,0){4}}
\put(10,29){$N_1^{(1)}$}
%
%
%
\put(24,35){$X_0$}
\put(29,29){\vector(-1,-1){8}} 
\put(29,29){\vector(1,-1){8}} 
\put(29,   29.25){\line(0,1){1.5}}
\put(28.25,30   ){\line(1,0){1.5}}
\put(29,   30   ){\circle{2.0}}
\put(34,30){\vector(-1,0){4}}
\put(34,29){$N_2^{(1)}$}
%
\put(18,24){$Y_2^{(1)}$}
\put(21,   19.25){\line(0,1){1.5}}
\put(20.25,20   ){\line(1,0){1.5}}
\put(21,   20   ){\circle{2.0}}
\put(25,20){\vector(-1,0){3}}
\put(25,19){$N_3^{(2)}$}
\put(21,19){\vector(0,-1){4}}
\put(19.5,12){$X_3$}
%
\put(35,24){$Y_2^{(1)}$}
\put(37,   19.25){\line(0,1){1.5}}
\put(36.25,20   ){\line(1,0){1.5}}
\put(37,   20   ){\circle{2.0}}
\put(41,20){\vector(-1,0){3}}
\put(41,19){$N_4^{(2)}$}
\put(37,19){\vector(0,-1){4}}
\put(35.5,12){$X_4$}
\end{picture}
\vspace*{-8mm}
\caption
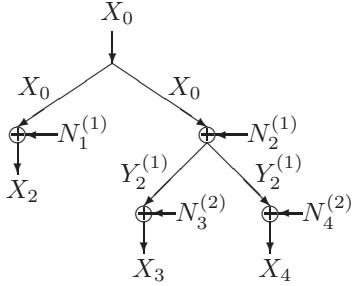
{BTS condition in the case of $L=4$, $\sigma_{N_1^{(2)}}\to\infty$ 
and $N_2^{(2)}=0$ is equivalent to the TS condition 
in the case of $L=3$ and $Z_1=0$.}
\label{fig:Fig6z}
\vspace*{-5mm}
\efig
%

The communication system treated by Tavildar {\it et al.} 
is shown in Fig. \ref{fig:Fig6}. It can be seen from this 
figure that their problem set up is slightly different 
from ours. In their communication system there is no encoder 
that can directly access to the source ${\vc X}_0$. 
Tavildar {\it et al.} studied a characterization of the rate distortion 
region ${\cal R}_L(D)\cap \{R_0=0\}$ for Gaussian sources 
with the binary tree structure and succeeded in it.
Their result is the following. 
\begin{Th}[Tavildar {\it et al.} \cite{tvw}] \label{Th:thTv}
When $L=2^k$ for some integer $k$ 
and 
$(X_0,X_1,$ $\cdots, X_L)$ satisfies the BTS condition, 
we have 
$$
{\cal R}_L(D)\cap \{R_0=0\}
=\tilde{\cal R}_L^{(\rm in)}(D)\cap \{R_0=0\}\,.
$$
\end{Th}

From the above theorem we have the following corollary.
\begin{co}[Tavildar {\it et al.} \cite{tvw}] \label{co:coZ} 
When $(X_0,X_1,$$\cdots,X_L$\\ $)$ satisfies 
the TS condition and ${Z_1=0}$, we have
$$
{\cal R}_L(D)\cap \{R_0=0\}
=\tilde{\cal R}_L^{(\rm in)}(D)\cap \{R_0=0\}\,.
$$
\end{co}

The BTS condition differs from the TS condition 
in its symmetrical property, which plays an essential 
role in the proof of Theorem \ref{Th:thTv}. 
We think that the method of Tavildar {\it et al.} \cite{tvw}
is applicable to the general case where $Z_1$ is not constant and $R_0>0$ 
and that 
${\cal R}_L(D)=$
$\tilde{\cal R}_L^{(\rm in)}(D)$
still holds in this general case. 

%

Unfortunately, our approach developed in \cite{oh7} and this paper can
not establish ${\cal R}_L(D)=\tilde{\cal R}_L^{(\rm in)}(D)$ for
Gaussian sources satisfying the TS condition without requiring the
condition on the variances of $Z_i, 2\leq i\leq L-1 $ and $N_i, 1\leq
i\leq L,$ specified with (\ref{eqn:xzx}) in Proposition
\ref{pro:pro1}. However, we think that our work in \cite{oh7} had
provided an important step toward the full characterization of the
rate distortion region established by Tavildar {\it et al.}
\cite{tvw}.
%
%
%

\bfig
\setlength{\unitlength}{1.05mm}
\begin{picture}(80,56)(8,-4)
\put(10,40){\framebox(6,6){$X_1$}}
\put(10,25){\framebox(6,6){$X_2$}}
\put(13,15){$\vdots$}
\put(10, 5){\framebox(6,6){$X_L$}}

\put(21,46){${\Xatn}$}
\put(16,43){\vector(1,0){13}}

\put(21,31){${\Xbtn}$}
\put(16,28){\vector(1,0){13}}

\put(23,19){$\vdots$}

\put(21,11){${\Xltn}$}
\put(16,8){\vector(1,0){13}}

\put(29,40){\framebox(6,6){$\varphi_1$}}
\put(37,46){$\varphi_1({\Xatn})$}

\put(29,25){\framebox(6,6){$\varphi_2$}}
\put(37,31){$\varphi_2({\Xbtn})$}

\put(32,15){\vdots}
\put(40,19){\vdots}

\put(29,5){\framebox(6,6){$\varphi_L$}}
\put(37,11){$\varphi_L({\Xltn})$}

\put(35,43){\line(1,0){15}}
\put(50,43){\vector(1,-1){15}}

\put(35,28){\vector(1,0){30}}

\put(35,8){\line(1,0){15}}
\put(50,8){\vector(3,4){15}}

\put(65,25){\framebox(6,6){$\psi$}}
\put(71,28){\vector(1,0){10}}
\put(82,27){$\hatXotn$}
\end{picture}
\vspace*{-5mm}
\caption{
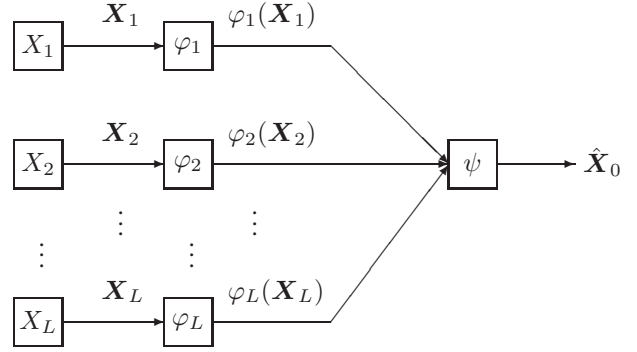
Communication system that Tavildar {\it et al.} treated.
}\label{fig:Fig6}
\efig

\section{Sum Rate Part of the Rate Distortion Region}
\label{sec:SumRate}

In this section we state our result on the rate 
sum part of ${\cal R}_L(D)$. Set 
\beqa
R_{{\rm sum},L}^{({\rm l})}(D,R_0)
&\defeq&\min_{\scs r^L: f_0(r^L)
\atop{\scs \geq g_0(D,R_0)}
}
J_{\Lambda}(D,R_0,r^{L-2},r^L)\,,
\nonumber\\
R_{{\rm sum},L}^{({\rm u}) }(D,R_0)
&\defeq&\min_{\scs r^L: f_0(r^L)
\atop{\scs \geq g_0(D,R_0)}
}
K_{\Lambda}(r^L)\,.
\nonumber
\eeqa
Let $\hat{R}_{ {\rm sum},L}^{({\rm l})}(D,R_0)$ be the minimum 
sum rate for $\hat{\cal R}^{(\rm out)}_L(D)$, that is, 
$$ 
\hat{R}_{ {\rm sum},L}^{({\rm l})}(D,R_0)
\defeq \min_{ (R_0,R_1,\cdots,R_L)
\in \hat{\cal R}_{L}^{(\rm out)}(D)}
\left\{\sum_{i=1}^{L}R_i \right\}\,.
$$
Then, it immediately follows from Theorem \ref{th:converse} 
that we have the following corollary. 
\begin{co}{\rm \label{co:coZz} 
For Gaussian sources with the TS condition  
\beqno
& & R_{{\rm sum},L}^{({\rm l})}(D,R_0)
    \leq \hat{R}_{{\rm sum},L}^{({\rm l})}(D,R_0)
\\
&\leq & R_{{\rm sum},L}(D,R_0)\leq R_{{\rm sum},L}^{({\rm u})}(D,R_0)\,.
\eeqno
}
\end{co}
On the other hand, we have the following lemma. 
\begin{lm}\label{lm:region1b} For Gaussian sources with the 
TS condition, we have 
$$
R_{{\rm sum},L}^{({\rm l})}(D,R_0)
\geq 
R_{{\rm sum},L}^{(\rm u)}(D,R_0)\,.
$$
\end{lm}

Proof of this lemma will be given in Section V. 
Combining Corollary \ref{co:coZz} and Lemma \ref{lm:region1b}, 
we have the following.
\begin{Th}\label{th:exth4}
 {\rm For Gaussian sources with the 
TS condition
\begin{align*}
& R_{{\rm sum},L}(D,R_0)  
 ={R}_{{\rm sum},L}^{(\rm u)}(D,R_0)
 =\hat{R}_{{\rm sum},L}^{(\rm l)}(D,R_0)
\\
&=R_{{\rm sum},L}^{\rm(l)}(D,R_0)
=-R_0+\frac{1}{2}\log\frac{\sigma_{X_0}^2}{D}
\\
&\qquad \quad +\min_{\scs r^L: f_0(r^L)
      \atop{\scs = g_0(D,R_0)}
      }
\left[\sum_{l=1}^Lr_l+\frac{1}{2}\log {F(r^L)}\right].
\end{align*}
}
\end{Th}

In \cite{ohitw09}, the author further derived an algorithm 
of computing $R_{{\rm sum}, L}(D,R_0)$. This algorithm, 
however, has a problem that it can not provide 
$R_{{\rm sum}, L}(D,R_0)$ for all $D\in (0,\sigma^2_{X_0}].$      
In fact, the function $R_{{\rm sum}, L}(D,R_0)$ is determined 
for relatively small value of $D$. In the remaining part 
of this subsection we present the algorithm given 
by \cite{ohitw09} and concretely explain the above problem. 

The algorithm of computing $R_{{\rm sum}, L}(D,R_0)$ given by 
the author \cite{ohitw09} is as follows.  
For $L\geq l\geq 1$, set
$\sigma_{N_l}^2=\sigma^2_l,$
$\sigma_{Z_l}^2=\epsilon_l\sigma^2_l\,.$ Furthermore, 
set $\tau_l=\sigma^2_l/\sigma^2_{l-1}$ for $L\geq l\geq 2$. 
Let $\omega\in [0,1)$. Define 
the sequence of functions 
$\{\theta_l(\omega)\}_{l=1}^L$ 
by the following recursion:
\beq
\left.
\ba{l}
\theta_{L}(\omega)=\omega,
\vspace{1mm}\\
\theta_{L-1}(\omega)=\ds
\frac{\frac{2\theta_{L}(\omega)-1}{\tau_L}+1}
{1+\epsilon_{L-1}
\left[\frac{2\theta_{L}(\omega)-1}{\tau_L}+1\right]}\,,
\vspace{1mm}\\
\theta_{l-1}(\omega)=\ds
\frac{\frac{1}{\tau_l}
      \left[
       2\theta_{l}(\omega)
       -\frac{1+ \frac{\theta_{l+1}(\omega)}{\tau_{l+1}}}
             {1+\epsilon_{l}
             \left(1+ \frac{\theta_{l+1}(\omega)}{\tau_{l+1}}\right)}
       +\tau_l \right]
     } 
     {
     1+\frac{ \epsilon_{l-1}}{\tau_l}
      \left[
       2\theta_{l}(\omega)
        -\frac{1+\frac{\theta_{l+1}(\omega)}{ \tau_{l+1} }}
             {1+\epsilon_{l}\left(
              1+ \frac{\theta_{l+1}(\omega)}{\tau_{l+1}}
              \right)}
      +\tau_l \right]
     }
\vspace{1mm}\\
\mbox{ for }L-1 \geq l \geq 2\,.
\ea
\right\} 
\label{eqn:SumEx006}
\eeq

\begin{Th}[Oohama \cite{ohitw09}]
\label{Th:SumRateOld}
Let $\{\theta_l(\omega)\}_{l=1}^L$ be a sequence 
of functions defined by (\ref{eqn:SumEx006}).
Suppose that the Gaussian source satisfies the TS condition 
and the condition
\beq
\left.
\ba{l}
\tau_L\geq 1,\mbox{ for }l=L,\\
\tau_l\geq \frac{1}{1+\epsilon_l},\mbox{ for }L-1\geq l \geq 2. 
\ea
\right\}
\label{eqn:CondAa}
\eeq
Then, we have the following parametric form 
of $R_{{\rm sum}, L}(D,R_0)$ 
with the parameter $\omega\in [0,1)$:
\begin{align*}
& D={\rm e}^{-2{R_0}}\frac{\sigma_1^2\sigma_{X_0}^2}{\sigma_{X_0}^2\theta_1(\omega)+\sigma_1^2},\nonumber\\
& R_{{\rm sum}, L}(D,R_0)
=-R_0+\frac{1}{2}\log\frac{\sigma_{X_0}^2}{D}
\\
& + \left(-\frac{1}{2}\right)
\Hugebl
\sum_{l=1}^{L-1}
\left\{\log 
\left(1-\frac{\theta_l(\omega)}{1-\epsilon_{l}\theta_l(\omega)}
+\frac{\theta_{l+1}(\omega)}{\tau_{l+1}}\right)\right.
\nonumber\\
& \qquad\qquad\qquad
    +\log\left(1-\epsilon_{l} \theta_l(\omega) \right)
     \huger+\log(1-\omega) \Hugebr.
\end{align*}
\end{Th}

In the following we state a problem that we have in the parametric expression of 
$(D,R_{{\rm sum}, L}(D,R_0))$ in the above theorem. 
We consider the case of $R_0=0$.     
From (\ref{eqn:SumEx006}), we can see that  when $\tau_l>1$ for $L\geq l\geq 1$, $\theta_1(\omega)$ is strictly positive function of $\omega \in [0,1]$. 
This implies that the parametric expression in Theorem \ref{Th:SumRateOld}   
can not provide $R_{{\rm sum}, L}(D,0)$ for all 
$D\in (0,\sigma^2_{{X_0}}]$. In this paper, we solve this problem to provide 
$R_{{\rm sum}, L}(D,0)$ for all $D\in (0,\sigma^2_{{X_0}}]$.

In the following argument, we consider the case of $R_0=0$. 
In this case we set $R_{{\rm sum},L}(D)=R_{{\rm sum},L}(D,0)$. 
Furthermore, set $g_0(D)=g_0(D,0)$.  
The optimal sum rate $R_{{\rm sum},L}(D)$ has a form 
of optimization problem. In the remaining part of this section 
we deal with this optimization problem.  
We let $\epsilon_L=0$. 
Then the recursion (\ref{eqn:recur0}) is 
\beq
\left.
\ba{rcl}
f_L(r_L)&=&\frac{1}{\sigma_L^2}\left(1-{\rm e}^{-2r_L}\right)\,,
\\
f_{l-1}(r_{l-1}^L)&=&\frac{f_l(r_l^L)}{1+\epsilon_{l}\sigma_{l}^2f_l(r_l^L)}
+\frac{1}{\sigma_{l-1}^2}\left(1-{\rm e}^{-2r_{l-1}}\right)
\\
& &\mbox{ for }L \geq l \geq 2,
\\
f_{0}(r^L)&=&\frac{f_1(r^L)}{1+\epsilon_{1}\sigma_{1}^2f_1(r^L)}\,.
\ea
\right\}
\label{eqn:SumEx000}
\eeq
The optimization problem presenting $R_{{\rm sum},L}(D)$ is
\beqno
& &R_{{\rm sum}, L}(D)=\frac{1}{2}\log^{+}\frac{\sigma_{X_0}^2}{D}\
\nonumber\\
& &+\min_{\scs r^L: f_0(r^L)
\atop{\scs = g_0(D)}
}
 \left[\sum_{l=1}^Lr_l
+\sum_{l=1}^{L-1}
\frac{1}{2}\log \left(1+\epsilon_{l}\sigma_l^2f_l(r_l^L)\right)
\right].
\eeqno
To describe an algorithm of computing 
$R_{{\rm sum},L}(D)$, for $1\leq l\leq L$, define 
$R_{{\rm sum}}^{(l)}(D)$ by 
\begin{align*}
& R_{{\rm sum}}^{(l)}(D)=\frac{1}{2}{\rm log}\frac{\sigma_{X_0}^2}{D}
\nonumber \\
& +\min_{\scriptstyle r^L:r_l>0, r_{l+1}^L=0
\atop{\scriptstyle  f_0(r^L) = g_0(D)}}
 \left[\sum_{i=1}^lr_i +\sum_{i=1}^{l-1}
\frac{1}{2}{\rm log} \left(1+\epsilon_{i}\sigma_i^2f_i(r_i^L)\right)
\right].
\end{align*}
By the above definition and an elementary computation we have that
for each $1\leq l\leq L$, $R=R_{{\rm sum}}^{(l)}(D)$ is monotone 
decreasing and convex function of $D>0$. 
\begin{align}
R_{{\rm sum},L}(D)&=\min_{1\leq l\leq L}R_{{\rm sum}}^{(l)}(D),
\label{eqn:RdFunc}
\end{align}
From (\ref{eqn:RdFunc}), we can see that 
$R_{\rm sum, L}(D)$ can be obtained by computing 
$R_{{\rm sum}}^{(l)}(D)$ for  $1\leq l\leq L$. 
In the following discussion we propose an algorithm to compute 
$\{(D, R_{{\rm sum}}^{(l)} (D)) \}_{l=1}^L$. 
To describe the algorithm, for each $1\leq l\leq L$, we define 
the sequence 
$\theta^{(l)}_\bullet(\omega)
=\{\theta_i^{(l)}(\omega)\}_{i=1}^l$
which consists of $l$ continuous functions of $\omega$.   
Concretely, for each $1\leq l\leq L$ and 
$\omega\in(0,(1+\epsilon_l)^{-1})$, define  
$\theta^{(l)}_\bullet(\omega)=\{\theta_i^{(l)}(\omega)\}_{i=1}^l$ by 
the following recursion:  
\begin{align}
& \theta_l^{(l)}(\omega)= \omega,
\nonumber\\
& \theta_{l-1}^{(l)}(\omega)=
\frac{ 
     \frac{ \theta_l^{(l)}(\omega)
            +\{(1+\epsilon_l)\theta_l^{(l)}(\omega)\}
                  \{1-\epsilon_l\theta_l^{(l)}(\omega)\}}
          {\tau_l}+1
}
{1+\epsilon_{l-1}
    \left[
    \frac{ \theta_l^{(l)}(\omega)
            +\{(1+\epsilon_l)\theta_l^{(l)}(\omega)\}
                  \{1-\epsilon_l\theta_l^{(l)}(\omega)\}}
          {\tau_l}+1
    \right]}       
\label{eqn:RecurEqaa}
\end{align}
\begin{align}
& \theta_{i-1}^{(l)}(\omega)=
 \frac{
 \frac{1}{\tau_i}\left[2\theta_i^{(l)}(\omega)
-\frac{1+\frac{\theta_{i+1}^{(l)}(\omega)}{\tau_{i+1}}}
  {1+\epsilon_i\left(1+\frac{\theta_{i+1}^{(l)}(\omega)} 
  {\tau_{i+1}}\right)}+\tau_i\right]
  }
  {1+\frac{\epsilon_{i-1}}{\tau_i} \left[2\theta_i^{(l)}(\omega)
   -\frac{1+\frac{\theta_{i+1}^{(l)}(\omega)}{\tau_{i+1}}} 
   {1+\epsilon_i\left(1+\frac{\theta_{i+1}^{(l)}(\omega)} 
    {\tau_{l+1}}\right)}+\tau_i\right]},
\nonumber\\
&\qquad l-1\geq i\geq 2
\label{eqn:RecurEqbb}
\end{align}

Our main result is the following: 
\begin{Th}\label{Th:ThSumRateComp}
{\rm 
Let $\theta^{(l)}_\bullet(\omega)=\{\theta_i^{(l)}(\omega)\}_{i=1}^l$
be a sequence of functions defined by (\ref{eqn:RecurEqaa}) 
and (\ref{eqn:RecurEqbb}).
Suppose that the Gaussian source satisfies the TS condition 
and the following condition: 
\beq
\tau_l=\sigma^2_l/\sigma^2_{l-1} \geq 1,\: L\geq l\geq 2. 
\label{eqn:CondZa}
\eeq
Under (\ref{eqn:CondZa}), we have the following parametric form of 
$(D,R_{{\rm sum},l}(D))$ using $\theta^{(l)}_\bullet(\omega)$:
\begin{align}
&D=\frac{\sigma_1^2\sigma_{X_0}^2}{\sigma_{X_0}^2\theta_1^{(l)}(\omega)+\sigma_1^2},\nonumber\\
&R_{{\rm sum}, l}(D)= \frac{1}{2}\log\frac{\sigma_{X_0}^2}{D}
\nonumber\\
&+ \left(-\frac{1}{2}\right)
  \left[
  \sum_{i=1}^{l-1}
\Biggl\{
\log \left(1-\frac{\theta_i^{(l)}(\omega)}
                  {1-\epsilon_{i}\theta_i^{(l)}(\omega)}
    +\frac{ \theta_{i+1}^{(l)}(\omega) } {\tau_{i+1}}\right)
\right.
\nonumber\\
&\left.
 + \log \left( 1- \epsilon_{i}\theta_i^{(l)}(\omega) \right)
\Biggr\} 
+\log\left(
          1-\frac{\theta_l^{(l)}(\omega)}
          {1-\epsilon_l\theta_l^{(l)}(\omega)}
     \right)
 \right] .
\nonumber
\end{align}
}
\end{Th}

Proof of Theorem \ref{Th:ThSumRateComp} is give in Section \ref{sec:ProofofResuls}. 
When $\epsilon_{l}=0$ for $L\geq l\geq 1$ and $\tau_l=1$ 
for $L\geq l\geq 2$, the recursion 
(\ref{eqn:RecurEqaa}) and (\ref{eqn:RecurEqaa}) defining 
$\theta^{(L)}_\bullet(\omega)$
becomes
the following: 
\beq
\left.
\ba{l}
\theta_{L}^{(L)}(\omega)=\omega, \theta_{L-1}^{(L)}(\omega)
=2\omega
\vspace{1mm}\\
\theta_{l-1}^{(L)}(\omega)=
2\theta_{l}^{(L)}(\omega)-\theta_{l+1}^{(L)}(\omega)
\vspace{1mm}\\
\mbox{ for }L-1 \geq l \geq 2\,.
\ea
\right\} 
\label{eqn:SumEx0077z}
\eeq
Solving (\ref{eqn:SumEx0077z}), we obtain 
$\theta_l^{(L)}(\omega)=(L-l+1)\omega$. 
The parametric form of $R_{{\rm sum}, L}(D)$ becomes
\beq
\left.
\ba{rcl}
\sigma_1^2g_0(D)&=&\theta_{1}(\omega)=L\omega\,,
\vspace{1mm}\\
R_{{\rm sum}, L}(D)
&=&\ds \left(-\frac{L}{2}\right)\log(1-\omega)
+\frac{1}{2}\log\frac{\sigma_{X_0}^2}{D}\,.
\ea
\right\}
\label{eqn:zaasx}
\eeq
From (\ref{eqn:zaasx}), we have 
\beqa
R_{{\rm sum}, L}(D)
&=&
\left(-\frac{L}{2}\right)
\log
\left(
1-\frac{\sigma_1^2}{L}g_0(D)
\right)
\nonumber\\
& &
+\frac{1}{2}\log\frac{\sigma_{X_0}^2}{D}\,.
\label{eqn:ssasz}
\eeqa
In particular, by letting $L\to \infty$ in (\ref{eqn:ssasz}), we have 
\beqa
\lim_{L\to\infty}R_{{\rm sum}, L}(D)
&=&\frac{1}{2}\sigma_1^2g_0(D)+\frac{1}{2}\log\frac{\sigma_{X_0}^2}{D}
\nonumber\\
&=&\frac{\sigma_1^2}{2\sigma_{X_0}^2}
   \left[\frac{\sigma_{X_0}^2}{D}-1\right]
    +\frac{1}{2}\log\frac{\sigma_{X_0}^2}{D}\,.
\nonumber
\eeqa
The above formula coincides with the rate distortion 
function for the quadratic Gaussian CEO problem 
obtained by Oohama \cite{oh2}. Hence, our solution to 
$R_{{\rm sum}, L}(D)$ includes the previous result 
on the Gaussian CEO problem as a special case.

\section{Proofs of the Results}
\label{sec:ProofofResuls}


In this section we prove Theorem \ref{th:converse} and Lemma 
\ref{lm:region1a} stated in Section III and prove 
Lemma \ref{lm:region1b} stated in Section \ref{sec:SumRate}. 
Furthermore, on the computation of $R_{{\rm sum},L}(D)$, we 
prove Theorem \ref{Th:ThSumRateComp} 
stated in Section \ref{sec:SumRate}. 

\subsection{Derivation of the Outer Bound}

In this subsection we prove $\hat{\cal R}_L^{({\rm out})}(D)$
$\subseteq$ ${\cal R}_L^{({\rm out})}(D)$ stated 
in Theorem \ref{th:converse}. 

{\it Proof of $\hat{\cal R}_L^{({\rm out})}(D)$
$\subseteq$ ${\cal R}_L^{({\rm out})}(D)$:} 
Set
\begin{align*}
& \hat{\rdf}_S (D,r_0,r^{L-2},r_S|r_{\coS},R_0)
\\
&\defeq  \left[\frac{1}{2} \ts \log^{+}
\left[
\frac{G(D,r_0,r^{L-2})\sigma_{X_0}^2}{
F(r_{\coS})\left\{1+ \sigma_{X_0}^2f_0(r_{\coS})\right\}D
}
\right]
+{\ds \sum_{i=1}^L{r_i}}-R_0\right]^{+}\,.
\end{align*}
We first observe that
\begin{align}
&  \hat{J}_{S}(D,r_0,r^{L-2},r_S|r_{\coS},r_0)
\nonumber\\
&= \frac{1}{2}\left[ \ts \log^{+}
\left[
\frac{G(D,r_0,r^{L-2})\sigma_{X_0}^2}{
F(r_{\coS})\left\{1+ \sigma_{X_0}^2 f_0(r_{\coS})\right\}
D}
\right]
+{\ds \sum_{i=1}^L{2r_i}}-2r_0\right]^+
\nonumber\\
&\geq  \frac{1}{2}\left[\ts \log
\left[
\frac{G(D,r_0,r^{L-2})\sigma_{X_0}^2}{
F(r_{\coS})\left\{1+ \sigma_{X_0}^2 f_0(r_{\coS})\right\}
D}
\right]
+{\ds \sum_{i=1}^L{2r_i}}-2r_0\right]^+
\nonumber\\
&={J}_{S}(D,r_0,r^{L-2},r_S|r_{\coS})\,.
\label{eqn:zaaa0}
\end{align}
Then, we have the following. 
\begin{align}
& \hat{\cal R}_L^{{(\rm out)}}(D)
\notag\\
&\MSub{a}\Bigl\{(R_0,R^L):\Bigr.
\ba[t]{l}
\mbox{There exists a nonnegative vector }
\vspace{1mm}\\
(r_0,r^L)\mbox{ such that }
\vspace{1mm}\\
R_0\geq r_0 \geq 
       \ds\frac{1}{2} \ts \log^{+}
        \left[\frac{\sigma_{X_0}^2}
              {\left\{
              1+ \sigma_{X_0}^2 f_0(r^L) 
              \right\}
              D}
        \right]\,,
\vspace{1mm}\\
\ds \sum_{i\in S}R_i 
\geq \hat{\rdf}_S(D,r_0,r^{L-2},r_S|r_{\coS}, R_0)
\vspace{1mm}\\
\mbox{ for any }S\subseteq \Lambda\,.\Bigr\}
\ea
\notag\\
&\MEq{b}\Bigl\{(R_0,R^L):\Bigr.
\ba[t]{l}
\mbox{There exists a nonnegative vector }
\vspace{1mm}\\
r^L\mbox{ such that }
\vspace{1mm}\\
R_0    \geq 
       \ds\frac{1}{2} \ts \log^{+}
        \left[\frac{\sigma_{X_0}^2}
              {\left\{
              1+ \sigma_{X_0}^2 f_0(r^L) 
              \right\}
              D}
        \right]\,,
\vspace{1mm}\\
\ds \sum_{i\in S}R_i 
\geq \hat{\rdf}_S(D,R_0,r^{L-2},r_S|r_{\coS}, R_0)
%
\ea
\notag \\
& \qquad \qquad \qquad \quad\Bigl.\mbox{ for any }S\subseteq      
  \Lambda\,.\Bigr\}.
\label{eqn:zdffa}
\end{align}
Step (a) follows from the definition of 
$\hat{\rdf}_S (D,R_0,r^{L-2},r_S$ $|r_{\coS}, R_0)$ 
and the nonnegative property of $R^L$. 
Step (b) follows from that 
$\hat{\rdf}_S(D,r_0,$ $r^{L-2},r_S|r_{\coS}, R_0)$ 
is a monotone decreasing function of $r_0$.
From (\ref{eqn:zdffa}), we continue to evaluate outer bounds 
of $\hat{\cal R}_L^{{(\rm out)}}(D)$ to obtain the following:
\begin{align*}
& \hat{\cal R}_L^{{(\rm out)}}(D)
\notag\\
&\subseteq\Bigl\{(R_0,R^L):\Bigr.
\ba[t]{l}
\mbox{There exists a nonnegative vector }
\vspace{1mm}\\
(r_0,r^L)\mbox{ such that }
\vspace{1mm}\\
R_0\geq r_0\geq 
        \ds\frac{1}{2} \ts \log
        \left[\frac{\sigma_{X_0}^2}
              {\left\{
              1+ \sigma_{X_0}^2 f_0(r^L) 
              \right\}
              D}
        \right]\,,
\vspace{1mm}\\
\ds \sum_{i\in S}R_i 
\geq \hat{\rdf}_S(D,r_0,r^{L-2},r_S|r_{\coS},r_0)
\vspace{1mm}\\
\mbox{ for any }S\subseteq \Lambda\,.\Bigr\}
\ea
\\
&\MSub{c}
\Bigl\{(R_0,R^L):\Bigr.
\ba[t]{l}
\mbox{There exists a nonnegative vector }
\vspace{1mm}\\
(r_0,r^L)\mbox{ such that }
\vspace{1mm}\\
R_0\geq r_0\geq 
        \ds\frac{1}{2} \ts \log
        \left[\frac{\sigma_{X_0}^2}
              {\left\{
              1+ \sigma_{X_0}^2 f_0(r^L) 
              \right\}
              D}
        \right]\,,
\vspace{1mm}\\
\ds \sum_{i\in S}R_i 
\geq {\rdf}_S(D,r_0,r^{L-2},r_S|r_{\coS})
\vspace{1mm}\\
\mbox{ for any }S\subseteq \Lambda\,.\Bigr\}
={\cal R}_{L}^{(\rm out)}(D)\,.
\ea
\end{align*}
Step (c) follows from (\ref{eqn:zaaa0}).
Thus 
$\hat{\cal R}_{L}^{(\rm out)}(D)$
$\subseteq$
${\cal R}_{L}^{(\rm out)}(D)$
is proved. \hfill\IEEEQED

\subsection{Derivation of the Inner Bound}

In this subsection we prove ${\cal R}_L^{({\rm in})}(D)$
$\subseteq$ $\tilde{\cal R}_L^{({\rm in})}(D)$ 
stated in Theorem \ref{th:converse}. 
We first derive a preliminary result on a form 
of ${\cal R}_L^{(\rm in)}(D)$. Fix 
$R_0 \geq r_0$ and set 
\beqno
{\cal R}_L^{({\rm in})}(r_0^L|R_0)
&\defeq &
\{(R_1,\cdots,R_L):
\\
&  &\ba[t]{l} (R_0,R_1,\cdots,R_L)
    \in {\cal R}_L^{({\rm in})}(r_0^L)\}\,.
    \ea
\eeqno
Let $(\Lambda,\tilde{\rho})$, 
$\tilde{\rho}=\{\tilde{\rho}_S(r_S|r_{\coS})\}_{S\subseteq \Lambda}$ 
be a co-polymatroid defined in Property \ref{pr:matroid}. 
Expression of ${\cal R}_L^{(\rm in)}(r_0^L|R_0)$ 
using $(\Lambda,\tilde{\rho})$ is
\beqno
{\cal R}_L^{({\rm in})}(r_0^L|R_0)
&=& \{(R_1,\cdots,R_L):
    \ba[t]{l}
    \ds \sum_{i \in S} R_i \geq \tilde{\rho}_{S}\left(r_S|r_{\coS} \right)
    \vspace{1mm}\\
    \mbox{ for any }S \subseteq \Lambda\,. \}\,.
    \ea
\eeqno
The set ${\cal R}_L^{(\rm in)}(r_0^L$ $|R_0)$ forms 
a kind of polytope, which is called a 
{\it co-polymatroidal polytope} in the terminology of 
matroid theory. 
It is well known as a property of this kind 
of polytope that the polytope 
${\cal R}_L^{(\rm in)}(r_0^L$ $|R_0)$ 
consists of $L!$ end-points whose components are given by 
\beq
\left.
\ba{rcl}
& &R_{\pi(i)}\\
&=&\tilde{\rho}_{\{ \pi(i),\cdots, \pi(L)\}}
  (r_{\{\pi(i),\cdots,\pi(L)\}}|r_{\{\pi(1),\cdots, \pi(i-1)\}})
\vspace*{1mm}\\
& &-\tilde{\rho}_{\{\pi(i+1), \cdots, \pi(L)\}}
 (r_{\{ \pi(i+1),\cdots,\pi(L)\}}|r_{\{ \pi(1), \cdots, \pi(i)\}})
\vspace*{1mm}\\
& &\mbox{$\quad$ for }i=1,2,\cdots, L-1\,,
\vspace*{1mm}\\
& &R_{\pi(L)}=\tilde{\rho}_{\{\pi(L)\}}
(r_{\pi(L)}|r_{\{ \pi(1), \cdots, \pi(L-1)\}})
\,,
\ea
\right\}
\label{eqn:za0}
\eeq
where 
$$
\pi=\left(
\ba{cccccc}
    1 &\cdots&    i &\cdots&    L\\ 
\pi(1)&\cdots&\pi(i)&\cdots&\pi(L)
\ea
\right)\in \Pi
$$
is an arbitrary permutation on $\Lambda$. 
For each $\pi \in \Pi$ and $r_0^L\in$ ${\cal B}_L(D)$, 
let ${\cal R}_{\piLam}^{({\rm in})}(r_0^L)$ be the set of 
nonnegative vectors $(R_0,R_1,\cdots,R_L)$ satisfying 
\beq
\left.
\ba{rcl}
& &R_{0}\geq r_0
\vspace*{1mm}\\
& &R_{\pi(i)}\\
&\geq&\tilde{\rho}_{\{ \pi(i),\cdots, \pi(L)\}}
  (r_{\{\pi(i),\cdots,\pi(L)\}}|r_{\{\pi(1),\cdots, \pi(i-1)\}})
\vspace*{1mm}\\
& &-\tilde{\rho}_{\{\pi(i+1), \cdots, \pi(L)\}}
 (r_{\{ \pi(i+1),\cdots,\pi(L)\}}|r_{\{ \pi(1), \cdots, \pi(i)\}})
\vspace*{1mm}\\
& &\mbox{$\quad$ for }i=1,2,\cdots, L-1\,,
\vspace*{1mm}\\
& &R_{\pi(L)}\geq
\tilde{\rho}_{\{\pi(L)\}}(r_{\pi(L)}|r_{\{ \pi(1), \cdots, \pi(L-1)\}})\,.
\ea
\right\}
\label{eqn:za0bb}
\eeq
Then, we have
\beqno
{\cal R}^{({\rm in})}(r_0^L)
=
{\rm conv}\left\{
 \bigcup_{\pi\in\Pi}
{\cal R}_{\piLam}^{({\rm in})}(r_0^L)
\right\}\,.
\label{eqn:za0bbz}
\eeqno

{\it Proof of ${\cal R}_L^{({\rm in})}(D)\subseteq \tilde{\cal R}_L^{({\rm in})}(D)$: }
Fix $\pi\in {\Pi}$ and $r_0^L\in {\cal B}_L(D)$ arbitrary.
By (\ref{eqn:za0bb}), it suffices to show that 
for $r_0^L\in {\cal B}_L(D)$, 
${\cal R}_{\piLam}^{({\rm in})}(r_0^L)$
$\subseteq$
$\tilde{\cal R}_{\piLam}^{({\rm in})}(D)$ 
to prove ${\cal R}_L^{({\rm in})}(D)
\subseteq \tilde{\cal R}_L^{({\rm in})}(D)$.
Let $V_i, i\in \left\{0\right\}\cup \Lambda$ be independent Gaussian 
random variables with mean 0 and variance $\sigma_{V_i}^2$. Suppose that 
$V_0^L$ is independent of $X_0^L$.
Define the Gaussian random variables 
$U_i, i\in \left\{0\right\}\cup \Lambda$ by 
\beqno
U_i \defeq X_i+V_i,\:\: i\in \left\{0\right\}\cup \Lambda.
\eeqno
From the above definition it is obvious that
\beq
\left.
\ba{l}
U^L\to X^L\to X_0 \to U_0,\\
U_S\to X_{S\cup\{0\}} \to X_{\coS} \to U_{\coS},\\
\mbox{ for any } S\subseteq \Lambda \,.  
\ea
\right\}
\label{eqn:gau00} 
\eeq
For given $r_i \geq 0, i\in S$ and $D>0$, set
$
\ts \frac{1}{\sigma_{V_i}^2 }=$ $\ts\frac{ {\rm e}^{2r_i} -1 }{\sigNi},
$
when $r_i>0$. When $r_i=0,$ we choose $U_i$ so that $U_i$ takes 
the constant value zero. Define the sequence of random variables 
$\{\Omega_{l}\}_{l=0}^{L}$ by 
\beq
\left.
\ba{rcl}
\Omega_{L-1}
&=& \ts \frac{1-{\rm e}^{-2r_{L-1}}}{\sigma_{N_{L-1}}^2}\cdot U_{L-1}
+ \ts \frac{1-{\rm e}^{-2r_L}}{\sigma_{N_L}^2}\cdot U_{L}
\vspace{1mm}\\
\Omega_{l}
&=&\ts \frac{1}{1+\sigma_{Z_{l+1}}^2f_{l+1}(r_{l+1}^L)}\cdot \Omega_{l+1}
   +\frac{1-{\rm e}^{-2r_{l}}}{\sigma_{N_l}^2}\cdot U_{l}
\vspace{1mm}\\
& &\mbox{ for }L-2 \geq l \geq 1
\vspace{1mm}\\
\Omega_{0}&=&\ts \frac{1}{1+\sigma_{Z_{1}}^2f_{1}(r^L)}\cdot \Omega_{1}\,.
\ea
\right\}
\eeq
Note that $\Omega_{0}=\Omega_{0}(U^L)$ is a linear function of   
$U^L$. Then, by an elementary computation, we have 
\beqa
   X_0
&=&\frac{1}{
 \frac{1}{\sigma_{X_0}^2}
+\frac{1}{\sigma_{V_0}^2}
+f_0(r^L)}
\left[\ts \frac{1}{\sigma_{V_0}^2}\cdot U_0+\Omega_{0}(U^L)\right]
\nonumber\\
& &+\tilde{N}_0\,,
\label{eqn:zaa00z}
\eeqa
where $\tilde{N}_0$ is a zero mean Gaussian random 
variable with variance 
$$
\left[\ts 
\frac{1}{\sigma_{X_0}^2}+\frac{1}{\sigma_{V_0}^2}+f_0(r^L)
\right]^{-1}\,. 
$$
$\tilde{N}_0$ is independent of $(U_0,U^L)$. 
Since $r_0^L \in {\cal B}_L(D)$, we have 
\beq
\ts \frac{{\rm e}^{-2r_0}}{D}-\frac{1}{\sigma_{X_0}^2} \leq f_0(r^L)\,.
\label{eqn:zzz0x}
\eeq
We put 
%
\beq
\ts \frac{1}{\sigma_{V_0}^2} =\frac{1-{\rm e}^{-2r_0}}{D}\,.
\label{eqn:z0z0x}
\eeq
Then, from (\ref{eqn:zzz0x}) and (\ref{eqn:z0z0x}), we have
\beqa
& &\left[
\ts 
\frac{1}{\sigma_{X_0}^2}+\frac{1}{\sigma_{V_0}^2}+f_0(r^L)
\right]^{-1} 
\nonumber\\
&=&\left[
\ts 
\frac{1}{\sigma_{X_0}^2}+\frac{1-{\rm e}^{-2r_0}}{D}
+f_0(r^L)
\right]^{-1}
\leq D\,.
\label{eqn:zaa00z0}
\eeqa
Based on (\ref{eqn:zaa00z}), (\ref{eqn:z0z0x}), 
and (\ref{eqn:zaa00z0}), define the linear function $\tilde{\psi}$ 
of $(U_0,U^L)$ by
\beqno
\tilde{\psi}(U_0, U^L)
&\defeq &\left[
\ts 
\frac{1}{\sigma_{X_0}^2}+\frac{1-{\rm e}^{-2r_0}}{D}+f_0(r^L)
\right]^{-1} 
\\
& & \times 
\left [\ts \frac{1-{\rm e}^{-2r_0}}{D}
\cdot U_0 + \Omega_{0}(U^L)
\right]\,.
\eeqno
Then, we obtain
\beqa
& &{\rm E}\left[X_0-\tilde{\psi}(U_0, U^L)\right]^2
={\rm \bf Var}\left[\tilde{N}_0\right]
\nonumber\\
&=&
\left[
\ts 
\frac{1}{\sigma_{X_0}^2}+\frac{1-{\rm e}^{-2r_0}}{D}+f_0(r^L)
\right]^{-1} 
\leq D\,.
\label{eqn:gau2} 
\eeqa
From (\ref{eqn:gau00}) and (\ref{eqn:gau2}), 
we have $(U_0,U^L)\in {\cal G}(D)$. 
By simple computations, we can show that 
\beq
\left.
\ba{rcl}
r_0&=&I(X_0;U_0|U^L)\,,
\\
r_i&=&I(X_i;U_i|X_0Y^{L-1})\,,
\\
   & &\mbox{ for any }i\in \Lambda\,,
\vspace{1mm}\\
& &\ts \frac{1}{2}\log \left[F_S(r_S)\cdot\{1+\sigma_{X_0}^2f_0(r_S)\}\right]
\vspace{1mm}\\
&=&I(X_0Y^{L-1};U_{S})\,,
\\
   & & \mbox{ for any }S\subseteq \Lambda\,. 
\ea
\right\}
\label{eqn:gau1} 
\eeq
Using (\ref{eqn:gau00}) and (\ref{eqn:gau1}), 
the $L+1$ inequalities of (\ref{eqn:za0bb}) 
are rewritten as 
\beqa
R_0&\geq&I(X_0;U_0|U^L)\,,
\nonumber\\
R_{\pisi}
&\geq& I(X_0Y^{L-1};U_{\pi(S_i)}|U_{\pi(S_i^{\rm c})})
\nonumber\\
&&   +I(X_{\pi(i)};U_{\pi(i)}|X_0Y^{L-1})
\nonumber\\
& & -I(X_0Y^{L-1};U_{\pi(S_{i+1})}|U_{\pi(S_{i+1}^{\rm c})}) 
\nonumber\\
&=& I(X_0Y^{L-1};U_{\pi(i)};|U_{\pi(S_i^{\rm c})}) 
\nonumber\\
& &+I(X_{\pi(i)};U_{\pi(i)}|X_0Y^{L-1}U_{\pi(S_i^{\rm c})}) 
\nonumber\\
&=&I(X_0Y^{L-1}X_{\pi(i)};U_{\pisi}|U_{\pi(S_i^{\rm c})}) 
\nonumber\\
&=&I(X_{\pi(i)};U_{\pisi}|U_{\pi(S_i^{\rm c})}) 
\nonumber\\
& &\quad \mbox{ for } i=1,2, \cdots, L\,.   
\nonumber
\eeqa
Thus, we conclude that $(R_0,R_{\pi(1)},$ $\cdots, R_{\pi(L)}) 
\in\tilde{\cal R}_{\piLam}^{({\rm in})}(D)$.\\
$\quad$\hfill\IEEEQED

\subsection{
Proofs of Lemmas \ref{lm:region1a} and \ref{lm:region1b}
} 

In this subsection we prove Lemmas \ref{lm:region1a} and \ref{lm:region1b}.
We first present a preliminary observation on ${\cal R}_L^{(\rm out)}(D)$. 
Fix $R_0 \geq r_0$ arbitrary and set 
\beqno
{\cal R}_L^{({\rm out})}(D,r_0^L|R_0)
&\defeq &
\{(R_1,\cdots,R_L):
\\
&  &\ba[t]{l} (R_0,R_1,\cdots,R_L)
    \in {\cal R}_L^{({\rm out})}(D,r_0^L)\}\,.
    \ea
\eeqno
Let 
$(\Lambda,\rho)$, 
$\rho=\{\rho_S(r_S|r_{\coS})\}_{S\subseteq \Lambda}$ 
be a co-polymatroid defined in Property \ref{pr:matroid}. 
Expression of ${\cal R}_L^{(\rm out)}(D_0,r_0^L|R_0)$ 
using $(\Lambda,\rho)$ is
\beqno
{\cal R}_L^{({\rm out})}(D,r_0^L|R_0)
&=& \{(R_1,\cdots,R_L):
    \ba[t]{l}
    \ds \sum_{i \in S} R_i \geq {\rho}_{S}\left(r_S|r_{\coS}\right)
    \\
    \mbox{ for any }S \subseteq \Lambda\,. \}\,.
    \ea
\eeqno
The set ${\cal R}_L^{(\rm out)}(D,r_0^L$ $|R_0)$ 
forms a {co-polymatroidal polytope}. The polytope 
${\cal R}_L^{(\rm out)}(D,r_0^L$ $|R_0)$ consists 
of $L!$ end-points whose components are given by 
\beq
\left.
\ba{rcl}
& &R_{\pi(i)}\\
&=& {\rho}_{\{\pi(i), \cdots, \pi(L)\}}
   (r_{\{\pi(i),\cdots, \pi(L)\}}|r_{\{\pi(1),\cdots, \pi(i-1)\}})
\vspace*{1mm}\\
& &-{\rho}_{\{\pi(i+1), \cdots, \pi(L)\}}
   (r_{\{\pi(i+1),\cdots,\pi(L)\}}|r_{\{ \pi(1), \cdots, \pi(i)\}})
\vspace*{1mm}\\
& &\mbox{$\quad$ for }i=1,2,\cdots, L-1\,,
\vspace*{1mm}\\
& &R_{\pi(L)}={\rho}_{\{\pi(L)\}}
(r_{\pi(L)}|r_{\{ \pi(1), \cdots, \pi(L-1)\}})\,,
\ea
\right\}
\label{eqn:zzza}
\eeq
where 
$$
\pi=\left(
\ba{ccccc}
    1 &\cdots&    i &\cdots&    L\\ 
\pi(1)&\cdots&\pi(i)&\cdots&\pi(L)
\ea
\right)\in \Pi\,.
$$
For each $\pi\in \Pi$ and $l=1,2,\cdots,L$, set
\beqno
{\cal B}_{\pi,l}(D)
&\defeq& 
\{r_0^L: 
\ba[t]{l} 
r_0^L\in {\cal B}_{L}(D)\mbox{ and }\\
r_{\pi(i)}=0\mbox{ for } i=l+1,\cdots,L \}\,, \\
\ea
\\
\partial{\cal B}_{\pi,l}(D)
&\defeq& 
\{r_0^L: 
\ba[t]{l} 
r_0^L\in \partial{\cal B}_{L}(D)\mbox{ and }\\
r_{\pi(i)}=0\mbox{ for } i=l+1,\cdots,L \}\,. \\
\ea
\eeqno
In particular, when $\pi$ is the identity map, 
we omit $\pi$ to write ${\cal B}_l(D)$ 
and $\partial{\cal B}_l(D)$. By Property \ref{pr:prz01z}, 
when $r_0^L \in {\cal B}_{\pi,l}(D)$, the end-point given 
by (\ref{eqn:zzza}) becomes
\beq
\left.
\ba{rcl}
& &R_{\pi(i)}\\
&=&{\rho}_{\{\pi(i), \cdots, \pi(l)\}}
   (r_{\{\pi(i),\cdots, \pi(l)\}}|r_{\{\pi(1),\cdots, \pi(i-1)\}})
\vspace*{1mm}\\
& &-{\rho}_{\{\pi(i+1), \cdots, \pi(l)\}}
   (r_{\{ \pi(i+1),\cdots,\pi(l)\}}|r_{\{ \pi(1), \cdots, \pi(i)\}})
\vspace*{1mm}\\
& &\mbox{$\quad$ for }i=1,2,\cdots, l-1\,,
\vspace*{1mm}\\
& &R_{\pi(l)}={\rho}_{\{\pi(l)\}}(r_{\pi(l)}|r_{\{ \pi(1), \cdots, \pi(l-1)\}})\,,
\vspace*{1mm}\\
& &R_{\pi(i)}=0,\mbox{ for }i=l+1,\cdots, L\,.
\ea
\right\}
\label{eqn:zza0}
\eeq
Next, we present a lemma on a property of $G(D,r_0,r^{L-1}).$
\begin{lm}\label{lm:lemlem} For $r_0^L\in {\cal B}_l(D)$, 
$G(D,r_0,r^{L-2})$ is computed as 
$$
\left. G(D,r_0,r^{L-2})\right|_{r_{l+1}^L={\lvc 0}}
=\prod_{k=1}^{l}\left[1+\sigma_{Z_k}^2g_k(D,r_0,r^{k-1})\right]\,.
$$
\end{lm}

{\it Proof:} 
By Property \ref{pr:pr01} part c), for $l+1\leq k\leq L$
$$
0 \leq g_k(D,r_0,r^{k-1})\leq f(r_k^L)=0\,.
$$
Hence, the result of Lemma \ref{lm:lemlem} follows.
\hfill\IEEEQED

{\it Proof of Lemma \ref{lm:region1a}:} Fix $\pi\in \Pi$ 
and $r_0^L \in {\cal B}_L(D)$ arbitrary. Let $(R_0,R^L)$ 
be a nonnegative rate vector such that 
$R_0\geq r_0$ and $L$ components of $R^L$ 
satisfy (\ref{eqn:zzza}). To prove Lemma \ref{lm:region1a}, 
it suffices to show that this nonnegative vector 
belongs to ${\cal R}_L^{(\rm in)}(D)$. 
For $l=1,2,\cdots,L$, we prove the claim that under 
the MI condition, 
if $r_0^L \in {\cal B}_{\pi,l}(D)$, then, the rate 
vector $(R_0,R^L)$ satisfying $R_0\geq r_0$ and 
(\ref{eqn:zza0}) belongs to ${\cal R}_L^{(\rm in)}(D)$. 
We prove this claim by induction with respect to $l$. 
When $l=1$, from (\ref{eqn:zza0}), we have 
\beq
\left.
\ba{rcl}
R_{\pi(1)}&=&\rho_{\{\pi(1)\}}(r_{\pi(1)})\,,
\vspace*{1mm}\\
R_{\pi(i)}&=&0,\mbox{ for }i=2,\cdots,L\,.
\ea
\right\}
\label{eqn:aaz}
\eeq
The function $\rho_{\{\pi(1)\}}(r_{\pi(1)})$ is computed as
\beqa
\hspace*{-4mm}& &\rho_{\{\pi(1)\}}(r_{\pi(1)})
\nonumber\\
\hspace*{-4mm}&=&J_{\{\pi(1)\}}
\left.(D,r_0,r^{L-2},r_{\pi(1)}|r_{\{\pi(1)\}^{\rm c}})
\right|_{r_{\{\pi(1)\}^{\rm c}}={\lvc 0}}
\nonumber\\
\hspace*{-4mm}&=&
\frac{1}{2} \ts \log^{+} 
        \left[
        \frac{
        \left. G(D,r_0,r^{L-2})\right|_{r_{\{\pi(1)\}^{\rm c}}={\lvc 0}}
        \sigma_{X_0}^2{\rm e}^{-2r_{0}}{\rm e}^{2r_{\pi(1)}}
        }{D}
\right]\,.
\label{eqn:za00a}
\eeqa
By the above form of $\rho_{\{\pi(1)\}}(r_{\pi(1)})$ and  
$$
\ts \frac{\sigma_{X_0}^2{\rm e}^{-2r_{0}}}{D}
\geq \frac{\sigma_{X_0}^2{\rm e}^{-2R_{0}}}{D}>1\,,
$$
$\rho_{\{\pi(1)\}}(r_{\pi(1)})$ 
is positive. Since $r_0^L \in {\cal B}_{\pi,l}(D)$, 
we can decrease $r_{\pi(1)}$ keeping 
$r_0^L\in {\cal B}_{\pi,1}(D)$ 
so that it arrives at $r_{\pi(1)}^{*}=0$ or 
a positive $r_{\pi(1)}^{*}$ satisfying
\beqa
& & (r_0,r_{\pi(1)}^{*},r_{\{\pi(1)\}^{\rm c}})
\nonumber\\
&=&(r_0,r_{\pi(1)}^{*},\underbrace{0,\cdots,0}_{L-1})
\in \partial{\cal B}_{\pi,1}(D)\,.
\label{eqn:z00}
\eeqa
Let $(R_0,$ $R_{\pi(1)}^{*},$ $\cdots, R_{\pi(L)}^{*})$
be a rate vector corresponding to 
$(r_0,$ $r_{\pi(1)}^{*},$ $r_{\{\pi(1)\}^{\rm c}})$.
If $r_{\pi(1)}^{*}=0$, then by Property \ref{pr:prz01z} part b), 
$\rho_{\{\pi(1)\}}(r_{\pi(1)})$ must be zero. This contradicts 
the fact that $\rho_{\{\pi(1)\}}(r_{\pi(1)})$ is positive. 
Therefore, $r_{\pi(1)}^{*}$ must be positive. Then, 
from (\ref{eqn:z00}), we have 
\beqno
& &(R_0,R_{\pi(1)}^{*},\cdots, R_{\pi(L)}^{*})
\\
&=&(R_0,R_{\pi(1)}^{*},\underbrace{0,\cdots,0}_{L-1})
\in {\cal R}_L^{(\rm in)}(D)\,. 
\eeqno
On the other hand, by Lemma \ref{lm:lemlem}, we have  
\beqa
& &\left. G(D,r_0,r^{L-2})\right|_{r_{\{\pi(1)\}^{\rm c}}={\lvc 0}}
\nonumber\\
&=&   \left. 
      G(D,r_0,r^{L-2})
       \right|_{r_{\pi(1)+1}^{L}={\lvc 0}, r^{\pi(1)-1}={\lvc 0}}
\nonumber\\
&=&\left. \prod_{k=1}^{ {\pi(1)} }
    \left[1+\sigma_{Z_l}^2g_k(D,r_0,r^{k-1}) \right]
    \right|_{r^{\pi(1)-1}={\lvc 0}}\,.
\label{eqn:ztt}
\eeqa
From (\ref{eqn:za00a}) and (\ref{eqn:ztt}), 
we can see that
$\left. G(D,r_0,r^{L-2})\right|_{r_{\{\pi(1)\}^{\rm c}}={\lvc 0}}$
does not depend on $r_{\pi(1)}$. This implies that 
$\rho_{\{\pi(1)\}}(r_{\pi(1)})$ is a monotone increasing 
function of $r_{\pi(1)}$. 
Then, we have $R_{\pi(1)} \geq $ $R_{\pi(1)}^{*}$.
Hence, we have 
\beqno
& &(R_0,R_{\pi(1)},\cdots, R_{\pi(L)})
\\
&=&(R_0,R_{\pi(1)},\underbrace{0,\cdots, 0}_{L-1})
   \in {\cal R}^{(\rm in)}_L(D)\,. 
\eeqno
Thus, the claim holds for $l=1$. 
We assume that the claim holds for $l-1$.
Since $f_0(r_0^L)$ is a monotone increasing function of 
$r_{\pi(l)}$ on ${\cal B}_{\pi,l}(D)$, we can 
decrease $r_{\pi(l)}$ keeping $r_0^L\in {\cal B}_{\pi,l}(D)$ 
so that it arrives at $r_{\pi(l)}^{*}=0$ or 
a positive $r_{\pi(l)}^{*}$ satisfying
\beq
(r_0,r_{\pi(l)}^{*},r_{\{\pi(l)\}^{\rm c}})
\in \partial{\cal B}_{\pi,l}(D)\,.
\label{eqn:zaaa}
\eeq
Let $(R_0,$ $R_{\pi(1)}^{*},$ $\cdots, R_{\pi(L)}^{*})$ 
be a rate vector corresponding to 
$(r_0,$ $r_{\pi(l)}^{*},$ $r_{\{\pi(l)\}^{\rm c}})$.
By Property \ref{pr:matroid} part b) and the MI condition, 
the $l$ functions  
\beqno
& &{\rho}_{\{\pi(i), \cdots, \pi(l)\}}
   (r_{\{\pi(i),\cdots, \pi(l)\}}|r_{\{\pi(1),\cdots, \pi(i-1)\}})
\\
& &-{\rho}_{\{\pi(i+1), \cdots, \pi(l)\}}
   (r_{\{ \pi(i+1),\cdots,\pi(l)\}}|r_{\{ \pi(1), \cdots, \pi(i)\}})
\\
& &\mbox{$\quad$ for }i=1,2,\cdots, l-1\,,
\vspace*{1mm}\\
& &{\rho}_{\{\pi(l)\}}(r_{\pi(l)}|r_{\{ \pi(1), \cdots, \pi(l-1)\}})\,,
\eeqno
appearing in the right members of (\ref{eqn:zza0}) are monotone 
increasing functions of $r_{\pi(l)}$. Then, from (\ref{eqn:zza0}), 
we have
\beq
\left.
\ba{rcl}
R_{\pi(i)} &\geq &R_{\pi(i)}^{*} 
\mbox{ for }i=1,2,\cdots,l\,,
\\
R_{\pi(i)} &= &R_{\pi(i)}^{*}=0
\mbox{ for }i=l+1,\cdots,L\,.
\ea
\right\}
\label{eqn:zazaa}
\eeq
When $r_{\pi(l)}^{*}=0$, we have 
$
(r_0,r_{\pi(l)}^{*},r_{\{\pi(l)\}^{\rm c}})
\in {\cal B}_{\pi,l-1}(D)\,.
$
Then, by induction hypothesis, we have
$$
(R_0,R_{\pi(1)}^{*},\cdots, R_{\pi(L)}^{*})
\in {\cal R}_L^{(\rm in )}(D)\,. 
$$
When $r_{\pi(l)}^{*}>0$, from (\ref{eqn:zaaa}), we have 
$$
(R_0,R_{\pi(1)}^{*},\cdots, R_{\pi(L)}^{*})
\in {\cal R}_L^{(\rm in)}(D)\,. 
$$
Hence, by $(\ref{eqn:zazaa})$, we have
\beqno
& &(R_0,R_{\pi(1)},\cdots, R_{\pi(L)})
\\
&=&(R_0,R_{\pi(1)}, \cdots, R_{\pi(l)},
\underbrace{0,\cdots, 0}_{L-l})
\in {\cal R}^{(\rm in)}_L(D)\,. 
\eeqno
Thus, the claim holds for $l$. This completes 
the proof of Lemma \ref{lm:region1a}.    
\hfill\IEEEQED

{\it Proof of Lemma \ref{lm:region1b}:} 
For $R_0>0$ and for $1\leq l\leq L$, set
\beqno
{\cal B}_{l}(D|R_0)
&\defeq& \{r^l: (R_0,r^L)\in {\cal B}_{l}(D)\}\,,
\\
\partial{\cal B}_{l}(D|R_0)
&\defeq& 
\{r^l: (R_0,r^L)\in \partial{\cal B}_{l}(D)\}\,.
\eeqno
We first observe that
\beqa
& &R_{{\rm sum},L}^{(\rm l)}(D, R_0)
\nonumber\\
&=&\min_{1\leq l\leq L}
\left[\min_{\scs r^l \in {\cal B}_l(D|R_0)}
\left. J_{\Lambda}(D, R_0,r^{L-2},r^L)
\right|_{r_{l+1}^{L}={\lvc 0}}
\right]\,,
\nonumber\\
& &R_{{\rm sum}, L}^{(\rm u)}(D, R_0)
\nonumber\\
&=&\min_{1\leq l\leq L}
\left[\min_{r^l\in \partial{\cal B}_l(D|R_0)}
K_{\Lambda}(r^l)\right]\,.
\nonumber
\eeqa
We compute 
$
J_{\Lambda}(D, R_0,$ 
$\left.r^{L-2},r^L)
\right|_{r_{l+1}^{L}={\lvc 0}}.
$
By Lemma \ref{lm:lemlem}, for $r^l\in {\cal B}_l(D|R_0)$
$$
\left. G(D, R_0,r^{L-2})\right|_{r_{l+1}^L={\lvc 0}}
=\prod_{k=1}^{l}\left[1+\sigma_{Z_l}^2g_k(D, R_0,r^{k-1})\right]\,.
$$
From the above formula, we can see that 
for $r^l\in {\cal B}_l(D|R_0)$, 
$G(D, R_0,r^{L-2})$
$|_{r_{l+1}^L={\lvc 0}}$ 
is a function of $r^{l-1}$. We denote this function by 
$G(D, R_0,r^{l-1})$, that is,
$$
G(D, R_0,r^{l-1})
\defeq \prod_{k=1}^{l}
\left[1+\sigma_{Z_l}^2g_k(D, R_0,r^{k-1})\right]\,.
$$
Then, for $r^l\in {\cal B}_l(D|R_0)$, 
\beqa
& & \left. J_{\Lambda}(D, R_0,r^{L-2},r^L)\right|_{r_{l+1}^{L}={\lvc 0}}
\nonumber\\
&=&\frac{1}{2}\log^{+}
\left[
G(D, R_0,r^{l-1})\cdot 
\frac{\sigma_{X_0}^2}{D}{\rm e}^{-2R_0}
\prod_{i=1}^l{\rm e}^{2r_i}
\right]\,.
\label{eqn:zas}
\eeqa
We denote the right member of (\ref{eqn:zas}) by 
$J_{\Lambda}(D, R_0,r^{l-1},r^l)$. 
Using this function, 
$R_{{\rm sum},L}^{(\rm l)}(D,R_0)$ can be written as
\beqno
R_{{\rm sum},L}^{(\rm l)}(D,R_0)
&=&\min_{1\leq l\leq L}
\left[\min_{\scs r^l \in {\cal B}_l(D|R_0)}
J_{\Lambda}(D, R_0,r^{l-1},r^l)\right]\,.
\eeqno
Note here that 
$J_{\Lambda}(D, R_0,r^{l-1},r^l)$ is a monotone increasing 
function of $r_l$. 
To prove 
$
R_{{\rm sum},L}^{(\rm l)}(D, R_0)
\geq
R_{{\rm sum},L}^{(\rm u)}(D, R_0)\,,
$
it suffices to show that for $1\leq l\leq L$,
\beqno
\min_{\scs r^l \in {\cal B}_l(D|R_0)}
J_{\Lambda}(D, R_0,r^{l-1},r^l)
\geq 
\min_{r^l\in \partial{\cal B}_l(D|R_0)}
K_{\Lambda}(r^l)\,.
\eeqno
We prove this claim by induction with respect to $l$.
When $l=1$, the function $J_{\Lambda}(D, R_0,r_1)$ 
is computed as
\beqno
J_{\Lambda}(D, R_0,r_1)
&=&
\frac{1}{2} \ts \log^{+} 
        \left[
        \frac{\left\{1+\sigma_{Z_1}^2g_1(D, R_0)\right\}
         \sigma_{X_0}^2
         {\rm e}^{-2R_{0}}{\rm e}^{2r_{1}}
        }{D}
\right]
\\
&=&
\frac{1}{2} \ts \log^{+} 
        \left[
        \frac{
         \sigma_{X_0}^2
         {\rm e}^{-2R_{0}}{\rm e}^{2r_{1}}
        }{\left\{1-\sigma_{Z_1}^2g_0(D, R_0)\right\}D}
\right]\,.
\eeqno
Since 
$ 
\frac{\sigma_{X_0}^2{\rm e}^{-2R_{0}}}{D}>1\,,
$
$J_{\Lambda}(D, R_0,r_1)$ is positive. Since 
$J_{\Lambda}(D, R_0,r_1)$ is a monotone increasing 
function of $r_1$, the minimum of this function is attained by
$r_{1}^{*}=0$ or a positive $r_{1}^{*}$ satisfying
$
r_{1}^{*}\in \partial{\cal B}_{1}(D|R_0)\,.
$
If $r_{1}^{*}=0$, then, by Property \ref{pr:prz01z} part b), 
$J_{\Lambda}(D, R_0,r_1)$ must be zero. This contradicts 
that $J_{\Lambda}(D, R_0,r_1)$ is positive. 
Therefore, $r_{1}^{*}$ must be positive. Then, 
by $r_1^{*}\in \partial{\cal B}_{1}(D|R_0)$, we have
\beqno   
J_{\Lambda}(D, R_0,r_1)&\geq& J_{\Lambda}(D, R_0,r_1^{*})
\\
&=&K_{\Lambda}(r_1^{*})
\geq \min_{r_1\in \partial{\cal B}_1(D|R_0)}K_{\Lambda}(r_1)\,.
\eeqno
Thus, the claim holds for $l=1$. We assume that the claim holds 
for $l-1$. Since $J_{\Lambda}(D, R_0,r^{l-1},r^l)$ 
is a monotone increasing function of $r_l$, the minimum 
of this function is attained by
$r_{l}^{*}=0$ or a positive $r_{l}^{*}$ satisfying
$
(r^{l-1},r_{l}^{*})\in \partial{\cal B}_{l}(D|R_0)\,.
$
When $r_{l}^{*}=0$, we have 
$
r^{l-1}\in {\cal B}_{l-1}(D|R_0)
$
and 
\beq
     J_{\Lambda}(D, R_0,r^{l-1},r^{l})
\geq J_{\Lambda}(D, R_0,r^{l-1},r^{l-1}r_l^{*})\,.
\label{eqn:abab}
\eeq
Computing $J_{\Lambda}(D, R_0,r^{l-1},r^{l-1}r_l^{*})$, we obtain
\beqa
& &J_{\Lambda}(D, R_0,r^{l-1},r^{l-1}r_l^{*})
\nonumber\\
&=& \left. J_{\Lambda}(D, R_0,r^{L-2},r^L) \right|_{r_{l}^{L}={\lvc 0}}
\nonumber\\
&=&\frac{1}{2}\log^{+}
\left[
G(D, R_0,r^{l-2})\cdot 
\frac{\sigma_{X_0}^2}{D}{\rm e}^{-2R_0}
\prod_{i=1}^{l-1}{\rm e}^{2r_i}
\right]
\nonumber\\
&=&J_{\Lambda}(D, R_0,r^{l-2},r^{l-1})\,.
\label{eqn:zass}
\eeqa
Combining (\ref{eqn:abab}) and (\ref{eqn:zass}), we have 
\beq
J_{\Lambda}(D, R_0,r^{l-1},r^{l})
\geq
J_{\Lambda}(D, R_0,r^{l-2},r^{l-1})\,.
\label{eqn:zass2}
\eeq
On the other hand, by induction hypothesis, we have
\beq
J_{\Lambda}(D, R_0,r^{l-2},r^{l-1})
\geq 
\min_{r^{l-1}\in \partial{\cal B}_{l-1}(D|R_0)}
K_{\Lambda}(r^{l-1})\,.
\label{eqn:ababc}
\eeq
Combining (\ref{eqn:zass2}) and (\ref{eqn:ababc}), we have
\beqno
J_{\Lambda}(D, R_0,r^{l-1},r^{l})
&\geq & 
\min_{r^{l-1}\in \partial{\cal B}_{l-1}(D|R_0)}
K_{\Lambda}(r^{l-1})
\\
&\geq & \min_{r^{l}\in \partial{\cal B}_{l}(D|R_0)}
         K_{\Lambda}(r^{l}).
\eeqno
When $r_{l}^{*}>0$, we have 
\beqno
        J_{\Lambda}(D, R_0,r_{l-1},r^{l})
&\geq & J_{\Lambda}(D, R_0,r_{l-1},r^{l-1}r_l^{*})
\\
&=& K_{\Lambda}(r_{l-1}r_l^{*})
\\
&\geq&\min_{r^l\in \partial{\cal B}_l(D|R_0)}
      K_{\Lambda}(r^l)\,,
\eeqno
where the second equality follows from 
$
(r^{l-1},r_{l}^{*})\in \partial{\cal B}_{l}($ 
$D|R_0)\,.
$
Thus, the claim holds for $l$, completing the proof.
\hfill\IEEEQED

%

\subsection{Computation of $\{R_{\rm sum}^{(l)}(D)\}_{l=1}^L$}

When $r_{l+1}^L=0$, by (\ref{eqn:SumEx000}), we can prove the following:
\beq
\left.
\ba{rcl}
f_i(r_i^L)&=&0,l+1\le i\le L,
\\
f(r_{l}^L)&=&\frac{1}{\sigma_l^2}(1-{\rm e}^{-2r_l}),
\\
f_{i-1}(r_{i-1}^L)&=&\frac{f_i(r_i^L)}{1+\epsilon_{i}\sigma_{i}^2f(r_i^L)}
+\frac{1}{\sigma_{i-1}^2}\left(1-{\rm e}^{-2r_{i-1}}\right),
\\
& &\mbox{ for }l \geq i \geq 2,
\\
f_{0}(r^L)&=&\frac{f_1(r^L)}{1+\epsilon_{1}\sigma_{1}^2f_1(r^L)}\,.
\ea
\right\}
\label{eqn:SumEx000'}
\eeq
Define the sequence $\{f_i(r_i^l)\}_{i=1}^l$ of $l$ functions 
and the function $f_0(r^l)$ by
\begin{align}
f_i(r_i^l)\triangleq f_i(r_i^l,r_{l+1}^L)&=f_i(r_i^L)|_{r_{l+1}^L=0},\mbox{ for }l\ge i\ge 1\nonumber
\\
f_0(r^l)&\triangleq f_0(r_l^L)|_{r_{l+1}^L=0}\nonumber.
\end{align}
Then, by (\ref{eqn:SumEx000'}) and the definitions of $f_i(r_i^l),l\ge i\ge 1$ and $f_0(r^l)$, we have 
\beq
\left.
\ba{rcl}
f_l(r_l)&=&\frac{1}{\sigma_l^2}\left(1-{\rm e}^{-2r_l}\right)\,,
\\
f_{i-1}(r_{i-1}^l)&=&\frac{f_i(r_i^l)}{1+\epsilon_{i}\sigma_{i}^2f_i(r_i^l)}
+\frac{1}{\sigma_{i-1}^2}\left(1-{\rm e}^{-2r_{i-1}}\right)
\\
& &\mbox{ for }l \geq i \geq 2,
\\
f_{0}(r^l)&=&\frac{f_1(r^l)}{1+\epsilon_{1}\sigma_{1}^2f_1(r^l)}\,.
\ea
\right\}
\label{eqn:SumEx000zb}
\eeq
We define the transformation of the vector $r^l$ into 
the vector $\alpha^l$ by
\begin{align}
\alpha_i=\frac{\sigma_i^2f_i(r_i^l)}{1+\epsilon_i\sigma_i^2f_i(r_i^l)},l\ge i\ge 1.
\label{eqn:SumEx001}
\end{align}
From (\ref{eqn:SumEx001}), we have  
\beq
f_i=f_i(r_i^l)=\frac{1}{\sigma_i^2}
\cdot\frac{\alpha_i}{1-\epsilon_{i}\alpha_i}\,, 
\mbox{ for }l\geq i\geq 1\,.
\label{eqn:SumEx002}
\eeq
Note that for $l\geq i\geq 1$, $f_i\geq 0$. From  
(\ref{eqn:SumEx002}), $\alpha_i$, $l\geq i\geq 1$ must satisfy 
$0\leq \alpha_i <\epsilon_{i}^{-1}$. For $l\geq i\geq 2$,
set $\tau_i$$\defeq \sigma_{i}^2/\sigma_{i-1}^2$. Considering 
(\ref{eqn:SumEx000zb}) and (\ref{eqn:SumEx002}), we have
\begin{align}
&{\rm e}^{-2r_{l}}= 1-\frac{\alpha_{l}}{1-\epsilon_{l}\alpha_{l}},
\label{eqn:zaaQ0}\\
& {\rm e}^{-2r_{i-1}}=1-\frac{\alpha_{i-1}}{1-\epsilon_{i}\alpha_{i-1}}
+\frac{\alpha_{i}}{\tau_i},\: l\geq i\geq 2.
\label{eqn:zaaQ}
\end{align}
Since $r_l\ge0$ and (\ref{eqn:zaaQ0}), $\alpha_l$ must be 
\beq
0< \alpha_l<(1+\epsilon_{l})^{-1}.
\label{eqn:SumEx004o}
\eeq
Furthermore, since $r_{i-1}\geq 0$ for $l-1\geq i\geq 2$ 
and (\ref{eqn:zaaQ}), $\alpha_i, l\geq i \geq 2$ must satisfy the following: 
\beq
\left.
\ba{l}
0\leq \alpha_{i}\leq 
\frac{\tau_i\alpha_{i-1}}{1-\epsilon_{i-1}\alpha_{i-1}}\,,
\vspace{1mm}\\
\tau_i\left(\frac{\alpha_{i-1}}{1-\epsilon_{i-1}\alpha_{i-1}}-1\right)
     <\alpha_{i} < \epsilon_{i}^{-1}\,.
\ea
\right\}
\label{eqn:SumEx004}
\eeq
We next express the objective function in the 
optimization problem defining $R_{{\rm sum}}^{(l)}(D)$
using $\alpha^l$. Set 
\begin{align*}
& \zeta^{(l)}=\zeta^{(l)}(\alpha_2^l)
\defeq 
\left(-\frac{1}{2}\right)\sum_{i=1}^{l-1}
\biggl\{\log 
\left(1-\frac{\alpha_i}{1-\epsilon_{i}\alpha_i}
+\frac{\alpha_{i+1}}{\tau_{i+1}}\right)
\nonumber\\
& \qquad +\log\left(1-\epsilon_{i} \alpha_i \right)
\biggr\}
  +\log\left(1-\frac{\alpha_l}{1-\epsilon_l\alpha_l}
         \right). 
\end{align*}
Then we have  
\begin{align*}
&
\sum_{l=1}^Lr_l
+\sum_{l=1}^{L-1}
\frac{1}{2}\log \left(1+\epsilon_{l}\sigma_l^2f_l(r_l^L)\right)
=\zeta^{(l)}(\alpha_2^l).
\end{align*}
Let ${\cal A}_l$ be a domain 
of the objective function in the optimization problem 
defining $R_{\rm sum}^{(l)}(D)$. Considering a form of the objective 
function and (\ref{eqn:SumEx004}), the domain ${\cal A}_l$ is 
a set of $\alpha^l$ such that for $l-1\geq i\geq 2$, 
$\alpha^l$ satisfies (\ref{eqn:SumEx004}) and 
\beq
\left.
\ba{l}
0\leq \alpha_{l}\leq 
\frac{\tau_l\alpha_{l-1}}{1-\epsilon_{l-1}\alpha_{l-1}}\,,
\vspace{1mm}\\
\tau_l\left(\frac{\alpha_{l-1}}{1-\epsilon_{l-1}\alpha_{l-1}}-1\right)
     <\alpha_{l} <(1+\epsilon_{l})^{-1}\,.
\ea
\right\}
\label{eqn:SumEx004b}
\eeq
Summarizing the above arguments, we can obtain an expression of
$R_{{\rm sum}}^{(l)}(D)$ using $\alpha^{l}$. This expression
is given by
\begin{align*}
& R_{{\rm sum}}^{(l)}(D)
=\frac{1}{2}\log\frac{\sigma_{X_0}^2}{D}+
\min_{\scs 
\alpha_2^l\in {\cal A}_l(\alpha_1)\,,
\atop{\scs \alpha_1=\sigma_1^2
g_0(D)
}}
\zeta^{(l)}(\alpha_2^l).
\end{align*}
Here we set 
$
{\cal A}_l(\alpha_1)\defeq
\{\alpha_2^l:\alpha^l=(\alpha_1,\alpha_2^l)\in {\cal A}_l\}. 
$

Then we have the following lemma.
\begin{lm}\label{lm:convlm0}
For $\alpha_2^l\in {\cal A}_l(\alpha_1)$, 
$(-2)\zeta_l(\alpha_2^l)$ $\alpha_2^l$ is strictly concave 
with respect to $\alpha_2^l$
\end{lm}

Proof of this lemma will be given in Appendix C.

The following lemma is a key result to establish 
recursive algorithms of computing 
$R_{\rm sum}^{(l)}(D)$ for $1\leq l\leq L$.
\begin{lm}\label{lm:prlmzsz1}{\rm We assume that 
\beq
\tau_l=\sigma_l^2/\sigma^2_{l-1}\geq 1,\: L\geq l\geq 2.
\label{eqn:CondZ}
\eeq
Under this assumption, the sequence
$\theta_{\bullet}^{(l)}=\{\theta_i^{(l)}(\omega)\}_{i=1}^l$ of
$l$ continuous functions defined by (\ref{eqn:RecurEqaa})
and (\ref{eqn:RecurEqbb}) satisfies the following three 
properties:

}\end{lm}

\begin{itemize}
\item[a)] We have 
\begin{align}
& \left.
\ba{l}
0\leq \theta_l^{(l)}(\omega)
\leq 
\frac{\tau_l\theta_{l-1}^{(l)}(\omega)}
{1-\epsilon_{l-1}\theta_{l-1}^{(l)}(\omega)}\,,
\vspace{1mm}\\
\tau_l\left(\frac{\theta_{l-1}^{(l)}(\omega)}
{1-\epsilon_{l-1}\theta_{l-1}^{(l)}(\omega)}-1
\right)
<\theta_{l}^{(l)}(\omega)<(1+\epsilon_{l})^{-1}
\ea
\right\}. 
\label{eqn:cond0} 
\end{align}
Furthermore, for $l-1\geq i\geq 2$, we have  
\begin{align}
&
\left.
\ba{l}
0\leq \theta_i^{(l)}(\omega)
\leq 
\frac{\tau_i\theta_{i-1}^{(l)}(\omega)}
{1-\epsilon_{i-1}\theta_{i-1}^{(l)}(\omega)}\,,
\vspace{1mm}\\
\tau_i\left(\frac{\theta_{i-1}^{(l)}(\omega)}
{1-\epsilon_{i-1}\theta_{i-1}^{(l)}(\omega)}-1
\right)
<\theta_{i}^{(l)}(\omega)<\epsilon_{i}^{-1}.
\ea
\right\}
\label{eqn:condl} 
\end{align}
The conditions ${\rm (\ref{eqn:cond0})}$ and ${\rm (\ref{eqn:condl})}$
imply $(\theta_i^{(l)})_{i=2}^{l}(\omega)\in {\cal A}_l
(\theta_1^{(l)}(\omega))$.
\item[b)]
$$
\left. \nabla \zeta_l 
\right|_{\alpha_2^l=(\theta_i^{(l)}(\omega))_{i=2}^{l}}
={\vc 0}.
$$
\item[c)] 
For each $l-1\geq i\geq 1$, $\theta_i^{(l)}(\omega)$ 
is differentiable with respect to $\omega\in [0,(1+\epsilon_l)^{-1})$ 
and satisfies the following:
\begin{align*}
\frac{{\rm d}\theta_{i}^{(l)} } {{\rm d}\omega}
\geq&
(l-i+1)\cdot\frac{\sigma_i^2}{\sigma_l^2}
\nonumber\\
&\times
\prod_{j=i+1}^{l}
{
\ts \frac{1}{
\left\{1+\epsilon_{j-1}
\left(\frac{\theta_{j}^{(l)}(\omega)}{\tau_j}+1\right)\right\}^2}
} > 0\,.
\end{align*}
This implies that for each $l\geq i\geq 1$, the mapping 
$\omega\in[0,1)$$\mapsto$ $\theta_i^{(l)}(\omega)$
is an injection.
\end{itemize}

Proof of this lemma is given in Appendix D. From this 
lemma, we immediately obtain Theorem \ref{Th:ThSumRateComp}.

\section{Conclusions}
\noindent

We have considered the Gaussian \maho problem and given a partial
solution to this problem by deriving explicit outer bound of the rate
distortion region for the case where information sources satisfy the
TS condition. Furthermore, we established a sufficient condition under
which this outer bound is tight. We have determined the sum rate part
of the rate distortion region for the case where information sources
satisfy the TS condition.

For the case that information sources do not satisfy the TS condition
we can not derive an outer bound having a similar form of ${\cal
  R}^{(\rm out)}(D)$ since the proof of the converse coding theorem
depends heavily on this property of information sources. Hence the
complete solution is still lacking for Gaussian information sources
with general correlation.

\section*{\empty}
\appendix

\subsection{
Proof of Proposition \ref{pro:pro1}
}

In this appendix we prove Proposition \ref{pro:pro1}.
To prove this proposition we give some preparations.
For $0\leq l\leq L-2$, we set 
\beqno
\hspace*{-2mm}& &\eta_l=\eta_l(D,r_0,r^{l}) 
\vspace{1mm}\\
\hspace*{-2mm}& \defeq & 
\left\{\ba{l}
g_0(D,r_0)\,,\mbox{ for }l=0\,,
\vspace{1mm}\\
g_l(D,r_0,r^{l-1})
-\ts \frac{1}{\sigma_{N_{l}}^2}\left(1-{\rm e}^{-2r_{l}}\right),
\mbox{ for }1\leq l\leq L-2\,.
\ea
\right.
\eeqno
For $1 \leq l \leq L-2$, and $ a < \frac{1}{\sigma_{Z_l}^2}$, define
\beqno
\tau_l(a)
\defeq\ts 
\frac{[a]^{+}}{1-\sigma_{Z_l}^2[a]^{+}}
-\frac{1}{\sigma_{N_l}^2}
\left(1-{\rm e}^{-2r_{l}}\right)\,.
\eeqno
Then, $\{\eta_{l} \}_{l=0}^{L-2}$ 
satisfies the following:
\beqa
\eta_{l}(D,r_0,r^{l})
&=& \tau_l \left( \eta_{l-1}(D,r_0,r^{l-1}) \right)
\nonumber\\
& & \mbox{ for }1 \leq l \leq L-2\,.
\label{eqn:zaxz}
\eeqa
Fix $a<\frac{1}{\sigma_{Z_{k+1}}^2}$ and set
\beqno
p_k(a)&\defeq & 
\sup\biggl\{p:\biggr.  
\ba[t]{l}
  \log \frac{1-\sigma_{Z_{k+1}}^2[a]^{+}}{1-\sigma_{Z_{k+1}}^2[b]^{+}}
  \geq p(b-a)
  \vspace{1mm}\\
  \left.
  \mbox{ for any }b< \frac{1}{\sigma_{Z_{k+1}}^2} 
  \:\right\}\,.
\ea
\eeqno
By a simple computation we have
\beqa
p_k(a)&=&
\left\{
\ba{cl}
\frac{\sigma_{Z_{k+1}}^2}{1-\sigma_{Z_{k+1}}^2a}\,,
    &\mbox{ for } 0\leq a  < \frac{1}{\sigma_{Z_{k+1}}^2}\,,\\
0\,,&\mbox{ for } a  < 0
\ea \right.
\nonumber\\
&\leq&{\ts \frac{\sigma_{Z_{k+1}}^2 }{1-\sigma_{Z_{k+1}}^2a}}
\mbox{ for } a  < \ts \frac{1}{\sigma_{Z_{k+1}}^2}\,.
\label{eqn:qqaa1}
\eeqa
Fix $a<\frac{1}{\sigma_{Z_{j}}^2}$ and set
\beqno
q_j(a)&\defeq & 
\sup\biggl\{q: \biggr. 
\ba[t]{l}
    \tau_j(b)- \tau_j(a) \geq q(b-a)
\vspace{1mm}\\
  \left.\mbox{ for any }b<\frac{1}{\sigma_{Z_{j}}^2} 
\:\right\}\,.
\ea
\eeqno
By a simple computation we have
\beqa
q_j(a)&=&
\left\{
\ba{cl}
\frac{1}{(1-\sigma_{Z_{j}}^2a)^2}\,,
&\mbox{ for } 0\leq a < \frac{1}{\sigma_{Z_{j}}^2}\,,\\
0\,,&\mbox{ for } a  < 0
\ea \right.
\nonumber\\
&\leq&{\ts \frac{1}{(1-\sigma_{Z_{j}}^2a)^2}}\,,
\mbox{ for } \ts a  < \frac{1}{\sigma_{Z_{j}}^2}\,.
\label{eqn:qqaa2}
\eeqa

{\it Proof of Proposition \ref{pro:pro1}:}
Let ${\cal L}$ be a set of integers $l$   
such that $\eta_{l}(D,r_0,r^{l})$ 
is positive for some $r_0^L\in {\cal B}_L(D)$. 
From (\ref{eqn:zaxz}), there exists a unique integer 
$1\leq L^{*}\leq L-2$ such that ${\cal L}$ 
$=\{0,1,\cdots, L^{*}\}$. 
%
Using $\{\eta_{l} \}_{l=1}^{L-2}$ 
and $L^{*}$, 
$\log G(D,r_0,r^{L-2})$ can be rewritten as 
\beqa 
\log G(D,r_0,r^{L-2})
&=&\sum_{k=1}^{L-1}
\log\left\{1+\sigma_{Z_k}^2g_l(D,r_0,r^{k-1})\right\}
\nonumber\\
&=&\sum_{k=1}^{L-1}
\log\left[ {\ts \frac{1}{1-\sigma_{Z_k}^2[\eta_{k-1}(D,r_0,r^{k-1})]^{+}}}\right]
\nonumber\\
&=&\sum_{k=0}^{L-2}
\log\left[{\ts \frac{1}{1-\sigma_{Z_{k+1}}^2[\eta_{k}(D,r_0,r^{k})]^{+}}} \right]
\nonumber\\
&=&\sum_{k=0}^{L^{*}}
\log\left[{\ts \frac{1}{1-\sigma_{Z_{k+1}}^2[\eta_{k}(D,r_0,r^{k})]^{+}}} \right]
\,.
\label{eqn:zaa}
\eeqa
Fix nonnegative vector $r^L$. 
For each $s_l \geq r_l$, $1\leq l$ $\leq$ $L-2$, 
let $G(s_l)$ be a function obtained by replacing $r_l$ in 
$G(D,r_0,r^{L-2})$ with $s_l$, that is 
$$
G(s_l) \defeq G(D,r_0,r^{l-1},s_l,r_{l+1}^{L-2})\,.
$$
It is obvious that when $s_l=r_l$, 
$$
G(r_l)=G(D,r_0,r^{l-1},r_l,r_{l+1}^{L-2})=G(D,r_0,r^{L-2})\,.
$$
By Property \ref{pr:prz001z} part b), we have
$G(s_l) \leq G(r_l)$ for $1\leq l\leq L-2$.
For each $s_k \geq r_k$, $l\leq k$ $\leq L-2$, 
let $\eta_k(s_l)$ be a function obtained by replacing $r_l$ 
in $\eta_k(D,r_0,r^{k})$ with $s_l$, that is 
$$
\eta_{k}(s_l) \defeq \eta_{k}(D,r_0,r^{l-1},s_l,r_{l+1}^k)\,.
$$
It is obvious that when $s_l=r_l$, 
$$
\eta_{k}(r_l) = \eta_{k}(D,r_0,r^{l-1},r_l,r_{l+1}^k)
=\eta_{k}(D,r_0,r^k)\,.
$$
By Property \ref{pr:pr01} part b), we have
$\eta_{k}(s_l) \leq \eta_{k}(r_l)$ for $l\leq k\leq L-2$. 
For each $l$ $=1,$ $\cdots,L^{*}$, we evaluate an upper bound of 
$\log G(s_l)$ $-$ $\log G(r_l)$. 
Using (\ref{eqn:zaa}), we have 
\beqa
\log \frac{G(s_l)}{G(r_l)}
&=&\sum_{k=0}^{L^{*}}
 \log \left[{\ts 
\frac{1-\sigma_{Z_{k+1}}^2[\eta_{k}(r_{l})]^{+}}
     {1-\sigma_{Z_{k+1}}^2[\eta_{k}(s_{l})]^{+} }}
      \right]
\nonumber\\
&=&\sum_{k=l}^{L^{*}}
 \log \left[{\ts 
\frac{1-\sigma_{Z_{k+1}}^2[\eta_{k}(r_{l})]^{+}}
     {1-\sigma_{Z_{k+1}}^2[\eta_{k}(s_{l})]^{+} }}
      \right]
\,. 
\label{eqn:zzxc}
\eeqa
By definition of $p_k(\cdot)$, we have
\beqa
& & 
\log\left[{\ts \frac{1-\sigma_{Z_{k+1}}^2[\eta_{k}(r_{l})]^{+}}
                    {1-\sigma_{Z_{k+1}}^2[\eta_{k}(s_{l})]^{+}}
          }\right]
\geq  p_k(\eta_k(r_l))
\left[\eta_{k}(s_{l})-\eta_{k}(r_{l})\right]
\nonumber\\
&\geq& 
{\ts \frac{\sigma_{Z_{k+1}}^2 }{1-\sigma_{Z_{k+1}}^2\eta_{k}(r_{l})}}
\left[\eta_{k}(s_{l})-\eta_{k}(r_{l})\right]\,,
\label{eqn:zaass}
\eeqa
where the last inequality follows from 
$\eta_{k}(s_{l})$$\leq$ $\eta_{k}(r_{l})$
and (\ref{eqn:qqaa1}).
From (\ref{eqn:zzxc}) and (\ref{eqn:zaass}), we have
\beqa 
\log \frac{G(s_l)}{G(r_l)}
&\geq &\sum_{k=l}^{L^{*}} 
{\ts \frac{\sigma_{Z_{k+1}}^2}{1-\sigma_{Z_{k+1}}^2\eta_{k}(r_l)}}
\left(\eta_k(s_l)-\eta_k(r_l)\right)\,.
\label{eqn:azaxz}
\eeqa
By definition of $q_j(\cdot)$ and 
(\ref{eqn:zaxz}), for $l+1\leq$ $j\leq k$, we have
\beqa
& & \eta_j(s_l)-\eta_j(r_l)
\geq  q_j(\eta_{j-1}(r_l))
\left[\eta_{j-1}(s_{l})-\eta_{j-1}(r_{l})\right]
\nonumber\\
&\geq& 
{\ts \frac{1}{
\left(1-\sigma_{Z_{j}}^2\eta_{j-1}(r_{l})\right)^2}}
\left[\eta_{j-1}(s_{l})-\eta_{j-1}(r_{l})\right]\,,
\label{eqn:zaasss}
\eeqa
where the last inequality follows from 
$\eta_{j-1}(s_{l})$ $\leq$$\eta_{j-1}(r_{l})$
and (\ref{eqn:qqaa2}).
Using (\ref{eqn:zaasss}) iteratively for $l+1\leq$ $j\leq k$, 
we obtain
\beqa
&    &
\eta_k(s_l)-\eta_k(r_l)
\nonumber\\
&\geq & \left(\eta_l(s_l)-\eta_l(r_l)\right)
\prod_{j=l+1}^{k}{\ts
\frac{1}{\left(1-\sigma_{Z_{j}}^2\eta_{j-1}\right)^2}}\,.
\label{eqn:xzzx0}
\eeqa
Observe that 
\beqa
\eta_l(s_l)-\eta_l(r_l)
&=& \ts \frac{1}{\sigma_{N_{l}}^2}
\left[{\rm e}^{-2s_l}-{\rm e}^{-2r_l}\right]
\nonumber\\
&\geq&-{\ts \frac{2{\rm e}^{-2r_l}}{\sigma_{N_{l}}^2}}(s_l-r_l)\,.
\label{eqn:xzzx1}
\eeqa
From (\ref{eqn:xzzx0}) and (\ref{eqn:xzzx1}), we have
\beqa
&  &
\eta_k(s_l)-\eta_k(r_l)
\geq -{\ts \frac{2{\rm e}^{-2r_l}}{\sigma_{N_{l}}^2}}(s_l-r_l)
\prod_{j=l+1}^{k}
{\ts
\frac{1}{\left(1-\sigma_{Z_{j}}^2\eta_{j-1}\right)^2}
}
\nonumber\\
&\geq &-{\ts \frac{2}{\sigma_{N_{l}}^2}}(s_l-r_l)
\prod_{j=l+1}^{k}
{\ts
\frac{1}{\left(1-\sigma_{Z_{j}}^2\eta_{j-1}\right)^2}
}\,.
\label{eqn:zzzxc}
\eeqa
From (\ref{eqn:azaxz}) and (\ref{eqn:zzzxc}), we have 
\beqa 
\hspace*{-6mm}& &
\frac{1}{2}\log \frac{G(s_l)}{G(r_l)}
\nonumber\\
\hspace*{-6mm}&\geq &
-(s_l-r_l)\sum_{k=l}^{L^{*}}
 {\ts \frac{\sigma_{Z_{k+1}}^2}{\sigma_{N_{l}}^2}
 \frac{1}{1-\sigma_{Z_{k+1}}^2\eta_{k}}}
\prod_{j=l+1}^{k}
{\ts
\frac{1}{\left(1-\sigma_{Z_{j}}^2\eta_{j-1}\right)^2}
}\,.
\label{eqn:aza0}
\eeqa
By Property \ref{pr:pr01} part b) and 
the definition of $\eta_j$, we have 
$$
\ts \eta_j \leq f_j -\frac{1}{\sigma_{N_{j}}^2}
(1-{\rm e}^{-2r_j})
=\frac{f_{j+1}}{1+\sigma_{Z_{j+1}}^2f_{j+1}}
\,, 
$$
from which we have
\beq 
\ts \frac{1}{1- \sigma_{Z_{j+1}}^2\eta_j}
\leq 1+\sigma_{Z_{j+1}}^2f_{j+1}
\leq 1+\sigma_{Z_{j+1}}^2f_{j+1}^{*}\,.
\label{eqn:zoz}
\eeq
From (\ref{eqn:aza0}) and (\ref{eqn:zoz}), we have 
\beqa 
\frac{1}{2}\log \frac{G(s_l)}{G(r_l)}
&\geq &-(s_l-r_l)\sum_{k=l}^{L^{*}}
{\ts \frac{\sigma_{Z_{k+1}}^2}{\sigma_{N_l}^2}}
\left(1+\sigma_{Z_{k+1}}^2f_{k}^{*}\right)
\nonumber\\
& & \qquad\qquad \times 
\prod_{j=l+1}^{k}\left( 1+\sigma_{Z_j}^2f_j^{*}\right)^2
\,.
\label{eqn:aza1}
\eeqa
If 
\beqa
\hspace*{-4mm}& &\sum_{k=l}^{L-2}
{\ts \frac{\sigma_{Z_{k+1}}^2}{\sigma_{N_{l}}^2}}
\left(
1+\sigma_{Z_{k+1}}^2f_{k+1}^{*}
\right)\hspace*{-1mm}
\prod_{j=l+1}^{k}
\left(1+\sigma_{Z_{j}}^2f_{j}^{*}\right)^2
\leq 1
\label{eqn:xzx000}
\eeqa
hold for $l=1,2,\cdots, L-2$, then, by (\ref{eqn:aza1}), 
we have 
\beqno 
 \frac{1}{2}\log \frac{G(s_l)}{G(r_l)}
&\geq & -(s_l-r_l)
\eeqno
or equivalent to 
\beqno 
 s_l + \frac{1}{2}\log G(s_l)
&\geq & r_l+ \frac{1}{2}\log G(r_l) 
\eeqno
for $l=1,2,\cdots, L-2$. Hence, (\ref{eqn:xzx000})
is a sufficient condition for the MI condition.
\hfill\IEEEQED 

\subsection{
Proof of ${\cal R}_L(D) \subseteq {\cal R}_L^{\rm (out)}(D)$
}


In this appendix we prove ${\cal R}_L(D)$
$\subseteq$ ${\cal R}_L^{({\rm out})}(D)$ stated 
in Theorem \ref{th:converse}. We first present 
a lemma necessary for the proof of this inclusion.

\begin{lm}{\label{lm:prelm}\rm 
\beqno
I(\Xotn;\hatXotn)
\geq\frac{n}{2} \log 
\ts 
\left(
\frac{\sigma_{X_0}^2}
{\Delta({\lvc X}_0,\hat{\lvc X}_0)}
\right)\,.
\eeqno
}
\end{lm}


\begin{IEEEproof}
See the proof of Lemma~1 in Oohama \cite{oh2}. 
\end{IEEEproof}

Next, we present an important lemma which is a mathematical 
core of the converse coding theorem. Let the encoded outputs 
of $\Xitn, i=0,1,\cdots, L$ by encoder functions $\varphi_i$ 
be denoted by $\varphi_i(\Xitn)=W_i$. Set 
\beqno
r_0 &\defeq & \frac{1}{n}I({\vc X}_0;W_0|W^L)\,,
\\
r_i &\defeq &\frac{1}{n}I(\Xitn;W_i|{\vc Y}^{L-1})
\\
&=&\left\{\ba{l}
\ds \frac{1}{n}I(\Xitn;W_i|{\vc Y}_{i})\,, 
\mbox{ for }1\leq i \leq L-1,
\vspace*{1mm}\\
\ds \frac{1}{n}I({\vc X}_L;W_L|{\vc Y}_{L-1})\,, 
\mbox{ for }i=L,
\ea
\right.
\\
\xi &\defeq& \sigma_{X_0}^2 
{\rm e}^{-\frac{2}{n}I({\lvc X}_0;W_0W^L)}\,. 
\eeqno
Then, we have the following lemma.
\begin{lm}{\label{lm:corelm}\rm 
\beqno
I(\Xotn ;W^L)
&\leq &
\frac{n}{2}\log
           \left[
               {
               1 + \sigma_{X_0}^2f_{0}(r^L) 
               }
           \right]\,.
\eeqno
For $1\leq l\leq L-1$, we have 
\beqno
& &\frac{n}{2}
 \log\left[
    1+\sigma_{Z_l}^2g_{l}(r_{0},r^{l-1},\xi)
    \right]
\\
&\leq &I({\vc Y}_l;W_{l}^L| {\vc Y}_{l-1})
\leq 
\frac{n}{2}\log
           \left[
               {
               1+\sigma_{Z_l}^2f_{l}(r_{l}^{L}) 
               }
           \right]\,.
\eeqno
}
\end{lm}

From the above lemma we immediately obtain the following. 
\begin{lm}{\label{lm:corelm2}\rm 
\beqno
I(\Xotn;W_{S}) &\leq &
\frac{n}{2}\log
           \left[
               {
               1 + \sigma_{X_0}^2f_{0}(r_S) 
               }
           \right]\,,
\\
I({\vc Y}^{L-1};W_{S}|\Xotn)&\leq &
\frac{n}{2}\log F(r_S),\: S\subseteq\Lambda, 
\\
I({\vc Y}^{L-1};W^L|\Xotn)
&\geq &
\frac{n}{2}\log G(\xi,r_0,r^{L-2})\,.
\eeqno
}
\end{lm}

We prove ${\cal R}_L(D)$ $\subseteq$ ${\cal R}_L^{({\rm out})}(D)$ 
by Lemmas \ref{lm:prelm} and \ref{lm:corelm2} and standard 
arguments for the proof of the converse coding theorem. 

{\it Proof of ${\cal R}_{\Iset}(D) 
\subseteq {\cal R}_{\Iset}^{({\rm out})}(D)$:} 
We first observe that by virtue of the TS condition, 
\beq
W_S \to \Xtn_S\to (\Xotn, {\vc Z}^{L-1})\to \Xtn_{\coS} \to W_{\coS}
\label{eqn:MkPr1}
\eeq
hold for any subset $S$ of $\Lambda$. 
Assume $(R_0, R_1,$ $\!\cdots, R_L) \in {\cal R}_{\Iset}(D)$. 
Then, for any $\delta>0$, there exists an integer 
$n_0(\delta)$ such that for $n\geq n_0(\delta)$ and 
for $i\in \Lambda$, we obtain the following chain of inequalities:
\beqa
n(R_0+\delta)
&\geq&\log M_0 
\geq H(W_0) \geq H(W_0|W^L)
\nonumber\\
&=&I({\vc X}_0;W_0|W^L)=nr_0\,.
\label{eqn:zzz}
\eeqa
Furthermore, for any subset $S \subseteq \Lambda$, we obtain
\beqa
\hspace*{-5mm}& &nr_0+\sum_{i\in S}n(R_i+\delta)
\nonumber\\
\hspace*{-5mm}&\geq & I(\Xotn;W_0|W^L) + \sum_{i\in S}H(W_i) 
\nonumber\\
\hspace*{-5mm}&=& H(W_0|W^L)+\sum_{i\in S}H(W_i) 
\nonumber\\
\hspace*{-5mm}&\geq &H(W_0|W_SW_{\coS})+H(W_S|W_{\coS})
=H(W_0W_S|W_{\coS})
\nonumber\\
\hspace*{-5mm}&=&I(\Xotn{\vc Z}^{L-1};W_0W_S|W_{\coS})
\nonumber\\
\hspace*{-5mm}& &\qquad\qquad +H(W_0W_S|W_{\coS}\Xotn {\vc Z}^{L-1})
\nonumber\\
\hspace*{-5mm}&\MEq{\rm a}&I(\Xotn{\vc Z}^{L-1};W_0W_S|W_{\coS}) 
   +\sum_{i\in S}H(W_i|\Xotn{\vc Z}^{L-1}) 
\nonumber\\
\hspace*{-5mm}&=&I(\Xotn{\vc Y}^{L-1};W_0W_S|W_{\coS}) 
+\sum_{i\in S}I(\Xitn; W_i|{\vc Y}^{L-1}). 
\label{eqn:conv1}
\eeqa
Step (a) follows from (\ref{eqn:MkPr1}). On the other 
hand, by Lemma \ref{lm:prelm}, we have for $n\geq n_0(\delta)$,
\beqa
& &I(\Xotn;W_0W^L)=\frac{n}{2}\ts 
\log\left(\frac{\sigma_{X_0}^2}{\xi}\right)
\nonumber \\
& \geq & I(\Xotn;\hatXotn)
\geq \frac{n}{2}\ts 
\log\left(\frac{\sigma_{X_0}^2}{D+\delta}\right)\,,
\nonumber 
\eeqa
which together with (\ref{eqn:zzz}), (\ref{eqn:conv1}), 
and Lemma \ref{lm:corelm2} yields the following 
lower bounds of $I(\Xotn;W_0|W^L)$ 
and $I(\Xotn{\vc Y}^{L-1}$ $;W_0W_S|W_{\coS})$:
\beqa 
& &I(\Xotn ;W_0|W^L )
=I({\vc X}_0;W_0W^L)-I({\vc X}_0;W^L)
\nonumber\\
&\geq &\frac{n}{2}\ts \log\left[
\frac{\sigma_{X_0}^2}{ \left\{1+\sigma_{X_0}^2f_0(r^L)\right\}\xi}
        \right]
\nonumber\\
&\geq &\frac{n}{2}\ts \log\left[
\frac{\sigma_{X_0}^2}{ \left\{1+\sigma_{X_0}^2f_0(r^L)\right\}(D+\delta)}
        \right]\,,
\label{eqn:zzuv}
\\
& &I(\Xotn {\vc Y}^{L-1};W_0W_S|W_{\coS})
\nonumber\\   
&=&I(\Xotn{\vc Y}^{L-1};W_0W_SW_{\coS})-I(\Xotn{\vc Y}^{L-1};W_{\coS})   
\nonumber\\   
&=&I(\Xotn ;W_0W^L)+ I({\vc Y}^{L-1} ;W^L|\Xotn)
\nonumber\\   
& &-I(\Xotn;W_{\coS}) -I({\vc Y}^{L-1};W_{\coS}|\Xotn)   
\nonumber\\
&\geq& \frac{n}{2}\ts \log
       \left[
        \frac{\sigma_{X_0}^2G(\xi,r_0,r^{L-2})}
             {
              F(r_{\coS})
              \left\{1+ \sigma_{X_0}^2 f_0(r_{\coS})\right\}\xi
             }
        \right]
\nonumber\\
&\geq& \ds \frac{n}{2}\ts \log
       \left[
        \frac{\sigma_{X_0}^2G(D+\delta,r_0,r^{L-2})}
             {F(r_{\coS})
              \left\{1+ \sigma_{X_0}^2 f_0(r_{\coS})\right\}
              (D+\delta)}
              \right]\,.
\label{eqn:ava}
\eeqa
From (\ref{eqn:conv1}) and (\ref{eqn:ava}), we have
\beqa
\sum_{i\in S}(R_i+\delta)
&\geq& \ds 
       \frac{1}{2}\ts \log
       \left[
        \frac{\sigma_{X_0}^2G(D+\delta,r_0,r^{L-2})}
             {
              F(r_{\coS})(D+\delta)
              \left\{1+ \sigma_{X_0}^2 f_0(r_{\coS})\right\}
              }
        \right]
\nonumber\\
       & &\ds +\sum_{i\in S}r_i-r_0\,.
\label{eqn:ava0}
\eeqa
Note here that $\sum_{i\in S}(R_i+\delta)$
are nonnegative. Hence, from (\ref{eqn:zzz}), (\ref{eqn:zzuv}) 
and (\ref{eqn:ava0}), we obtain
\beq
R_0+\delta \geq r_0\geq 
\frac{1}{2}\ts \log\left[
\frac{\sigma_{X_0}^2}{ \left\{1+\sigma_{X_0}^2f_0(r^L)\right\}(D+\delta)}
        \right]
\label{eqn:z0z}
\eeq
and for $S\subseteq \Lambda$
\beqno
\sum_{i\in S}(R_i+\delta) 
&\geq &J_S(D+\delta,r_0,r^{L-2},r_S|r_{\coS})\,. 
\eeqno
The inequality (\ref{eqn:z0z}) implies that 
$r_0^L\in {\cal B}_L($ $D+\delta)$.
Thus, by letting $\delta\to 0$, we obtain 
$(R_0,R_1,$ $\cdots, R_L)$$\in {\cal R}_{\Iset}^{({\rm out})}(D)$. 
$\quad$\hfill\IEEEQED

Finally, we prove Lemma \ref{lm:corelm}. 
For $n$ dimensional random vector ${\vc U}$ 
with density, let $h({\vc U})$ be 
a differential entropy of ${\vc U}$. 
The following two lemmas are some variants of 
the entropy power inequality.

\begin{lm}\label{lm:lm5} Let ${\vc U}_i,i=1,2,3$ be $n$ dimensional 
random vectors with densities and let $T$ be a random variable 
taking values in a finite set. 
We assume that ${\vc U}_3$ is independent of ${\vc U}_1$, 
${\vc U}_2$, and $T$. Then, we have 
\beqno
\EP{ {\lvc U}_2+{\lvc U}_3|{\lvc U}_1T}
\geq 
\EP{{\lvc U}_2|{\lvc U}_1T}+\EP{{\lvc U}_3}\,.
\eeqno
\end{lm}
\begin{lm}\label{lm:lm5b} Let ${\vc U}_i$, $i=1,2,3$ be 
$n$ random vectors with densities. Let $T_1, T_2$ be random variables 
taking values in finite sets. We assume that those five random variables
form a Markov chain 
$
T_1\to {\vc U}_1\to {\vc U}_3 \to {\vc U}_2 \to T_2
$
in this order. Then, we have    
\beqno
& &\EP{{\lvc U}_1+{\lvc U}_2|{\lvc U}_3T_1T_2}
\\
&\geq& \EP{{\lvc U}_1|{\lvc U}_3T_1}
      +\EP{{\lvc U}_2|{\lvc U}_3T_2}\,.
\eeqno
\end{lm}

{\it Proof of Lemma \ref{lm:corelm}:}
Define the sequence of $n$ dimensional random vectors
$\{$ ${\vc S}_{l}\}_{l=1}^{L-1}$ by 
\beq
{\vc S}_{l}= \ts  \frac{1}{\sigma_{N_{l}}^2 }{\vc X}_{l}
                 +\frac{1}{\sigma_{Z_{l+1}}^2}{\vc Y}_{l+1},\: 
1 \leq l \leq L-1.
\label{eqn:prlm00}
\eeq
By an elementary computation, we obtain  
\beq
\left.
\ba{lcl}
{\vc X}_0
&=&\frac{\sigma_{\hat{N}_0}^2}
  {\sigma_{{Z}_1}^2} {\vc Y}_{1}
+\hat{\vc N}_0\,,
\\
{\vc Y}_l
&=&
\frac{\sigma_{\hat{N}_l }^2}{ \sigma_{Z_{l}}^2 } {\vc Y}_{l-1} 
+ \sigma_{\hat{N}_l }^2
{\vc S}_l
+\hat{\vc N}_l\,,
1\leq l \leq L-1\,.
\ea
\right\}
\label{eqn:prlm0z}
\eeq
where $\hat{{\vc N}}_l,$ $0\leq l\leq L-1$ 
is an $n$ dimensional random vector 
whose components are $n$ independent copies of a 
Gaussian random variable with mean 0 and variance 
$\sigma_{\hat{N}_l}^2$. $\hat{{\vc N}_0}$ is 
independent of ${\vc Y}_{1}$. For each $ 1\leq l \leq L-1$, 
$\hat{{\vc N}_l}$ is independent of ${\vc Y}_{l-1}$ 
and ${\vc S}_{l}$. 
The variance $\sigma_{\hat{N}_l}^2$, $0\leq l\leq L-1$ 
have the following form:
\beq
\left.
\ba{lcl}
\frac{1}{\sigma_{\hat{N}_0}^2}
&=& \ts 
    \frac{1}{\sigma_{X_{0}}^2}
   +\frac{1}{\sigma_{Z_{1}}^2}
\,, \\
 \frac{1}{\sigma_{\hat{N}_l}^2}
&=&\ts \frac{1}{\sigma_{Z_{l}}^2}
      +\frac{1}{\sigma_{N_{l}}^2}
      +\frac{1}{\sigma_{Z_{l+1}}^2}\,,
1\leq l \leq L-1\,.
\ea
\right\}
\eeq
Set
\beqno
\lambda_{0}&\defeq& {\ts \frac{1}{2\pi{\rm e}}}
{\rm e}^{\frac{2}{n}h({{\lvc X}}_{0} | W^{L}) }\,,\:
\tilde{\mu}_{0} \defeq 
{\ts \frac{1}{2\pi{\rm e}}}
{\rm e}^{\frac{2}{n}h({{\lvc Y}}_{1} | W^{L}) }\,,
\\
{\mu}_{0} &\defeq &
{\ts \frac{1}{2\pi{\rm e}}}
{\rm e}^{\frac{2}{n}h({{\lvc Y}}_{1}|{\lvc X}_0W^{L})}\,,
\\
\lambda_{l}&\defeq&
{\ts \frac{1}{2\pi{\rm e}}}
{\rm e}^{\frac{2}{n}h({\lvc Y}_{l}|{\lvc Y}_{l-1}W_l^{L})},
\:1\leq l \leq L\,,
\\
\tilde\mu_{l}&\defeq& 
{\ts \frac{1}{2\pi{\rm e}}}
{\rm e}^{\frac{2}{n}h({{\lvc S}}_{l}|{\lvc Y}_{l-1}W_l^{L})},
\:
{\mu}_{l}\defeq 
{\ts \frac{1}{2\pi{\rm e}}}
{\rm e}^{\frac{2}{n}h({{\lvc S}}_{l}|{\lvc Y}_{l}W_l^{L})}\,,
\\& & 1\leq l \leq L-1.
\eeqno
We can easily verify that 
\beqa
\tilde{\mu}_{0}&=&\mu_{0}\lambda_{0}\ts \frac{1}{\sigma_{\hat{N}_0}^2}
\,,
\tilde{\mu}_{l}=\mu_{l}\lambda_{l}\ts \frac{1}{\sigma_{\hat{N}_l}^2}
\,,\:1 \leq l \leq L-1\,.
\label{eqn:prlm21b}
\eeqa
Applying Lemma \ref{lm:lm5} to (\ref{eqn:prlm0z}), we obtain 
\beq
\left.
\ba{lll}
\lambda_{0} &\geq & 
\frac{\sigma_{\hat{N}_{0}}^4}
     {\sigma_{Z_1}^4}
\tilde{\mu}_{0} + \sigma_{\hat{N}_{0}}^2\,,
\\
\lambda_{l} &\geq &
{\sigma_{\hat{N}_{l}}^4}
 \tilde{\mu}_{l}+\sigma_{\hat{N}_{l}}^2\,,\:
1 \leq l\leq L-1\,.
\ea
\right\}
\label{eqn:prlm21d}
\eeq
From (\ref{eqn:prlm21b}) and (\ref{eqn:prlm21d}), we obtain
\beq
\left.
\ba{lll}
\lambda_0^{-1}&\leq &
\frac{1}{\sigma_{X_0}^2}+ \frac{1}{\sigma_{Z_1}^2}
\left(1- \frac{\lambda_1}{\sigma_{Z_1}^2 }\right)
\,,
\\
\lambda_l^{-1}&\leq &
\ts \frac{1}{\sigma_{Z_l}^2}
  + \frac{1}{\sigma_{N_l}^2}
  + \frac{1}{\sigma_{Z_{l+1}}^2}-\mu_l
\,,
1\leq l\leq L-1.
\ea
\right\}
\label{eqn:prlmbb}
\eeq
On the other hand, we note that for each 
$1 \leq l \leq L-1$, the five random variables 
$W_l$, ${\vc X}_l$, ${\vc Y}_l$, 
${\vc Y}_{l+1}$, and $W_{l+1}^L$ form a Markov chain 
$ 
W_l\to {\vc X}_l\to {\vc Y}_l\to {\vc Y}_{l+1} \to W_{l+1}^L
$
in this order. Then, applying Lemma \ref{lm:lm5b} to 
(\ref{eqn:prlm00}), we obtain
\beq
\mu_l 
 \geq 
\ts \frac{1}{\sigma_{N_l}^2}{\rm e}^{-2r_l}
 + \frac{1}{\sigma_{Z_{l+1}}^4}\lambda_{l+1}\,,
1\leq l\leq L-1\,.
\label{eqn:prlmaa}
\eeq
Combining (\ref{eqn:prlmbb}) and (\ref{eqn:prlmaa}),
we obtain for $1\leq l\leq L-1$,
\beqa
\lambda_l^{-1} 
& \leq &
\ts \frac{1}{ \sigma_{Z_l}^2    }
+ \frac{1}{ \sigma_{N_l}^2 }(1-{\rm e}^{-2r_{l}})
+ \frac{1}{ \sigma_{Z_{l+1}}^2}
\left(1- \frac{\lambda_{l+1}}{\sigma_{Z_{l+1}}^2} 
\right) \,.
\label{eqn:za}
\eeqa
Set 
$
\nu_0\defeq \ts \lambda_0^{-1}-\frac{1}{\sigma_{X_0}^2},
$
$
\nu_l\defeq \ts \lambda_l^{-1}-\frac{1}{\sigma_{Z_l}^2}\,,
1\leq l\leq L-1\,. 
$
Then, we have 
\beqno
I({\vc X}_0,W^L)&=&
\frac{n}{2}\ts\log(1+\sigma_{X_0}^2\nu_0),
\\
I({\vc Y}_l,W_l^L|{\vc Y}_{l-1})
&=& 
\frac{n}{2}\ts \log(1+\sigma_{Z_l}^2\nu_l)
\,,1\leq l\leq L-1,
\\
I({\vc Y}_L,W_L|{\vc Y}_{L-1})
&=&\frac{n}{2}\ts \log(1+\sigma_{Z_L}^2\nu_L)
=nr_L
\eeqno
Note that $\nu_l, 0\leq l\leq L-1$ are nonnegative. 
From (\ref{eqn:prlmbb}) and (\ref{eqn:za}), 
$\{\nu_l\}_{l=0}^{L}$ satisfies 
the following recursion:
\beqa
\hspace*{-3mm}\nu_{L}&=&
\ts \frac{1}{\sigma_{Z_{L}}^2}\left({\rm e}^{2r_{L}}-1\right)\,,
\label{eqn:recur1}\\
\hspace*{-3mm}\nu_{L-1}&\leq&
    \ts \frac{\nu_{L}}{1+ \sigma_{Z_{L}}^2 \nu_{L}}
   +\frac{1-{\rm e}^{-2r_{L-1}}}{\sigma_{N_{L-1}}^2}
\nonumber\\
\hspace*{-3mm}&=& 
   \ts \frac{1-{\rm e}^{-2r_L}}{\sigma_{N_L}^2}
       +\frac{1-{\rm e}^{-2r_{L-1}}}{\sigma_{N_{L-1}}^2}
\label{eqn:recur1b}\\
\hspace*{-3mm}\nu_{l}&\leq&
    \ts \frac{\nu_{l+1}}{1+ \sigma_{Z_{l+1}}^2 \nu_{l+1}}
   +\frac{1-{\rm e}^{-2r_{l}}}{\sigma_{N_{l}}^2}\,,
   L-2 \geq l \geq 1\,,
\label{eqn:recur2}\\
\hspace*{-3mm}\nu_{0}&\leq&\ts 
              \frac{\nu_{1}}{1+ \sigma_{Z_{1}}^2 \nu_{1}}\,,
              \nu_{0}=\frac{{\rm e}^{-2r_0}}{\xi}-\frac{1}{\sigma_{X_0}^2}\,.
\label{eqn:recur3}
\eeqa
From (\ref{eqn:recur1})-(\ref{eqn:recur3}), we obtain the upper bounds
of $I({\vc X}_0;W^L)$ and $I({\vc Y}_l;$ $W_l^L|{\vc Y}_{l-1}),$
$1\leq l\leq L-1$ in Lemma \ref{lm:corelm}. 
On the other hand, from (\ref{eqn:recur2}), (\ref{eqn:recur3}), 
and the nonnegative property of $\nu_l, 0\leq l\leq L-1$, 
we have 
\beqa
\hspace*{-3mm}\nu_{0}&=&\ts \left[
              \frac{{\rm e}^{-2r_0}}{\xi}-\frac{1}{\sigma_{X_0}^2}
              \right]^{+}\,, \nu_{1}
\geq \frac{\nu_0}{1-\sigma_{Z_{1}}^2\nu_0}\,,
\label{eqn:recur10}\\
\hspace*{-3mm}\nu_{l+1}&\geq& \ts
   \frac{
    \left[
   \nu_{l}-\frac{1}{\sigma_{N_{l}}^2}\left(1-{\rm e}^{-2r_{l}}\right)
    \right]^{+}}
   {1-\sigma_{Z_{l+1}}^2
    \left[
    \nu_{l}-
    \frac{1}{\sigma_{N_{l}}^2}\left(1-{\rm e}^{-2r_{l}}\right)
    \right]^{+}}\,,
    1 \leq l \leq L-1\,.
\label{eqn:recur11}
\eeqa
From (\ref{eqn:recur10}) and (\ref{eqn:recur11}), we obtain 
the lower bound of $I({\vc Y}_l;$ $W_l^L |{\vc Y}_{l-1}), $ 
$1\leq l\leq L-1$ in Lemma \ref{lm:corelm}. 
$\quad$\hfill \IEEEQED

\subsection{
Proof of Lemma \ref{lm:convlm0}
}

Let $\alpha_2^L$, $\beta_2^L \in {\cal A}_L(\alpha_1)$. Let 
$t\in [0,1]$ and $\bar{t}=1-t$.
Then, we have the following chain of inequalities:  
\begin{align*}
& t(-2)\zeta^{(l)}(\alpha_2^l)+\bar{t}(-2)\zeta^{(l)}(\beta_2^L)
\nonumber\\
&=\sum_{i=1}^{l-1}
         \left\{t\log 
         \left(1-\frac{\alpha_i}{i-\epsilon_i\alpha_i}
         +\frac{\alpha_{i+1}}{\tau_{i+1}}\right)\right. 
\nonumber\\
&\quad\qquad        +\bar{t}\log 
         \left(1-\frac{\beta_i}{1-\epsilon_i\beta_i}
         +\frac{\beta_{i+1}}{\tau_{i+1}}\right)\huger
\nonumber\\
&\quad  +\sum_{i=1}^{l-1}
     \left\{     t\log \left(1-\epsilon_i\alpha_i \right)
            +\bar{t}\log \left(1-\epsilon_i \beta_i \right)
     \right\}
\nonumber\\
& \quad       +t\log\left(1-\frac{\alpha_l}{1-\epsilon_l\alpha_l}\right)
    +\bar{t}\log\left(1-\frac{ \beta_l}{1- \epsilon_l\beta_l}\right)
\nonumber\\
&\MLeq{a} \sum_{i=1}^{l-1}
         \log 
         \left( 1-t\frac{\alpha_i}{1-\epsilon_i\alpha_i}
         +t\frac{\alpha_{i+1}}{\tau_{i+1}}
          -\bar{t}\frac{\beta_i}{1-\epsilon_i \beta_i}
          +\bar{t}\frac{\beta_{i+1}}{\tau_{i+1}}\right)
\nonumber\\
&\quad  +\sum_{i=1}^{l-1}
    \log \left(
          1-\epsilon_i[t\alpha_i+\bar{t}\beta_i] 
         \right)
\nonumber\\
& \quad 
+\log  \left(1
  - \frac{t \alpha_l } {1-\epsilon_l\alpha_l}
   -\frac{\bar{t}\beta_l}{1-\epsilon_l\beta_l}
           \right)
\nonumber\\
&\MLeq{b}\sum_{i=1}^{l-1}
         \log 
         \left( 1-\frac{t\alpha_i+\bar{t}\beta_i}
           {1-\epsilon_i[t\alpha_i+\bar{t}\beta_i] }
         \right. 
+\frac{t\alpha_{i+1}+\bar{t}\beta_{i+1}}
{\tau_{i+1}}\hugecr
\nonumber\\
&\quad  +\sum_{i=1}^{l-1}
    \log \left\{ 1-\epsilon_i[t \alpha_i
                    +\bar{t}\beta_i] \right\}
\nonumber\\
&\quad  
+\log  \left(1
  - \frac{t\alpha_l+\bar{t}\beta_l}
     {1-\epsilon_l[t\alpha_l+\bar{t}\beta_l]}
          \right)
= (-2)\zeta \left( t\alpha_2^l+\bar{t}\beta_2^l
            \right)\,.
\end{align*}
Step (a) follows from  the strict concavity of the logarithm function.  
Step (b) follows from  the strict concavity of $\frac{-a}{1-\epsilon a}$ 
for $a>0$.  
\hfill\IEEEQED

\subsection{
Proof of Lemma \ref{lm:prlmzsz1}
}

{\it Proof of Lemma \ref{lm:prlmzsz1} part a):} 
For the proof we use the following inequality:
\beqa
\frac{1+a}{1+\epsilon(1+a)}
    -\frac{a}{1+\epsilon a}
&\leq&\frac{1}{1+\epsilon}\,.
\label{eqn:zsza00}
\eeqa
The recursion (\ref{eqn:RecurEqbb}) is equivalent to
\beq
\frac{\tau_i\theta_{i-1}^{(l)}(\omega)}
{1-\epsilon_{i-1}\theta_{i-1}^{(l)}(\omega)}
=2\theta_{i}^{(l)}(\omega)-
\frac{1+\frac{\theta_{i+1}^{(l)}(\omega)}{\tau_{i+1}}}
{1+\epsilon_{i}
\left[1+\frac{\theta_{i+1}^{(l)}(\omega)}{\tau_{i+1}}\right]}+\tau_i
\label{eqn:zsza}
\eeq
for $l-1\geq i \geq 2$. Applying (\ref{eqn:zsza00}) 
to the second term in the right members of (\ref{eqn:zsza}) 
and considering the assumption $\tau_l\geq \frac{1}{1+\epsilon_l}$ 
for $L-1\geq l\geq 2$, we have 
\beqno
\frac{\tau_i\theta_{i-1}^{(l)}(\omega)}{1-\epsilon_{i-1}
\theta_{i-1}^{(l)}(\omega)}
&\geq &
2\theta_i^{(l)}(\omega)-\frac{\frac{\theta_{i+1}^{(l)}(\omega)}{\tau_{i+1}}}
{1+\epsilon_{i} \frac{\theta_{i+1}^{(l)}(\omega)}{\tau_{i+1}}}
\eeqno
or equivalent to
\beq
\frac{\tau_i\theta_{i-1}^{(l)}(\omega) }
{1+\epsilon_{i-1}\theta_{i-1}^{(l)}(\omega)}
-\theta_{i}^{(l)}(\omega)
\geq
\theta_{i}^{(l)}(\omega)
-\frac{\frac{\theta_{i+1}^{(l)}(\omega)}{\tau_{i+1}} }
{1+\epsilon_{i}\frac{\theta_{i+1}^{(l)}(\omega)}{\tau_{i+1}}}
\label{eqn:abba}
\eeq
for $l-1\geq i\geq 2$. We first prove (\ref{eqn:cond0}) for $i=l$.
The equality
\begin{align}
& \theta_{l-1}^{(l)}(\omega)
\nonumber\\
& =
\frac{
     \frac{ \theta_l^{(l)}(\omega)
     +\{(1+\epsilon_l)\theta_l^{(l)}(\omega)\}
      \{1-\epsilon_l\theta_l^{(l)}(\omega)\}
     }{\tau_l}+1 }
{1+\epsilon_{l-1}
     \left[
      \frac{ \theta_l^{(l)}(\omega)
            +\{(1+\epsilon_l)\theta_l^{(l)}(\omega)\}
           \{1-\epsilon_l\theta_l^{(l)}(\omega)\}}
          {\tau_l}+1
    \right]}     
\label{eqn:zzz0}
\end{align}
is equivalent to the following two equalities:
\begin{align}
& \tau_l
 \left(
\frac{\theta_{l-1}^{(l)}(\omega)}
{1-\epsilon_{l-1}\theta_{l-1^{(l)}}(\omega)}
-1\right)
-\theta_{l}^{(l)}(\omega)
\nonumber\\
& =\{(1+\epsilon_l)\theta_l^{(l)}(\omega)\}
           \{1-\epsilon_l\theta_l^{(l)}(\omega)\}
 \label{eqn:z01}\\
& =\theta_{l}^{(l)}(\omega)-1
+ \epsilon_l\theta_l^{(l)}(\omega)
\{2-(1+\epsilon_l)\theta_l^{(l)}(\omega)\}
\label{eqn:z02}
\end{align}
From (\ref{eqn:z01}), we have
\begin{align}
&\tau_l
\left(
\frac{\theta_{l-1}^{(l)}(\omega)}
{1-\epsilon_{l-1}\theta_{l-1}^{(l)}(\omega)}
-1\right)
-\theta_{l}^{(l)}(\omega)
\nonumber\\
&=\{(1+\epsilon_l)\theta_l^{(l)}(\omega)\}
           \{1-\epsilon_l\theta_l^{(l)}(\omega)\}
           \MsL{a}0.
\label{eqn:z03}
\end{align}
Step (a) follows from the original assumption 
$\theta_l^{(l)}(\omega)\in (0,(1+\epsilon_l)^{-1})$.
From (\ref{eqn:z02}), we have 
\begin{align}
&\frac{\tau_l\theta_{l-1}^{(l)}(\omega)}
{1-\epsilon_{l-1}\theta_{l-1}^{(l)}(\omega)}
-\theta_{l}^{(l)}(\omega)
\nonumber\\
& =\tau_l-1+\theta_{l}^{(l)}(\omega)
+ \epsilon_l\theta_l^{(l)}(\omega)
\{2-(1+\epsilon_l)\theta_l^{(l)}(\omega)\}
\nonumber\\
& \geq \tau_l-1 \MGeq{b}0.
\label{eqn:z04}
\end{align}
Step (b) follows from the assumption $\tau_l\geq 1$.
From (\ref{eqn:z03}) and (\ref{eqn:z04}), we have 
\beqa
& &
\left.
\ba{l}
0\leq \theta_l^{(l)}(\omega)
\leq 
\frac{\tau_l\theta_{l-1}^{(l)}(\omega)}
{1-\epsilon_{l-1}\theta_{l-1}^{(l)}(\omega)}\,,
\vspace{1mm}\\
\tau_l\left(\frac{\theta_{l-1}^{(l)}(\omega)}
{1-\epsilon_{l-1}\theta_{l-1}^{(l)}(\omega)}-1
\right)
<\theta_{l}^{(l)}(\omega)<(1+\epsilon_{l})^{-1}
\ea
\right\} 
\label{eqn:cond2} 
\eeqa
Thus, (\ref{eqn:condl}) holds for $i=l$. 
We next assume that (\ref{eqn:condl}) holds for 
some $i+1$ with $l\geq i+1$, that is,  
\beqa
\left.
\ba{l}
0\leq \theta_{i+1}^{(l)}(\omega)
\leq 
\frac{\tau_{i+1}\theta_{i}^{(l)}(\omega)}
{1-\epsilon_{ i }\theta_{i}^{(l)}(\omega)}\,,
\vspace*{1mm}\\
\tau_{i+1}
\left(\frac{\theta_{i}^{(l)}(\omega)}
{1-\epsilon_{ i }\theta_{i}^{(l)}(\omega)}-1\right)
< \theta_{i+1}^{(l)}(\omega)
<\epsilon_{i+1}^{-1} \,.
\ea
\right\}
\label{eqn:lzxz0}
\eeqa
Then, from (\ref{eqn:lzxz0}), we obtain 
\beq
\left.
\ba{l}
\epsilon_{ i }^{-1}>
\theta_{i}^{(l)}(\omega)
\geq \frac{\frac{\theta_{i+1}^{(l)}(\omega)}{\tau_{i+1}}}
{1+\epsilon_{i}\frac{\theta_{i+1}^{(l)}(\omega)}{\tau_{i+1}}}>0\,,
\\
\theta_{i}^{(l)}(\omega)
< \frac{1+\frac{\theta_{i+1}^{(l)}(\omega)}{\tau_{i+1}}}
{1+\epsilon_{ i }\left(
1+\frac{\theta_{i+1}^{(l)}(\omega)}{\tau_{i+1}} \right)}\,.
\ea
\right\}
\label{eqn:IeQforlzxz2z}
\eeq
Using (\ref{eqn:zsza}), we have 
\beqa
& &
\frac{\tau_i\theta_{i-1}^{(l)}(\omega)}{1-\epsilon_{i-1}\theta_{i-1}^{(l)}(\omega)}
- \theta_{i}^{(l)}(\omega)
\nonumber\\
&=&
\theta_{i}^{(l)}(\omega)-
\frac{1+\frac{\theta_{i+1}^{(l)}(\omega)}{\tau_{i+1}}}
{1+\epsilon_{i}\left(
1+\frac{\theta_{i+1}^{(l)}(\omega)}{\tau_{i+1}}\right)}+\tau_i
\MsL{a}\tau_i\,.
\nonumber
\eeqa
Step (a) follows from the second inequality of
(\ref{eqn:IeQforlzxz2z}). Using (\ref{eqn:abba}), we have  
\beqa
\frac{\tau_i\theta_{i-1}^{(l)}(\omega)}{1-\epsilon_{i-1}\theta_{i-1}^{(l)}(\omega)}
-\theta_{i}^{(l)}(\omega)
&\geq&
\theta_{i}^{(l)}(\omega)
-\frac{ \frac{\theta_{i+1}^{(l)}(\omega)}{\tau_{i+1}}}
 {1+\epsilon_{ i }\frac{\theta_{i+1}^{(l)}(\omega)}{\tau_{i+1}}}
\MGeq{a}0\,.
\nonumber
\eeqa
Step (a) follows from the first inequality 
of (\ref{eqn:IeQforlzxz2z}). 
Thus, (\ref{eqn:condl}) holds for $i$, completing the proof.  
\hfill\IEEEQED 

{\it Proof of Lemma \ref{lm:prlmzsz1} part b):} 
We first observe that
\begin{align}
& 
(-2)\zeta^{(l)}(\alpha_2^l)
\nonumber\\
&=\sum_{i=1}^{l-1}
         \left\{\log 
         \left(1-\frac{\alpha_i}{1-\epsilon_{ i }\alpha_i}
         +\frac{\alpha_{i+1}}{\tau_{i+1}}\right)
         +\log\left(1-\epsilon_{i}\alpha_i\right)
         \right\} 
\nonumber\\         
&\quad+\log\left(1-\frac{\alpha_l}{1-\epsilon_l\alpha_l}\right)   
\nonumber\\
&=\sum_{i=2}^{l}
   \log 
    \left\{
1-\epsilon_{i-1}\alpha_{i-1}-\alpha_{i-1}
  +(1-\epsilon_{i-1}\alpha_{i-1})
   \frac{\alpha_{i}}{\tau_i}
    \right\}
   \nonumber\\         
&\quad+\log\left(1-\frac{\alpha_l}{1-\epsilon_l\alpha_l}\right)   
\nonumber\\
&=\sum_{i=1}^{l-1}
       \log \left[
        1+\frac{\alpha_{i+1}}{\tau_{i+1}}
         -\left\{1+\epsilon_{ i }
\left(1+\frac{\alpha_{i+1}}{\tau_{i+1}}\right)\right\}\alpha_i
        \right]
     \nonumber\\         
&\quad+\log\left(1-\frac{\alpha_l}{1-\epsilon_l\alpha_l}\right).       
\end{align}
Computing $(-2)\frac{\partial}{\partial \alpha_i}
\zeta^{(l)}(\alpha_2^l)$, we obtain
\beq
\left.
\ba{lcl}
(-2)\frac{\partial}{\partial \alpha_l}\zeta^{(l)}(\alpha_2^l)
&=&\frac{1}{\alpha_l- \tau_{l}
\left(\frac{\alpha_{l-1}}{1-\epsilon_{l-1}\alpha_{l-1}}-1\right)}
\vspace{1mm}\\
& &\qquad-\frac{1}{\{1-(1+\epsilon_l)\alpha_l\}(1-\epsilon_{l}\alpha_l)}\,,
\vspace{1mm}\\
(-2)\frac{\partial}{\partial \alpha_i}\zeta^{(l)}(\alpha_2^l)
&=&
\frac{1}{\alpha_i-\tau_i
\left(\frac{\alpha_{i-1}}{1-\epsilon_{i-1} \alpha_{i-1}}-1\right)}      \vspace{1mm}\\
& & \qquad   -\frac{1}{
         \frac{1+\frac{\alpha_{i+1}}{\tau_{i+1}}}
              {1+\epsilon_{i}\left[1+\frac{\alpha_{i+1}}{\tau_{i+1}}\right]}
          -\alpha_i
          } 
\vspace{1mm}\\
& &\mbox{ for }l-1\geq i \geq 2\,.
\ea
\right\}
\label{eqn:argmin0}
\eeq
From (\ref{eqn:argmin0}), when
$\nabla \zeta^{(l)}(\alpha_2^l)={\vc 0}$, $\alpha_2^l$ must satisfy 
\beq
\left.
\ba{l}
\alpha_l
-\tau_l\left(\frac{\alpha_{l-1}}{1-\epsilon_{l-1}\alpha_{l-1}}-1\right)
=\{1-(1+\epsilon_l)\alpha_l\}(1-\epsilon_{l}\alpha_l)\,,
\vspace{1mm}\\
\frac{1+\frac{\alpha_{i+1}}{\tau_{i+1}}}
{1+\epsilon_{i}
\left[1+\frac{\alpha_{i+1}}{\tau_{i+1}}
\right]
}
-2\alpha_i 
+\tau_i\left(\frac{\alpha_{i-1}}
{1-\epsilon_{i-1} \alpha_{i-1}}-1\right)=0\,,
\vspace{1mm}\\
\mbox{ for }l-1\geq i \geq 2\,.
\ea
\right\}
\label{eqn:argmin3}
\eeq
From (\ref{eqn:argmin3}), we obtain  
\beq
\left.
\ba{l}
\alpha_{l-1}=\ds
\frac{ 
\frac{\alpha_l+\{1-(1+\epsilon_l)\alpha_l\}(1-\epsilon_{l}\alpha_l)}
{\tau_l}+1     }
{1+\epsilon_{l-1}
\left[
\frac{\alpha_l+\{1-(1+\epsilon_l)\alpha_l\}(1-\epsilon_{l}\alpha_l)}
{\tau_l}+1
\right] 
}
\vspace{1mm}\\
\alpha_{i-1}=\ds
\frac{
      \frac{1}{\tau_i}
       \left[
       2\alpha_{i}
       -\frac{1+\frac{\alpha_{i+1}}{\tau_{i+1}}}
             {1+\epsilon_{i}
        \left(1+\frac{\alpha_{i+1}}{\tau_{i+1}}\right)
             }
       +\tau_i \right]
     } 
     {
     1+\frac{\epsilon_{i-1}}{\tau_i}
       \left[
       2\alpha_{i}
       -\frac{1+\frac{\alpha_{i+1}}{\tau_{i+1}}}
             {1+\epsilon_{i}
              \left(1+\frac{\alpha_{i+1}}{\tau_{i+1}}\right)
             }
      +\tau_i
      \right]
     }
\vspace{1mm}\\
\mbox{ for }l-1 \geq i \geq 2\,.
\ea
\right\} 
\label{eqn:SumEx0077}
\eeq
The relation (\ref{eqn:SumEx0077}) implies that
$
\left. \nabla \zeta^{(l)}\right|_{\alpha_2^L
=(\theta_i^{(l)}(\omega))_{i=2}^l}
={\vc 0}$ 
\hfill\IEEEQED 

{\it Proof of Lemma \ref{lm:prlmzsz1} part c):} 
For the proof we use the following recursion 
for $l\geq i\geq 2$:  
\beq
\frac{\tau_i\theta_{i-1}^{(l)}(\omega) }
{1-\epsilon_{i-1}\theta_{i-1}^{(l)}(\omega)}
=2\theta_i^{(l)}(\omega)-
\frac{1+\frac{\theta_{i+1}^{(l)}(\omega)}{\tau_{i+1}}}
{1+\epsilon_{i}\left[1+
    \frac{\theta_{i+1}^{(l)}(\omega) } {\tau_{i+1}}
\right] }\,.
\label{eqn:zsza000}
\eeq
Taking the derivative of both sides of 
(\ref{eqn:zsza000}) with respect to $\omega$, we obtain
\begin{align}
& \frac{1}{\left\{1-\epsilon_{l-1} \theta_{i-1}^{(l)}(\omega)\right\}^2}
\frac{{\rm d}\theta_{i-1}^{(l)}}{{\rm d}\omega}\cdot\tau_i
\nonumber\\
&=2\frac{{\rm d}\theta_i^{(l)}}{{\rm d}\omega}
-\frac{1}{\left\{1+\epsilon_{l}
\left[\frac{\theta_{i+1}^{(l)}(\omega)}{\tau_{i+1}}+1\right]
\right\}^2}
 \frac{{\rm d}\theta_{i+1}^{(l)}}{{\rm d}\omega}\cdot\tau_{i+1}^{-1}\,.
\label{eqn:zaa0}
\end{align}
Since 
$\theta_{2}^l(\omega)\in {\cal A}_l\left(\theta_1^{(l)}(\omega)\right)$, 
we have
$$
\tau_i\left(\frac{\theta_{i-1}^{(l)}(\omega)}{1-\epsilon_{i-1} 
\theta_{i-1}^{(l)}(\omega)}-1\right) <\theta_i^{(l)}(\omega)\,.
$$
The above inequality is equivalent to 
\beq
{1+\epsilon_{ i-1}\left(\frac{\theta_{i}^{(l)}(\omega)}{\tau_i}+1\right)} 
>\frac{1}{1-\epsilon_{i-1} 
\theta_{i-1}^{(l)}(\omega)}\,. 
\label{eqn:zaa1}
\eeq
From (\ref{eqn:zaa0}) and (\ref{eqn:zaa1}) 
we have 
\begin{align}
& \left\{1+\epsilon_{i-1}
\left(\frac{\theta_{i}^{(l)}(\omega)}{\tau_i}+1\right)\right\}^2
\frac{{\rm d}\theta_{i-1}^{(l)}}{{\rm d}\omega}\cdot \tau_l
\nonumber\\
&\geq2\frac{{\rm d}\theta_{i}^{(l)}}{{\rm d}\omega}
-\frac{1}{\left\{1+\epsilon_{i}\left(
\frac{\theta_{i+1}^{(l)}(\omega)}{\tau_{l+1}}+1\right)\right\}^2}
 \frac{{\rm d}\theta_{i+1}^{(l)} }{{\rm d}\omega}\cdot \tau_{i+1}^{-1}\,.
\label{eqn:zaa0-2}
\end{align}
The above inequality is equivalent to 
\begin{align}
& \left\{1+\epsilon_{i-1}
\left(\frac{\theta_{i}^{(l)}(\omega)}{\tau_i}+1\right)\right\}^2
\left(\frac{1}{\sigma_{i-1}^2}
\frac{{\rm d}\theta_{i-1}^{(l)}}{{\rm d}\omega}\right)
\nonumber\\
&\geq 
2\left(
\frac{1}{\sigma_i^2}\frac{{\rm d}\theta_{i}^{(l)} }{{\rm d}\omega}
\right)
-\frac{1}{\left\{1+\epsilon_{i}\left(
\frac{\theta_{i+1}^{(l)}(\omega)}{\tau_{i+1}}+1\right)\right\}^2}
\nonumber\\
& \qquad\qquad\times 
\left(\frac{1}{\sigma_{i+1}^2}
\frac{{\rm d}\theta_{i+1}^{(l)}}{{\rm d}\omega}\right).
\label{eqn:zaa2}
\end{align}
For $l\geq i\geq 1$, set
\begin{align*}
&\Phi_i^{(l)}(\omega) 
\\
& \defeq 
\left\{
\ba{l}
\ds \left(\frac{1}{\sigma_{i}^2}
\frac{{\rm d}\theta_{i}^{(l)}}{{\rm d}\omega}\right)
\prod_{j=2}^{i}{
\ts \frac{1}{\left\{1+\epsilon_{j-1}
\left(\frac{\theta_{j}^{(l)}(\omega)}{\tau_j}+1\right)\right\}^2}},\:
l\geq i\geq 2,
\vspace{1mm}\\
\ds  \frac{1}{\sigma_{1}^2}
\frac{{\rm d}\theta_{1}^{(l)}}{{\rm d}\omega},\: i=1.
\ea
\right.
\end{align*}
Then, by (\ref{eqn:zaa2}), we have
\beq
\Phi_{i-1}^{(l)}(\omega)\geq 
2\Phi_{i}^{(l)}(\omega)-\Phi_{i+1}^{(l)}(\omega)\,\mbox{ for } 
l-1\geq i \geq 2\,.
\label{eqn:zaa3}
\eeq
From (\ref{eqn:zaa3}) we have 
\begin{align}
&    \Phi_{i-1}^{(l)}(\omega)-\Phi_{i}^{(l)}(\omega)\geq 
\Phi_{i}^{(l)}(\omega)-\Phi_{i+1}^{(l)}(\omega)
\nonumber\\
&\geq \Phi_{l-1}^{(l)}(\omega)-\Phi_{l}^{(l)}(\omega)
\nonumber\\
& =
      \left[
        \tau_l 
        \cdot\frac{{\rm d}\theta_{l-1}^{(l)}}{{\rm d}\omega}
        - {\ts
          \frac{1}{\left\{1+\epsilon_{l-1}\left(\frac{\omega} 
          {\tau_l}+1\right)\right\}^2}}
         \right]\frac{1}{\sigma_l^2}
\nonumber\\
& \quad\: \times 
\prod_{j=2}^{l-1}{
\ts \frac{1}{\left\{1+\epsilon_{j-1}
\left(\frac{\theta_{j}(\omega)}{\tau_j}+1\right)\right\}^2}}
\nonumber\\
& \MEq{a} \left[{\ts
       \frac{2(1+\epsilon_l)(1-\epsilon_l\omega))}
       { \left\{1+\epsilon_{l-1}
           \left[\frac{\omega+\{(1+\epsilon_l)\omega-1\}
                      (1-\epsilon_{l}\omega)}
                      {\tau_l}+1
           \right]
        \right\}^2}  }
   \right.
\nonumber\\
&\:\quad \left.{\ts        
       -\frac{1}{\left\{1+\epsilon_{l-1}
       (\frac{\omega}{\tau_l}+1)\right\}^2}}
       \right]
\frac{1}{\sigma_l^2}\cdot 
\prod_{j=2}^{l-1}{
\ts\frac{1}{\left\{1+\epsilon_{j-1}
  \left(\frac{\theta_{j}(\omega)}{\tau_j}+1\right)\right\}^2}}
\nonumber\\
&\MGeq{b}{\ts        
       \frac{1}{\left\{1+\epsilon_{l-1}
       (\frac{\omega}{\tau_l}+1)\right\}^2}}
\frac{1}{\sigma_l^2}\cdot 
\prod_{j=2}^{l-1}{
\ts\frac{1}{\left\{1+\epsilon_{j-1}
  \left(\frac{\theta_{j}(\omega)}{\tau_j}+1\right)\right\}^2}}
\nonumber\\
&=\Phi_{l}^{(l)}(\omega).
\label{eqn:zaa4}
\end{align}
Step (a) follows from $\theta_l^{(l)}(\omega)=\omega$ and 
$$
\theta_{l-1}^{(l)}(\omega)=
{\ts \frac{   
           \epsilon_{l-1}
           \left[\frac{\omega+\{(1+\epsilon_l)\omega-1\}
                      (1-\epsilon_{l}\omega)}
                      {\tau_l}+1
           \right] }
       { 1+\epsilon_{l-1}
           \left[\frac{\omega+\{(1+\epsilon_l)\omega-1\}
                      (1-\epsilon_{l}\omega)}
                      {\tau_l}+1
           \right]}}.
$$
Step (b) follows from that for 
$\omega\in [0,(1+\epsilon)^{-1})$, we have 
\begin{align*}
& 2(1+\epsilon_l)(1-\epsilon_l\omega))>2, 
\\
& \omega+\{(1+\epsilon_l)\omega-1\}
                      (1-\epsilon_{l}\omega)<\omega.               
\end{align*}
By (\ref{eqn:zaa4}), we have 
\beqno
\Phi_{i}^{(l)}(\omega)&\geq& \Phi_{l}^{(l)}(\omega)
    +(l-i)\Phi_{l}^{(l)}(\omega)=(l-i+1)\Phi_{l}^{(l)}(\omega),
\eeqno
from which we obtain
\beqno
\frac{{\rm d}\theta_i^{(l)}}{{\rm d}\omega}
&\geq&
(l-i+1)\frac{\sigma_i^2}{\sigma_l^2}\cdot 
\prod_{j=i+1}^{l}{
\ts \frac{1}{
\left\{1+\epsilon_{j-1}
\left(\frac{\theta_{j}(\omega)}{\tau_j}+1\right)\right\}^2}
},
\eeqno
completing the proof.
\hfill\IEEEQED

{\it Proof of Lemma \ref{lm:prlmzsz1} part c):} 
For the proof we use the following recursion 
for $l\geq i\geq 2$:  
\beq
\frac{\tau_i\theta_{i-1}^{(l)}(\omega) }
{1-\epsilon_{i-1}\theta_{i-1}^{(l)}(\omega)}
=2\theta_i^{(l)}(\omega)-
\frac{1+\frac{\theta_{i+1}^{(l)}(\omega)}{\tau_{i+1}}}
{1+\epsilon_{i}\left[1+
    \frac{\theta_{i+1}^{(l)}(\omega) } {\tau_{i+1}}
\right] }\,.
\label{eqn:zsza000}
\eeq
Taking the derivative of both sides of 
(\ref{eqn:zsza000}) with respect to $\omega$, we obtain
\begin{align}
& \frac{1}{\left\{1-\epsilon_{l-1} \theta_{i-1}^{(l)}(\omega)\right\}^2}
\frac{{\rm d}\theta_{i-1}^{(l)}}{{\rm d}\omega}\cdot\tau_i
\nonumber\\
&=2\frac{{\rm d}\theta_i^{(l)}}{{\rm d}\omega}
-\frac{1}{\left\{1+\epsilon_{l}
\left[\frac{\theta_{i+1}^{(l)}(\omega)}{\tau_{i+1}}+1\right]
\right\}^2}
 \frac{{\rm d}\theta_{i+1}^{(l)}}{{\rm d}\omega}\cdot\tau_{i+1}^{-1}\,.
\label{eqn:zaa0}
\end{align}
Since 
$\theta_{2}^l(\omega)\in {\cal A}_l\left(\theta_1^{(l)}(\omega)\right)$, 
we have
$$
\tau_i\left(\frac{\theta_{i-1}^{(l)}(\omega)}{1-\epsilon_{i-1} 
\theta_{i-1}^{(l)}(\omega)}-1\right) <\theta_i^{(l)}(\omega)\,.
$$
The above inequality is equivalent to 
\beq
{1+\epsilon_{ i-1}\left(\frac{\theta_{i}^{(l)}(\omega)}{\tau_i}+1\right)} 
>\frac{1}{1-\epsilon_{i-1} 
\theta_{i-1}^{(l)}(\omega)}\,. 
\label{eqn:zaa1}
\eeq
From (\ref{eqn:zaa0}) and (\ref{eqn:zaa1}) 
we have 
\begin{align}
& \left\{1+\epsilon_{i-1}
\left(\frac{\theta_{i}^{(l)}(\omega)}{\tau_i}+1\right)\right\}^2
\frac{{\rm d}\theta_{i-1}^{(l)}}{{\rm d}\omega}\cdot \tau_l
\nonumber\\
&\geq2\frac{{\rm d}\theta_{i}^{(l)}}{{\rm d}\omega}
-\frac{1}{\left\{1+\epsilon_{i}\left(
\frac{\theta_{i+1}^{(l)}(\omega)}{\tau_{l+1}}+1\right)\right\}^2}
 \frac{{\rm d}\theta_{i+1}^{(l)} }{{\rm d}\omega}\cdot \tau_{i+1}^{-1}\,.
\label{eqn:zaa0-2}
\end{align}
The above inequality is equivalent to 
\begin{align}
& \left\{1+\epsilon_{i-1}
\left(\frac{\theta_{i}^{(l)}(\omega)}{\tau_i}+1\right)\right\}^2
\left(\frac{1}{\sigma_{i-1}^2}
\frac{{\rm d}\theta_{i-1}^{(l)}}{{\rm d}\omega}\right)
\nonumber\\
&\geq 
2\left(
\frac{1}{\sigma_i^2}\frac{{\rm d}\theta_{i}^{(l)} }{{\rm d}\omega}
\right)
-\frac{1}{\left\{1+\epsilon_{i}\left(
\frac{\theta_{i+1}^{(l)}(\omega)}{\tau_{i+1}}+1\right)\right\}^2}
\nonumber\\
& \qquad\qquad\times 
\left(\frac{1}{\sigma_{i+1}^2}
\frac{{\rm d}\theta_{i+1}^{(l)}}{{\rm d}\omega}\right).
\label{eqn:zaa2}
\end{align}
For $l\geq i\geq 1$, set
\begin{align*}
&\Phi_i^{(l)}(\omega) 
\\
& \defeq 
\left\{
\ba{l}
\ds \left(\frac{1}{\sigma_{i}^2}
\frac{{\rm d}\theta_{i}^{(l)}}{{\rm d}\omega}\right)
\prod_{j=2}^{i}{
\ts \frac{1}{\left\{1+\epsilon_{j-1}
\left(\frac{\theta_{j}^{(l)}(\omega)}{\tau_j}+1\right)\right\}^2}},\:
l\geq i\geq 2,
\vspace{1mm}\\
\ds  \frac{1}{\sigma_{1}^2}
\frac{{\rm d}\theta_{1}^{(l)}}{{\rm d}\omega},\: i=1.
\ea
\right.
\end{align*}
Then, by (\ref{eqn:zaa2}), we have
\beq
\Phi_{i-1}^{(l)}(\omega)\geq 
2\Phi_{i}^{(l)}(\omega)-\Phi_{i+1}^{(l)}(\omega)\,\mbox{ for } 
l-1\geq i \geq 2\,.
\label{eqn:zaa3}
\eeq
From (\ref{eqn:zaa3}) we have 
\begin{align}
&    \Phi_{i-1}^{(l)}(\omega)-\Phi_{i}^{(l)}(\omega)\geq 
\Phi_{i}^{(l)}(\omega)-\Phi_{i+1}^{(l)}(\omega)
\nonumber\\
&\geq \Phi_{l-1}^{(l)}(\omega)-\Phi_{l}^{(l)}(\omega)
\nonumber\\
& =
      \left[
        \tau_l 
        \cdot\frac{{\rm d}\theta_{l-1}^{(l)}}{{\rm d}\omega}
        - {\ts
          \frac{1}{\left\{1+\epsilon_{l-1}\left(\frac{\omega} 
          {\tau_l}+1\right)\right\}^2}}
         \right]\frac{1}{\sigma_l^2}
\nonumber\\
& \quad\: \times 
\prod_{j=2}^{l-1}{
\ts \frac{1}{\left\{1+\epsilon_{j-1}
\left(\frac{\theta_{j}(\omega)}{\tau_j}+1\right)\right\}^2}}
\nonumber\\
& \MEq{a} \left[{\ts
       \frac{2(1+\epsilon_l)(1-\epsilon_l\omega))}
       { \left\{1+\epsilon_{l-1}
           \left[\frac{\omega+\{(1+\epsilon_l)\omega-1\}
                      (1-\epsilon_{l}\omega)}
                      {\tau_l}+1
           \right]
        \right\}^2}  }
   \right.
\nonumber\\
&\:\quad \left.{\ts        
       -\frac{1}{\left\{1+\epsilon_{l-1}
       (\frac{\omega}{\tau_l}+1)\right\}^2}}
       \right]
\frac{1}{\sigma_l^2}\cdot 
\prod_{j=2}^{l-1}{
\ts\frac{1}{\left\{1+\epsilon_{j-1}
  \left(\frac{\theta_{j}(\omega)}{\tau_j}+1\right)\right\}^2}}
\nonumber\\
&\MGeq{b}{\ts        
       \frac{1}{\left\{1+\epsilon_{l-1}
       (\frac{\omega}{\tau_l}+1)\right\}^2}}
\frac{1}{\sigma_l^2}\cdot 
\prod_{j=2}^{l-1}{
\ts\frac{1}{\left\{1+\epsilon_{j-1}
  \left(\frac{\theta_{j}(\omega)}{\tau_j}+1\right)\right\}^2}}
\nonumber\\
&=\Phi_{l}^{(l)}(\omega).
\label{eqn:zaa4}
\end{align}
Step (a) follows from $\theta_l^{(l)}(\omega)=\omega$ and 
$$
\theta_{l-1}^{(l)}(\omega)=
{\ts \frac{   
           \epsilon_{l-1}
           \left[\frac{\omega+\{(1+\epsilon_l)\omega-1\}
                      (1-\epsilon_{l}\omega)}
                      {\tau_l}+1
           \right] }
       { 1+\epsilon_{l-1}
           \left[\frac{\omega+\{(1+\epsilon_l)\omega-1\}
                      (1-\epsilon_{l}\omega)}
                      {\tau_l}+1
           \right]}}.
$$
Step (b) follows from that for 
$\omega\in [0,(1+\epsilon)^{-1})$, we have 
\begin{align*}
& 2(1+\epsilon_l)(1-\epsilon_l\omega))>2, 
\\
& \omega+\{(1+\epsilon_l)\omega-1\}
                      (1-\epsilon_{l}\omega)<\omega.               
\end{align*}
By (\ref{eqn:zaa4}), we have 
\beqno
\Phi_{i}^{(l)}(\omega)&\geq& \Phi_{l}^{(l)}(\omega)
    +(l-i)\Phi_{l}^{(l)}(\omega)=(l-i+1)\Phi_{l}^{(l)}(\omega),
\eeqno
from which we obtain
\beqno
\frac{{\rm d}\theta_i^{(l)}}{{\rm d}\omega}
&\geq&
(l-i+1)\frac{\sigma_i^2}{\sigma_l^2}\cdot 
\prod_{j=i+1}^{l}{
\ts \frac{1}{
\left\{1+\epsilon_{j-1}
\left(\frac{\theta_{j}(\omega)}{\tau_j}+1\right)\right\}^2}
},
\eeqno
completing the proof.
\hfill\IEEEQED

\end{document}